%% file: thesis.tex
\documentclass[12pt,oneside,onecolumn]{report}
\input{texfiles/header}
\input{texfiles/tables.tex}             
\begin{document}
\input{texfiles/title}
\pagenumbering{roman} 
\mbox{}
\input{texfiles/ackn}

\mbox{}
\tableofcontents
\mbox{}
\listoffigures
\addcontentsline{toc}{chapter}{List of Figures}
\mbox{}
\listoftables
\addcontentsline{toc}{chapter}{List of Tables}
\mbox{}
\newpage
\bc
{\Large \underline {This thesis is partially based on the following publications}}
\ec
\input{texfiles/pub}
\vspace*{10cm}
\pagenumbering{arabic}
\chapter{ Introduction}
\input{texfiles/intro}

\chapter{ Probing Warped Extra dimension at the LHC }
\input{texfiles/hggc}

\chapter{ Extra-dimensional relaxation of the upper limit 
of the lightest supersymmetric neutral Higgs mass}
\input{texfiles/rshg}

\newpage
\chapter{ A phenomenological study of 5d supersymmetry}
\input{texfiles/rg5d}

\newpage
\chapter{Summary and Conclusions}
\input{texfiles/con}

\newpage

\end{document}

%% file: texfiles/header
\usepackage{graphicx,bm,floatflt,amssymb,epsf,psfig,rotate,slashed,axodraw} 
\usepackage{amsfonts,makeidx,latexsym,setspace}
\usepackage{graphics,subfigure} 
\usepackage{times,cite,color}
\newsavebox{\fmbox}                       

\def\l{\lambda}
\def\L{\Lambda}

\def\g{\gamma}

\def\D{\Delta}

\def\a{\alpha}

\def\e{\epsilon}

\def\t{\theta}

\def\ra{\rightarrow}

\def\bc{\begin{center}}
\def\ec{\end{center}}
\def\be{\begin{equation}}
\def\ee{\end{equation}}
\def\bi{\begin{itemize}}
\def\ei{\end{itemize}}

\newcommand{\br}{\begin{array}}
\newcommand{\er}{\end{array}}
\newcommand{\ba}{\begin{eqnarray}}
\newcommand{\ea}{\end{eqnarray}}

\newcommand{\CL}{{\cal L}}

\def\bpmat{\begin{pmatrix}}
\def\epmat{\end{pmatrix}}

\def\ra{\rightarrow}
\def\stilde{\widetilde}

\def\lim{\raisebox{-.9ex}{\rlap{\tiny $\e \rightarrow {\tiny 0}$}} \raisebox{.9ex}{lim}}

\def\ltap{\raisebox{-.4ex}{\rlap{$\sim$}} \raisebox{.4ex}{$<$}} 
\def\gtap{\raisebox{-.4ex}{\rlap{$\sim$}} \raisebox{.4ex}{$>$}}

%% file: texfiles/tables.tex
\newbox\hdbox%
\newcount\hdrows%
\newcount\multispancount%
\newcount\ncase%
\newcount\ncols
\newcount\nrows%
\newcount\nspan%
\newcount\ntemp%
\newdimen\hdsize%
\newdimen\newhdsize%
\newdimen\parasize%
\newdimen\spreadwidth%
\newdimen\thicksize%
\newdimen\thinsize%
\newdimen\tablewidth%
\newif\ifcentertables%
\newif\ifendsize%
\newif\iffirstrow%
\newif\iftableinfo%
\newtoks\dbt%
\newtoks\hdtks%
\newtoks\savetks%
\newtoks\tableLETtokens%
\newtoks\tabletokens%
\newtoks\widthspec%
\immediate\write15{%
CP SMSG GJMSINK TEXTABLE --> TABLE MACROS V. 851121 JOB = \jobname%
}%
\tableinfotrue%
\catcode`\@=11
\def\tstrut{\vrule height3.1ex depth1.2ex width0pt}%
\def\and{\char`\&}
\def\tablerule{\noalign{\hrule height\thinsize depth0pt}}%
\def\thickrule{\noalign{\hrule height\thicksize depth0pt}}%
\def\ctr#1{\hfil\ #1\hfil}%
\tablewidth=-\maxdimen%
\spreadwidth=-\maxdimen%
\def\tabskipglue{0pt plus 1fil minus 1fil}%
\centertablestrue%
\gdef\ARGS{########}
\gdef\headerARGS{####}
\def\@mpersand{&}
{\catcode`\|=13
\gdef\letbarzero{\let|0}
\gdef\letbartab{\def|{&&}}%
\gdef\letvbbar{\let\vb|}%
}
{\catcode`\&=4
\def\ampskip{&\omit\hfil&}
\catcode`\&=13
\let&0
\xdef\letampskip{\def&{\ampskip}}%
\gdef\letnovbamp{\let\novb&\let\tab&}
}
\def\begintable{
   \begingroup%
   \catcode`\|=13\letbartab\letvbbar%
   \catcode`\&=13\letampskip\letnovbamp%
   \def\multispan##1{
      \omit \mscount##1%
      \multiply\mscount\tw@\advance\mscount\m@ne%
      \loop\ifnum\mscount>\@ne \sp@n\repeat%
   }
   \def\|{%
      &\omit\widevline&%
   }%
   \ruledtable
}
\long\def\ruledtable#1\endtable{%
   \offinterlineskip
   \tabskip 0pt
   \def\widevline{\vrule width\thicksize}
   \def\endrow{\@mpersand\omit\hfil\crnorm\@mpersand}%
   \def\crthick{\@mpersand\crnorm\thickrule\@mpersand}%
   \def\crthickneg##1{\@mpersand\crnorm\thickrule
          \noalign{{\skip0=##1\vskip-\skip0}}\@mpersand}%
   \def\crnorule{\@mpersand\crnorm\@mpersand}%
   \def\crnoruleneg##1{\@mpersand\crnorm
          \noalign{{\skip0=##1\vskip-\skip0}}\@mpersand}%
   \let\nr=\crnorule
   \def\endtable{\@mpersand\crnorm\thickrule}%
   \let\crnorm=\cr
   \edef\cr{\@mpersand\crnorm\tablerule\@mpersand}%
   \def\crneg##1{\@mpersand\crnorm\tablerule
          \noalign{{\skip0=##1\vskip-\skip0}}\@mpersand}%
   \let\ctneg=\crthickneg
   \let\nrneg=\crnoruleneg
   \the\tableLETtokens
   \tabletokens={&#1}
   \countROWS\tabletokens\into\nrows%
   \countCOLS\tabletokens\into\ncols%
   \advance\ncols by -1%
   \divide\ncols by 2%
   \advance\nrows by 1%
   \iftableinfo %
      \immediate\write16{[Nrows=\the\nrows, Ncols=\the\ncols]}%
   \fi%
   \ifcentertables
      \ifhmode \par\fi
      \line{
      \hss
   \else %
      \hbox{%
   \fi
      \vbox{%
         \makePREAMBLE{\the\ncols}
         \edef\next{\preamble}
         \let\preamble=\next
         \makeTABLE{\preamble}{\tabletokens}
      }
      \ifcentertables \hss}\else }\fi
   \endgroup
   \tablewidth=-\maxdimen
   \spreadwidth=-\maxdimen
}
\def\makeTABLE#1#2{
   {
   \let\ifmath0
   \let\header0
   \let\multispan0
   \ncase=0%
   \ifdim\tablewidth>-\maxdimen \ncase=1\fi%
   \ifdim\spreadwidth>-\maxdimen \ncase=2\fi%
   \relax
   \ifcase\ncase %
      \widthspec={}%
   \or %
      \widthspec=\expandafter{\expandafter t\expandafter o%
                 \the\tablewidth}%
   \else %
      \widthspec=\expandafter{\expandafter s\expandafter p\expandafter r%
                 \expandafter e\expandafter a\expandafter d%
                 \the\spreadwidth}%
   \fi %
   \xdef\next{
      \halign\the\widthspec{%
      #1
      \noalign{\hrule height\thicksize depth0pt}
      \the#2\endtable
      }
   }
   }
   \next
}
\def\makePREAMBLE#1{
   \ncols=#1
   \begingroup
   \let\ARGS=0
   \edef\xtp{\widevline\ARGS\tabskip\tabskipglue%
   &\ctr{\ARGS}\tstrut}
   \advance\ncols by -1
   \loop
      \ifnum\ncols>0 %
      \advance\ncols by -1%
      \edef\xtp{\xtp&\vrule width\thinsize\ARGS&\ctr{\ARGS}}%
   \repeat
   \xdef\preamble{\xtp&\widevline\ARGS\tabskip0pt%
   \crnorm}
   \endgroup
}
\def\countROWS#1\into#2{
   \let\countREGISTER=#2%
   \countREGISTER=0%
   \expandafter\ROWcount\the#1\endcount%
}%
\def\ROWcount{%
   \afterassignment\subROWcount\let\next= %
}%
\def\subROWcount{%
   \ifx\next\endcount %
      \let\next=\relax%
   \else%
      \ncase=0%
      \ifx\next\cr %
         \global\advance\countREGISTER by 1%
         \ncase=0%
      \fi%
      \ifx\next\endrow %
         \global\advance\countREGISTER by 1%
         \ncase=0%
      \fi%
      \ifx\next\crthick %
         \global\advance\countREGISTER by 1%
         \ncase=0%
      \fi%
      \ifx\next\crnorule %
         \global\advance\countREGISTER by 1%
         \ncase=0%
      \fi%
      \ifx\next\crthickneg %
         \global\advance\countREGISTER by 1%
         \ncase=0%
      \fi%
      \ifx\next\crnoruleneg %
         \global\advance\countREGISTER by 1%
         \ncase=0%
      \fi%
      \ifx\next\crneg %
         \global\advance\countREGISTER by 1%
         \ncase=0%
      \fi%
      \ifx\next\header %
         \ncase=1%
      \fi%
      \relax%
      \ifcase\ncase %
         \let\next\ROWcount%
      \or %
         \let\next\argROWskip%
      \else %
      \fi%
   \fi%
   \next%
}
\def\counthdROWS#1\into#2{%
\dvr{10}%
   \let\countREGISTER=#2%
   \countREGISTER=0%
\dvr{11}%
\dvr{13}%
   \expandafter\hdROWcount\the#1\endcount%
\dvr{12}%
}%
\def\hdROWcount{%
   \afterassignment\subhdROWcount\let\next= %
}%
\def\subhdROWcount{%
   \ifx\next\endcount %
      \let\next=\relax%
   \else%
      \ncase=0%
      \ifx\next\cr %
         \global\advance\countREGISTER by 1%
         \ncase=0%
      \fi%
      \ifx\next\endrow %
         \global\advance\countREGISTER by 1%
         \ncase=0%
      \fi%
      \ifx\next\crthick %
         \global\advance\countREGISTER by 1%
         \ncase=0%
      \fi%
      \ifx\next\crnorule %
         \global\advance\countREGISTER by 1%
         \ncase=0%
      \fi%
      \ifx\next\header %
         \ncase=1%
      \fi%
\relax%
      \ifcase\ncase %
         \let\next\hdROWcount%
      \or%
         \let\next\arghdROWskip%
      \else %
      \fi%
   \fi%
   \next%
}%
{\catcode`\|=13\letbartab
\gdef\countCOLS#1\into#2{%
   \let\countREGISTER=#2%
   \global\countREGISTER=0%
   \global\multispancount=0%
   \global\firstrowtrue
   \expandafter\COLcount\the#1\endcount%
   \global\advance\countREGISTER by 3%
   \global\advance\countREGISTER by -\multispancount
}%
\gdef\COLcount{%
   \afterassignment\subCOLcount\let\next= %
}%
{\catcode`\&=13%
\gdef\subCOLcount{%
   \ifx\next\endcount %
      \let\next=\relax%
   \else%
      \ncase=0%
      \iffirstrow
         \ifx\next& %
            \global\advance\countREGISTER by 2%
            \ncase=0%
         \fi%
         \ifx\next\span %
            \global\advance\countREGISTER by 1%
            \ncase=0%
         \fi%
         \ifx\next| %
            \global\advance\countREGISTER by 2%
            \ncase=0%
         \fi
         \ifx\next\|
            \global\advance\countREGISTER by 2%
            \ncase=0%
         \fi
         \ifx\next\multispan
            \ncase=1%
            \global\advance\multispancount by 1%
         \fi
         \ifx\next\header
            \ncase=2%
         \fi
         \ifx\next\cr       \global\firstrowfalse \fi
         \ifx\next\endrow   \global\firstrowfalse \fi
         \ifx\next\crthick  \global\firstrowfalse \fi
         \ifx\next\crnorule \global\firstrowfalse \fi
         \ifx\next\crnoruleneg \global\firstrowfalse \fi
         \ifx\next\crthickneg  \global\firstrowfalse \fi
         \ifx\next\crneg       \global\firstrowfalse \fi
      \fi
\relax
      \ifcase\ncase %
         \let\next\COLcount%
      \or %
         \let\next\spancount%
      \or %
         \let\next\argCOLskip%
      \else %
      \fi %
   \fi%
   \next%
}%
\gdef\argROWskip#1{%
   \let\next\ROWcount \next%
}
\gdef\arghdROWskip#1{%
   \let\next\ROWcount \next%
}
\gdef\argCOLskip#1{%
   \let\next\COLcount \next%
}
}
}
\def\spancount#1{
   \nspan=#1\multiply\nspan by 2\advance\nspan by -1%
   \global\advance \countREGISTER by \nspan
   \let\next\COLcount \next}%
\def\dvr#1{\relax}%
\def\header#1{%
\dvr{1}{\let\cr=\@mpersand%
\hdtks={#1}%
\counthdROWS\hdtks\into\hdrows%
\advance\hdrows by 1%
\ifnum\hdrows=0 \hdrows=1 \fi%
\dvr{5}\makehdPREAMBLE{\the\hdrows}%
\dvr{6}\getHDdimen{#1}%
{\parindent=0pt\hsize=\hdsize{\let\ifmath0%
\xdef\next{\valign{\headerpreamble #1\crnorm}}}\dvr{7}\next\dvr{8}%
}%
}\dvr{2}}
\def\makehdPREAMBLE#1{
\dvr{3}%
\hdrows=#1
{
\let\headerARGS=0%
\let\cr=\crnorm%
\edef\xtp{\vfil\hfil\hbox{\headerARGS}\hfil\vfil}%
\advance\hdrows by -1
\loop
\ifnum\hdrows>0%
\advance\hdrows by -1%
\edef\xtp{\xtp&\vfil\hfil\hbox{\headerARGS}\hfil\vfil}%
\repeat%
\xdef\headerpreamble{\xtp\crcr}%
}
\dvr{4}}
\def\getHDdimen#1{%
\hdsize=0pt%
\getsize#1\cr\end\cr%
}
\def\getsize#1\cr{%
\endsizefalse\savetks={#1}%
\expandafter\lookend\the\savetks\cr%
\relax \ifendsize \let\next\relax \else%
\setbox\hdbox=\hbox{#1}\newhdsize=1.0\wd\hdbox%
\ifdim\newhdsize>\hdsize \hdsize=\newhdsize \fi%
\let\next\getsize \fi%
\next%
}%
\def\lookend{\afterassignment\sublookend\let\looknext= }%
\def\sublookend{\relax%
\ifx\looknext\cr %
\let\looknext\relax \else %
   \relax
   \ifx\looknext\end \global\endsizetrue \fi%
   \let\looknext=\lookend%
    \fi \looknext%
}%
\def\tablelet#1{%
   \tableLETtokens=\expandafter{\the\tableLETtokens #1}%
}%
\catcode`\@=12

%% file: texfiles/title
\thispagestyle{empty}


\vskip4cm
\begin{center}
{\huge \bf{Beyond The Standard Model:\\ Some Aspects of \\ Supersymmetry and Extra Dimension}}

\end{center}

\vspace{3cm}

\begin{center}

{\bf{       
 \large{Thesis Submitted to \\
  The University of Calcutta \\
       for The Degree of \\
Doctor of Philosophy (Science)} \\}}

\vspace{3cm}
            
{\large By}\\

{\bf {\Large Tirtha Sankar Ray}}\\
{\large {Theory Division,\\ Saha Institute of Nuclear Physics, \\ 1/AF Bidhan Nagar, \\ Kolkata 700 064, India
}}

\vspace{1cm}     
{\large Under the supervision of }\\

{\bf {\Large Prof. Gautam Bhattacharyya}}\\
{\large {Theory Division,\\ Saha Institute of Nuclear Physics, India
}}
 
\vspace{1cm}   
{\large  2010     }
\end{center}


%% file: texfiles/ackn
\newpage

\normalsize \begin{center} {\large {\bf ACKNOWLEDGMENTS}} \end{center} \vskip
2\baselineskip
 
\begin{normalsize}

I gratefully acknowledge the academic and personal support received from my
supervisor Prof. Gautam Bhattacharyya.  He has been the \textit{friend,
philosopher and guide}, in its truest sense.  I thank him for the infinite
patience, with which he discussed and addressed all my problems in physics and
beyond.  I shall remain indebted to Prof. Amitava Raychaudhuri and
Prof. Palash Baran Pal from whom I learned the basics of particle physics and
received vital guidance, that navigated me through the labyrinth of my early
research life. I also thank Prof. Probir Roy for his encouragement and
guidance.

It is my great pleasure to thank Prof. Tapan Kumar Das, Prof. Subinay Dasgupta
and Prof. Anirban Kundu from the University of Calcutta and Prof. Partha
Majumdar, Prof. Pijushpani Bhattacharjee and Prof. Kamalesh Kar from Saha
Institute of Nuclear Physics for their encouragement and support at different
stages of my PhD work. I thank Prof. Stephane Lavignac of IPhT, CEA-Saclay,
for reading the manuscript, pointing out mistakes and suggesting improvements.
I thank Dr. Swarup Kumar Majee, Biplob Bhattacharjee and Kirtiman Ghosh for
their collaborations and insightful discussions on the nuances of particle
physics. I am grateful to SINP, and all its members, for providing an
enriching and friendly environment to carry out my research. I would
specifically like to thank all the academic members, the non-academic members
and the research scholars of Theory Division at SINP for their continued
encouragements and support. I would like to acknowledge the Council of
Scientific and Industrial Research, India for the S. P. Mukherjee Fellowship.

When the going got tough in physics, my friends pulled me through. I thank
Bappa, Abhiroop, Gopal, Pappu, Sumon, Arnab, Samriddhi, Purbasha and Sourav,
for all the eagerly awaited \textit{adda} sessions. I fondly remember the
regular discussions, the occasional debates and the rare fights, that I had
with Abhishek , on science, politics, cricket, films and everything else.

This thesis is dedicated to my grand mother Late (Mrs.) Sadhana Ray.

This work would not have been possible without the inspiration and
encouragements of my father Prof. Siddhartha Ray and the unconditional love of
my mother Mrs. Dipali Ray. I shall remember with humility all the sacrifices
they made for me.  I am grateful for the love and affection bestowed on me by
my brother, Mr. Ananda Ray and his wife, Mrs. Arpita Ray. It remains to name
my fiercest critic and my closest aide; my life support system, my wife,
Srirupa.

\vskip3cm
\begin{flushright} 
Tirtha Sankar Ray\\ 
Kolkata, India ~~~~
\end{flushright}
\end{normalsize}

%% file: texfiles/pub
\begin{itemize}
\item
{\bf Radiative correction to the lightest neutral Higgs mass in warped
  supersymmetry} \\
Authors: {\sf G.~Bhattacharyya, S.~K.~Majee and \textbf{T.~S.~Ray}}\\
Published in: {Phys.\ Rev.\  D {\bf 78} (2008) 071701 (Rapid communication)}\\
e-Print: arXiv:0806.3672 [hep-ph]

\item
{\bf Probing warped extra dimension via ${gg \to h}$ and ${h \to
      \gamma \gamma}$ at LHC}\\
Authors: {\sf G.~Bhattacharyya and \textbf{T.~S.~Ray}}\\
Published in: {Phys.\ Lett.\  B {\bf 675} (2009) 222}\\
e-Print: arXiv:0902.1893 [hep-ph]

\item
{\bf A phenomenological study of 5d Supersymmetry}\\
Authors: {\sf G.~Bhattacharyya and \textbf{T.~S.~Ray}}\\
Published in: { JHEP 1005 (2010) 040}\\
e-Print: arXiv:1003.1276 [hep-ph]

\end{itemize}

%% file: texfiles/intro
Particle physics endeavors to provide a description of fundamental particles
and their interactions in the quantum realm. Intense experimental
investigations and clairvoyant theoretical innovations in the last century
culminated in the formulation of the \textit{Standard Model} of particle
physics. It bloomed from the ideas originally put forward by S. L. Glashow,
S. Wienberg and A. Salam~\cite{sm} in the 1960's. Decades of increasingly
intense experimental scrutiny has put this theory on strong footing. Today it
is believed that the Gauge Field Theoretic~\cite{QFT} language of the Standard
Model (SM) is the right path to describe quantum particle interactions. The
notion of theoretic consistancy, cosmological observations like the detection
of dark matter etc.  indicate that the SM only provides a partial picture of
the fundamental particles. Nevertheless, any extensions of this theory must
closely resemble the SM at the energies that have already been explored at
collider and other laboratory experiments.

\section{The Standard Model}
\label{theSM}

\begin{figure} 
\begin{minipage}{0.47\textwidth}
 \begin{center}
\includegraphics[width=0.9\textwidth,angle=0,keepaspectratio] {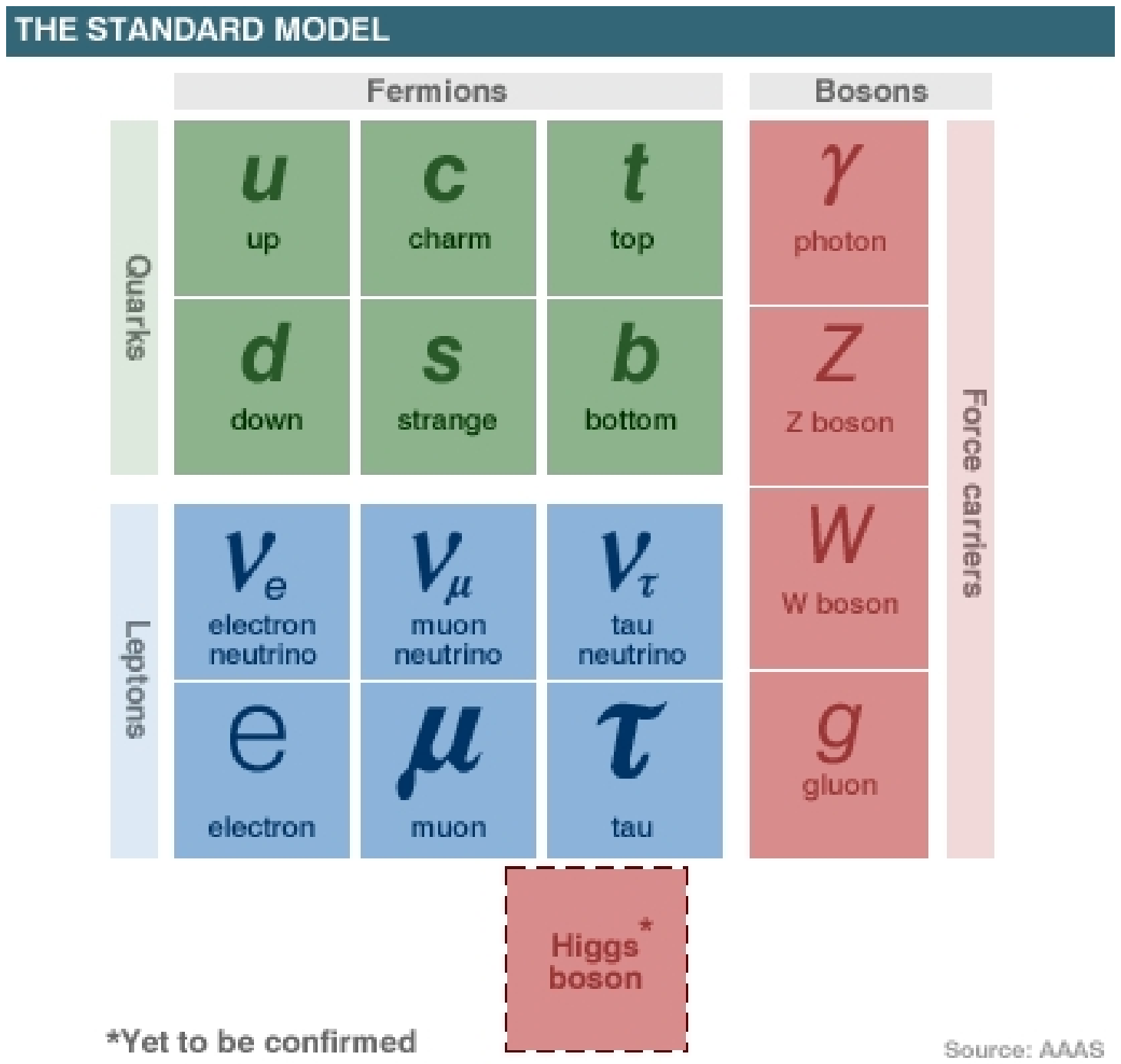}
\caption{ \small The particle content of the Standard Model.}  \label{zoo}
\end{center} 
\end{minipage} 
\hspace{7mm} 
\begin{minipage}{0.47\textwidth}
\begin{center} 
\includegraphics[width=0.8\textwidth,angle=0,keepaspectratio]
{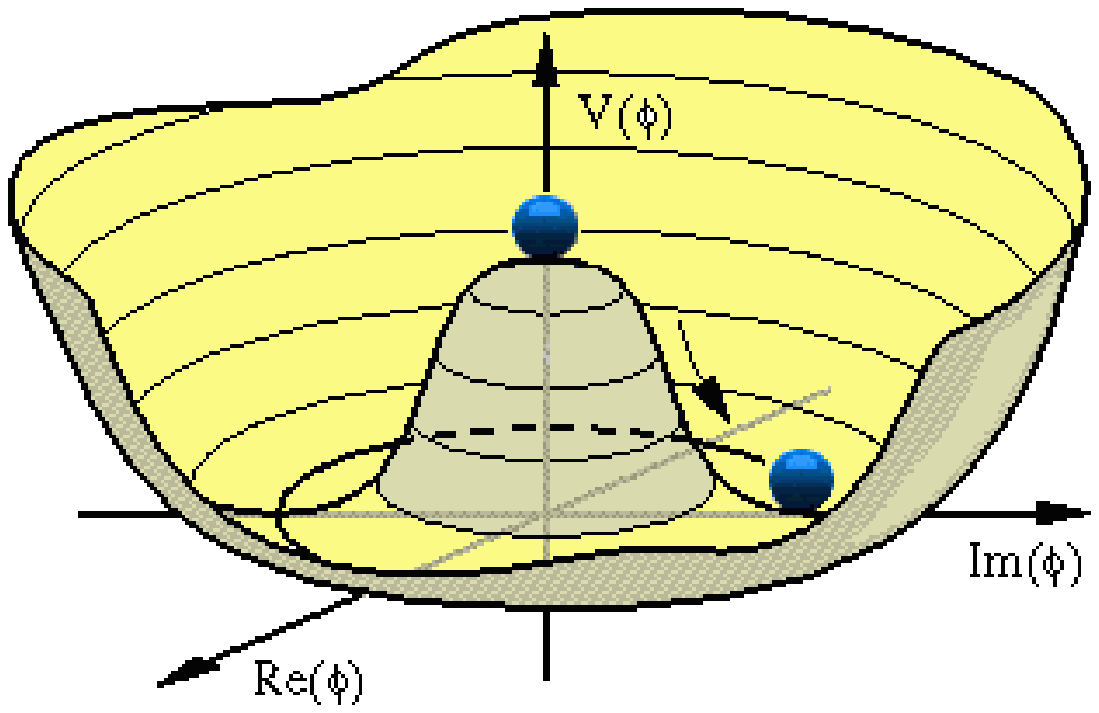} \caption{ \small The Higgs
potential in the Standard Model.}  
\label{higpot} 
\end{center} 
\end{minipage}
\end{figure}

The particle content of the Standard Model was discovered at the various
collider experiments. The last particle to be discovered is the Top quark,
discovered at the Tevatron.  The Higgs field which is an integral part of the
theory has evaded discovery till the date of writing this thesis. The so
called \textit{zoo} of fundamental particles is summarized in Figure \ref{zoo}.

The Standard Model (SM) is a specific form of a gauge field theory with a
gauge group of $SU(3)_c \times SU(2)_L \times U(1)_Y$. It provides a unified
picture of the strong, weak and electromagnetic interactions.  The $SU(3)_c$
part of the gauge group exclusively describes the strong interactions and is
independently called \textit{Quantum Chromo Dynamics} (QCD).  Whereas the
$SU(2)_L \times U(1)_Y$ part of the gauge group provides a unified picture of
the electromagnetic and the weak interactions and is called the
\textit{Electroweak} sector of the theory.

The strong interaction part, or the (QCD)~\cite{QCD} has an $SU(3)$ gauge
group. The Lagrangian density may be written as, \begin{equation}
\mathcal{L}_{QCD} = - \frac{1}{4} G^i_{\mu\nu} G^{i\mu\nu} + \sum_r
\bar{q}_{r\alpha} i \not{\!\!D}^\alpha_\beta \,q^\beta_r, \label{QCD}
\end{equation}

where, \begin{equation} G^i_{\mu\nu} = \partial_\mu G^i_\nu - \partial_\nu
G^i_\mu - g_s f_{ijk}\; G_\mu^j\; G_\nu^k \end{equation} is the field strength
tensor for the gluon fields $G^i_\mu, \; i = 1, \cdots, 8$, $g_s$ is the QCD
gauge coupling constant and the structure constants $f_{ijk}$ $ (i, j, k = 1,
\cdots, 8)$ are defined by \begin{equation} [\lambda^i, \lambda^j] = 2 i
f_{ijk} \lambda^k, \end{equation} where the $\lambda$ are the $SU(3)$
generator matrices normalized by $\mbox{Tr}\, \lambda^i \lambda^j = 2
\delta^{ij}$, so that $\mbox{Tr}\, [\lambda^i, \lambda^j] \lambda^k=4i
f_{ijk}$.

 The $G^2$ term leads to the self-interaction of gluons.  The second term in
$\mathcal{L}_{QCD}$ is the gauge covariant derivative for the quarks: $q_r$ is
the $r^{ th}$ quark flavor, $\alpha, \beta = 1,2,3$ are color indices, and \be
D^\alpha_{\mu \beta} = (D_\mu)_{\alpha \beta} = \partial_\mu \delta_{\alpha
\beta} + i g_s \;G^i_\mu\; L^i_{\alpha\beta}, \ee where the quarks transform
according to the triplet representation matrices $L^i ={\lambda^i}/{2}$.  The
color interactions are diagonal in the flavor indices, but in general change
the quark colors. These interactions are purely vector like and thus parity
conserving.  There are in addition, effective ghost and gauge-fixing terms
which enter into the quantization of both the $SU(3)$ and electroweak parts of
the theory. In the QCD part of the theory, there is the possibility of adding
a (unwanted) term which violates $CP$ invariance.  QCD has the property of
asymptotic freedom~\cite{asyfre}, i.e., the coupling becomes weak at high
energies enabling perturbative study at these energy scales or short
distances.  At low energies or large distances it becomes strongly
coupled~\cite{Fritzsch:1973pi} which is sometimes called \textit{infrared
slavery}, leading to the confinement of quarks and gluons. The confinement of
quarks and gluons is still an ill-understood facet of QCD as it is riddled
with the difficulty of being a non-perturbative phenomenon. Note that there
are no tree level mass terms for the quarks in the Lagrangian given in
Eq.~\ref{QCD}.  These would be allowed by QCD alone, but are forbidden by the
chiral symmetry of the electroweak part of the theory.  The quark masses are
generated by phenomenon of spontaneous electroweak symmetry breaking.

The theoretical picture of QCD described above was painstakingly verified
through various collider experiments. The scaling of structure functions in
the deep inelastic collisions of nucleons provided the first glimpse of
hadronic substructure, parton model of hadrons was invoked to explain this
phenomenon.  The scaling violations that were discovered later provided
indirect verification of perturbative QCD. Though QCD is a vast subject by
itself and is an integral part of present quest for a quantum description of
particle interaction, it is not the main subject of study in this thesis and
it will not be explored any further in what follows.

The gauge group of the electroweak sector is the $SU(2)_L \times U(1)_Y$.  The
constituents of the SM fall into valid representation of these groups. An
important feature of this model is the chiral nature of the
interactions. Unlike QCD, the left and right chiral parts of the fields behave
differently under the electroweak gauge transformation. This phenomenon can be
consistently described by using the following representations of the field.
We represent the leptonic sector of the electroweak theory by the left-handed
leptons \begin{equation} \mathsf{L}_{1} = \left( \begin{array}{c} \nu_{e} \\
e^{-} \end{array} \right)_{\mathrm{L}} \;\; \mathsf{L}_{2} = \left(
\begin{array}{c} \nu_{\mu} \\ \mu^{-} \end{array} \right)_{\mathrm{L}} \;\;
\mathsf{L}_{3} = \left( \begin{array}{c} \nu_{\tau} \\ \tau^{-} \end{array}
\right)_{\mathrm{L}} \;, \label{lleptons} \end{equation} with weak isospin $I
= 1/2$ and weak hypercharge $Y(\mathsf{L}_{i}) = -1$, corresponding to the
$SU(2)$ and $U(1)$ charges respectively. The right-handed weak-isoscalar
charged leptons are represented by \begin{equation} \mathsf{E}_{1,2,3} =
e_{\mathrm{R}}, \mu_{\mathrm{R}}, \tau_{\mathrm{R}}\;, \label{rleptons}
\end{equation} with weak hypercharge $Y(\mathsf{E}_{i}) = -2$. The right
handed fields are singlets under $SU(2)$.  The weak hypercharges are chosen to
reproduce the observed electric charges, through the connection $Q = I_{3} +
1/2Y$. The original Glashow-Wienberg-Salam model did not have a right chiral
neutrino, leaving the neutrinos massless.

The hadronic sector consists of the left-handed quarks \begin{equation}
\mathcal{Q}_{1} = \left( \begin{array}{c} u \\ d \end{array}
\right)_{\mathrm{L}} \;\; \mathcal{Q}_{2} = \left( \begin{array}{c} c \\ s
\end{array} \right)_{\mathrm{L}} \;\; \mathcal{Q}_{3} = \left(
\begin{array}{c} t \\ b \end{array} \right)_{\mathrm{L}} \;, \label{lquarks}
\end{equation} with weak isospin $I = 1/2$ and weak hypercharge
$Y(\mathcal{Q}_{i}) = 1/3$, and their right-handed weak-isoscalar counterparts
\begin{equation} \mathcal{U}_{(1,2,3)} = u_{\mathrm{R}}, c_{\mathrm{R}},
t_{\mathrm{R}}\hbox{ and } \mathcal{D}_{(1,2,3)} = d_{\mathrm{R}},
s_{\mathrm{R}}, b_{\mathrm{R}}\;, \label{rquarks} \end{equation} with weak
hypercharges $Y(\mathcal{U}_{i}) = 4/3$ and $Y(\mathcal{D}_{i}) = -2/3$.
According to the basic tenets of quantum physics, identical quantum numbers
can mix with each other.  It can be shown that all but one of these mixing
matrices can be absorbed into the redefinition of the fields.  As per
convention, the weak eigenstates in the lower component of the quark doublets
in Eq.~\ref{lquarks} are considered to be admixtures of the mass
eigenstates. This mixing of the fields may be represented by: \begin{equation}
\left(\begin{array}{c} d \\ s \\ b \end{array}\right) = \left(
\begin{array}{ccc} V_{ud} & V_{us} & V_{ub} \\ V_{cd} & V_{cs} & V_{cb} \\
V_{td} & V_{ts} & V_{tb} \end{array}\right) \left(\begin{array}{c} d^\prime \\
s^\prime \\ b^\prime \end{array}\right) \equiv \mathsf{V_{CKM}}
\left(\begin{array}{c} d^\prime \\ s^\prime \\ b^\prime \end{array}\right) ,
\label{ckmmatrix} \end{equation} where the $d^\prime, s^\prime,b^\prime$ are
the mass eigenstates. This kind of mixing leads to flavor violation i.e.,
mixing between different generations of quarks.  Experimental observations
have put strong constraints on \textit{flavor changing neutral current}.
Glashow-Iliopoulos-Maiami~\cite{Glashow:1970gm} demonstrated that if $V_{CKM}$
is constrained to be a unitary matrix, such flavor changing processes mediated
by neutral gauge bosons are suppressed.  Following
Cabibbo~\cite{Cabibbo:1963yz}--Kobayashi--Maskawa~\cite{Kobayashi:1973fv} a
simple parameter counting of a $n \times n$ unitary matrix reveals the
existence of $n(n-1)/2$ independent real mixing angles and $(n-1)(n-2)/2$
independent complex phases. It is clear that the $V_{CKM}$ contains three real
mixing angles and single complex phase. The complex phase leads to complex
gauge interactions that violates CP symmetry within the framework of the SM.
The unitarity of the CKM-matrix implies various relations between its
elements. In particular, we have \begin{equation}\label{2.87h}
V_{ud}^{}V_{ub}^* + V_{cd}^{}V_{cb}^* + V_{td}^{}V_{tb}^* =0.  \end{equation}
Phenomenologically this relation is very interesting as it involves
simultaneously the elements $V_{ub}$, $V_{cb}$ and $V_{td}$ which are under
extensive discussion at present.  The relation in Eq.~\ref{2.87h} can be
represented as a ``unitarity'' triangle in the complex
$(\bar\varrho,\bar\eta)$ plane.  Where $\frac{| V_{ud}^{}V^*_{ub}|}{|
V_{cd}^{}V^*_{cb}|} = \sqrt{\bar\varrho^2 +\bar\eta^2}$ and $\frac{|
V_{td}^{}V^*_{tb}|}{| V_{cd}^{}V^*_{cb}|} = \sqrt{(1-\bar\varrho)^2
+\bar\eta^2}$.  Eq.~\ref{2.87h} is invariant under any phase-transformations,
they are phase convention independent and are physical observables.
Consequently they can be measured directly in suitable experiments.  One can
construct additional five unitarity triangles corresponding to other
orthogonality relations, like the one in Eq.~\ref{2.87h}.  They are discussed
in \cite{Kayser}. Some of them should be useful when LHC-B experiment will
provide data.  The areas of all unitarity triangles are equal and related to
the measure of CP violation $J_{\rm CP}$ \cite{CJ}: \begin{equation} \mid
J_{\rm CP} \mid = 2\cdot A_{\Delta}, \end{equation} where $A_{\Delta}$ denotes
the area of the unitarity triangle.

The fact that each left-handed lepton doublet is matched by a left-handed
quark doublet guarantees that the theory is anomaly free, this is a
prerequisite for a theory to be renormalizable. It ensures that the higher
order contributions in the perturbation theory will respect the gauge symmetry
imposed at the zeroth (tree) order in the Lagrangian~\cite{Bouchiat:1972iq}.

The electroweak gauge group predicts two sets of gauge fields: a weak
isovector $\bm{W}_\mu$, with coupling constant $g$, and a weak isoscalar
${{B}}_\mu$, with its own coupling constant $g^\prime$. In order for the
Lagrangian to be gauge independent, these gauge fields must transform to
compensate the variation induced in the mass fields. This specifies the
transformation of the gauge fields to be, $\bm{W}_\mu \to \bm{W}_\mu -
\bm{\alpha} \times \bm{W}_\mu - (1/g)\partial_\mu \bm{\alpha}$ under an
infinitesimal weak-isospin rotation generated by $G = 1 + (i/2)\bm{\alpha}
\cdot \bm{\tau}$ (where $\bm{\tau}$ are the Pauli matrices) and ${B}_\mu \to
{B}_\mu - (1/g^\prime)\partial_\mu \alpha$ under an infinitesimal hypercharge
phase rotation.  The corresponding field-strength tensors are defined as,
\begin{equation} \mathcal{W}^{i}_{\mu\nu} \equiv \partial_{\nu}W^{i}_{\mu} -
\partial_{\mu}W^{i}_{\nu} + g\varepsilon_{jki}W^{j}_{\mu}W^{k}_{\nu}\; ,
\label{Wmunu} \end{equation} where $i = 1,2,3$ for the three components of the
weak isovector, and \begin{equation} \mathcal{B}_{\mu\nu} =
\partial_{\nu}{{B}}_\mu - \partial_{\mu}{{ B}}_\nu \; , \label{eq:fmunu}
\end{equation} for the weak-hypercharge symmetry.
 
We may summarize the SM electroweak interactions by the Lagrangian,
\begin{equation} \mathcal{L}_{ew} = \mathcal{L}_{\rm gauge} + \mathcal{L}_{\rm
leptons} + \mathcal{L}_{\rm quarks}\ , \end{equation} with \begin{equation}
\mathcal{L}_{gauge}=-\frac{1}{4}\sum_i {\mathcal{W}}^i_{\mu\nu}
{\mathcal{W}}^{i\,\mu\nu}
-\frac{1}{4}\mathcal{B}_{\mu\nu}\mathcal{B}^{\mu\nu}, \label{laggauge}
\end{equation} \begin{eqnarray} \mathcal{L}_{\rm leptons} & = & \sum_j
\overline{{\sf E}}_{j}\:i\gamma^\mu\!\left(\partial_\mu
+i\frac{g^\prime}{2}{B}_\mu Y\right)\!{\sf E}_{j} \label{laglepton} \\ & + &
\sum_j \overline{{\sf L}}_{j}\:i\gamma^\mu\!\left(\partial_\mu
+i\frac{g^\prime}{2}{ B}_\mu
Y+i\frac{g}{2}\bm{\tau}\cdot\bm{W}_\mu\right)\!{\sf L}_{j}\;, \nonumber
\end{eqnarray} where $j$ is the generational index and runs over $e, \mu,
\tau$, and \begin{eqnarray} \mathcal{L}_{\rm quarks} & = & \sum_n
\overline{{\sf \mathcal{U}}}_{n}\:i\gamma^\mu\!\left(\partial_\mu
+i\frac{g^\prime}{2}{B}_\mu Y\right)\!{\sf \mathcal{U}}_{n} \nonumber \\ & + &
\sum_n \overline{{\sf \mathcal{D}}}_{n}\:i\gamma^\mu\!\left(\partial_\mu
+i\frac{g^\prime}{2}{B}_\mu Y\right)\!{\sf \mathcal{D}}_{n} \label{lagquark}
\\ & + & \sum_n \overline{{\sf
\mathcal{Q}}}_{n}\:i\gamma^\mu\!\left(\partial_\mu +i\frac{g^\prime}{2}{
B}_\mu Y+i\frac{g}{2}\bm{\tau}\cdot\bm{W}_\mu\right)\!{\sf \mathcal{Q}}_{n}\;,
\nonumber \end{eqnarray} where the generation index $n$ runs over $1, 2,
3$. The objects in parentheses in Eq.~\ref{laglepton} and Eq.~\ref{lagquark}
are the \textit{gauge-covariant derivatives.}

 The gauge Lagrangian (Eq.~\ref{laggauge}) contains four massless electroweak
gauge bosons, \textit{viz.\/} $W^{1}_{\mu}$, $W^{2}_{\mu}$, $W^{3}_{\mu}$,
${{B}}_\mu$. They are massless because a mass term such as
$1/2m^2\mathcal{B}_\mu\mathcal{B}^\mu$ is prohibited by gauge
symmetry. Massless gauge fields manifest in interaction with infinite
range. In nature, only electromagnetism fits this bill and the corresponding
gauge field is called the \textit{photon}.  Moreover, the gauge symmetry
forbids fermion mass terms of the form $m\bar{f}\!f =
m(\bar{f}_{\mathrm{R}}f_{\mathrm{L}} + \bar{f}_{\mathrm{L}}f_{\mathrm{R}})$ in
Eq.~\ref{laglepton} and Eq.~\ref{lagquark}, because the left-chiral and
right-chiral components of the fields transform differently under gauge
symmetry.

To generate masses of the gauge bosons other than the photon and the chiral
fermions in a gauge invariant way, we need to break the gauge symmetry in a
very special way.  We consider that the gauge symmetries are respected
everywhere in the theory but are broken by the vacuum state. This procedure is
called the \textit{spontaneous breaking of gauge symmetry}\footnote{It is
curious to note that this phenomenon of spontaneous breaking of gauge symmetry
is possible only for space dimensions 2 and above. This is called the
Coleman-Mermin-Wagner theorem.}. It was first introduced in the context of
superconducting phase transition.  In particle physics what has come to be
called the Higgs mechanism~\cite{ewsb} is but a relativistic generalization of
the Ginzburg-Landau theory~\cite{Ginzburg:1950sr} of superconductivity.

In the standard model this is achieved by introducing a complex scalar that
transforms as a doublet under the $SU(2)$ gauge group. The $U(1)$ charge is
represented by its $+1$ hypercharge. The field is a color singlet. Let us
define the scalar doublet as, \be \Phi = \left( \begin{array}{c} \phi ^+ \\
\phi ^0 \end{array} \right) = \left( \begin{array}{c} \frac{1}{\sqrt{2}} (
\phi _1 - i \phi _2) \\ \frac{1}{\sqrt{2}}(\phi _3- i \phi _4 ) \end{array}
\right). \label{fhigg} \ee
 
The gauge invariant Lagrangian for the field $\Phi$ may be written as, \be
\mathcal{L}_\Phi =(D^\mu \Phi)^{\dag} D_\mu \Phi - V(\Phi), \label{laghigg}
\ee where, \be V(\Phi ) = \frac{1}{2} \mu^2 \left( \sum^4_{i=1} \phi ^2_i
\right) + \frac{1}{4} \lambda \left( \sum^4_{i=1} \phi ^2_i \right)^2 ,
\label{pothigg} \ee and \be D_\mu \Phi = \left( \partial_\mu + i g
\frac{\tau^i}{2} W_\mu^i + \frac{i g'}{2} B_\mu \right) \Phi .
\label{higgcovderi} \ee The Lagrangian has a global $SO(4)~(\equiv SU(2)\times
SU(2))$ symmetry.  For $\mu^2<0$, the Higgs potential\footnote{It should be
noted that in the quantized theory, there are going to be quantum corrections
to the classical Lagrangian. It can be shown that the phenomenon of
electroweak symmetry breaking is nonperturbative, i.e even after
incorporating higher order corrections, the vacuum structure of the potential
as depicted in Figure~\ref{higpot} will remain identical.} in
Eq.~\ref{pothigg} takes the form shown in Figure~\ref{higpot}. With this
configuration, clearly $\langle 0| \phi _i |0 \rangle \neq 0$. Rather it lies
on a four dimensional circle with radius $\nu$.  From the orbit structure
$\sum_i|\langle 0| \phi _i |0 \rangle|^2 = \nu^2$, we note that the vacuum has
a $SO(4)$ symmetry as mentioned above and as soon as we select a direction for
the vev it reduces to $SO(3)$.  The group $SO(3)$ is isomorphic to
$SU(2)$. Thus the original $SU(2)\times SU(2)$ global symmetry is now reduced
to a $SU(2)$. This residual global symmetry in the Higgs potential is called
the custodial symmetry. This remains unbroken even after the vev is generated,
and this unbroken symmetry implies the equality of all gauge boson masses
generated by spontaneous symmetry breaking, a phenomenon the we demonstrate
below.  Without loss of generality we can choose the axis in this
four-dimensional space so that $\langle 0| \phi _i |0 \rangle = 0, \;\ i = 1,
2, 4 $ and $\langle 0 | \phi _3 | 0 \rangle = \nu$.  This choice of the
physical vacuum results in the breaking of the gauge symmetry in the vacuum
state.
 
To quantize around the classical vacuum, we introduce the physical scalar
$\Phi'$ defined by the relation, $\Phi = \frac{1}{\sqrt{2}} \left(
\begin{array}{c} 0 \\ v \end{array} \right) + \Phi '$, where $\langle 0
|\Phi^\prime| 0 \rangle = 0$ . To proceed further it will be useful to rewrite
the four components of $\Phi^\prime$ in terms of a new set of variables
following Kibble~\cite{Kibble:1967sv} as, \be \Phi = \frac{1}{\sqrt{2}} e^{i
\sum \xi^i \frac{1}{2}\tau^i} \left( \begin{array}{c} 0 \\ v + h \end{array}
\right). \label{kibhigg} \ee where $h$ is a hermitian field which will turn
out to be the physical Higgs scalar. The $\xi^i$ are the massless pseudoscalars
Nambu-Goldstone bosons~\cite{NambuGoldstone} that are necessarily associated
with broken symmetry generators.  However, the $SU(2)$ gauge invariance of the
SM allows us to select a gauge where these fields disappear from the physical
spectrum. This so called unitary gauge is defined as, \be \Phi \rightarrow
\tilde{\Phi} = e^{-i \sum \xi^i L^i} \Phi = \frac{1}{\sqrt{2}} \left(
\begin{array}{c} 0 \\ v + h \end{array} \right), \label{higgaugset} \ee
where the Goldstone bosons disappear.  In this gauge, the scalar kinetic term
takes the form \begin{eqnarray} ({D}_\mu \tilde{\Phi} )^{\dag} ({D}^\mu
\tilde{\Phi}) & \sim & \frac{1}{2} (0\; v) \left[ \frac{g}{2} \tau^i W^i_\mu
+ \frac{g'}{2} B_\mu \right]^2 \left( \begin{array}{c} 0 \\ v \end{array}
\right) + h \mbox{ terms} \nonumber \\ & \sim & M^2_W W^{+\mu} W^-_\mu +
\frac{M_Z^2}{2} Z^\mu Z_\mu + h \mbox{ terms}, \label{kinhigg} \end{eqnarray}
where the terms involving the physical $h$ field have been clubbed
together as the `$h ~\mbox{terms}$'. The third component of the $SU(2)$ gauge
field $W^3_{\mu}$ and the $U(1)$ gauge field $B_{\mu}$ have identical quantum
numbers after the spontaneous breaking of the electroweak gauge group and 
they get mixed in the Higgs kinetic term. The mass diagonal fields
are related to these fields by the following relations, \begin{eqnarray}
W^{\pm}_{\mu}&=&\frac{1}{\sqrt{2}} (W^1_{\mu} \mp i W^2_{\mu}),\nonumber\\
Z_{\mu} &=& -\sin \theta_W B_{\mu} + \cos \theta_W W^3_{\mu}, \end{eqnarray}
and the orthogonal combination, \be A_{\mu} = \cos \theta_W B_{\mu} + \sin
\theta_W W^3_{\mu} \ee is the photon field that remains massless. Where the
weak angle $\theta_W$ is defined by \be \tan\theta_W\equiv \frac{g'}{g} \
\Rightarrow\ \sin^2 \theta_W= 1-\frac{M_W^2}{M_Z^2}.\label{weakang} \ee

Thus, spontaneous symmetry breaking generates mass terms for the $W$ and $Z$
 gauge bosons proportional to the Higgs vacuum expectation value $v$. They
 are given by, \be M_W = \frac{g v}{2} \label{wmass}, \ee and \be M_Z =
 \sqrt{g^2 + g^{\prime 2} } \frac{v}{2} = \frac{M_W}{\cos
 \theta_W}. \label{zmass} \ee Observe that $M_Z > M_W$ which is in
 contradiction to the argument of equal gauge boson mass we gave from the idea
 of custodial symmetry.  In the SM the custodial symmetry associated with the
 $SU(2)$ gauge group is broken, and it has been broken by hypercharge mixing,
 i.e. by expanding the gauge group to $SU(2) \times U(1)$. If we put the
 hypercharge gauge coupling $g'$ to zero, we recover the symmetric condition.
 We will define an important parameter: \begin{equation} \rho \equiv {M_W^2
 \over M_Z^2 \cos^2\t_W}.  \end{equation} With the $SU(2)$ doublet scalar
 representation (at tree level), one can easily show from Eq~\ref{zmass}
 that $\rho = 1$, which is a non-trivial prediction of the SM at the tree level.

The Goldstone bosons $\xi$'s, disappear from the theory as physical entities
but reemerge as the longitudinal degrees of freedom of massive vector boson
fields.

The gauge boson masses are related to the Fermi constant by the relation:
$G_F/\sqrt{2} = g^2/8 M^2_W$, where $G_F \simeq 1.16637 \times 10^{-5}$
GeV$^{-2}$, as determined from the muon lifetime measurements. The weak scale
$v$ is therefore \be v = 2M_W/g \simeq (\sqrt{2} G_F)^{-1/2} \simeq 246
\mbox{ GeV}.  \ee Where, $g = e/\sin \theta_W$, where $e$ is the electric
charge of the positron. Hence, to lowest order \be M_W = M_Z \cos \theta_W
\sim \frac{(\pi \alpha/\sqrt{2} G_F)^{1/2}}{ \sin \theta_W}, \label{wztree}
\ee where $\alpha \approx 1/137.036$ is the fine structure constant. Using the
measured value of $\sin^2 \theta_W \approx 0.23$ as obtained from from neutral
current scattering experiments, one expects $M_W \approx 78$ GeV, and $M_Z \approx
89$ GeV. (These predictions are increased by $2-4 \%$ by higher order
corrections.)

From symmetry considerations we are free to add gauge-invariant interactions
between the scalar fields and the fermions. These are called the Yukawa terms
in the Lagrangian and they are the means of generating fermion masses within
the framework of the SM\footnote{The Higgs mechanism in the SM not only breaks
the gauge symmetry but it also drives a breaking of the chiral symmetry in the
fermionic sector.}.  To generalize for all the matter fields, we can write the
Yukawa interaction term as, \be \CL_{Yukawa} = -
Y^u_{ij}\bar{\mathcal{Q}}_{i}{\mathcal{U}_{j}}\bar{\Phi} -
Y^d_{ij}\bar{\mathcal{Q}}_{i}{\mathcal{D}_{j}}{\Phi} -
Y^l_{ij}\bar{L}_{i}{E_{j}}\Phi + h.c \ee

where, $\bar{\Phi}=-i\sigma_2\Phi^*$ and $Y^u$, $Y^d$, $Y^l$ are the up-quark,
down-quark and charged lepton Yukawa coupling constant matrices respectively.
Once, the Higgs field gets a vev $v$, then the Lagrangian takes the form
$\overline{f_L}m_f f_R$ with the mass matrices \be {({\rm m}_u)}_{ij} \propto
Y^u_{ij}v,~~ {({\rm m}_d)}_{ij} \propto Y^d_{ij}v,~~ {({\rm m}_l)}_{ij}
\propto Y^l_{ij}v, \ee where, $i,j$ represent the generational index. These
mass matrices are in the flavor basis, and not in the mass basis.  It should
be noted that due to the absence of their right chiral components, the
neutrinos remain massless in the SM.

\subsection{Experimental status of the Standard Model}

\begin{figure}  \begin{minipage}{0.47\textwidth}
 \begin{center}
\includegraphics[width=0.8\textwidth,angle=0,keepaspectratio]
{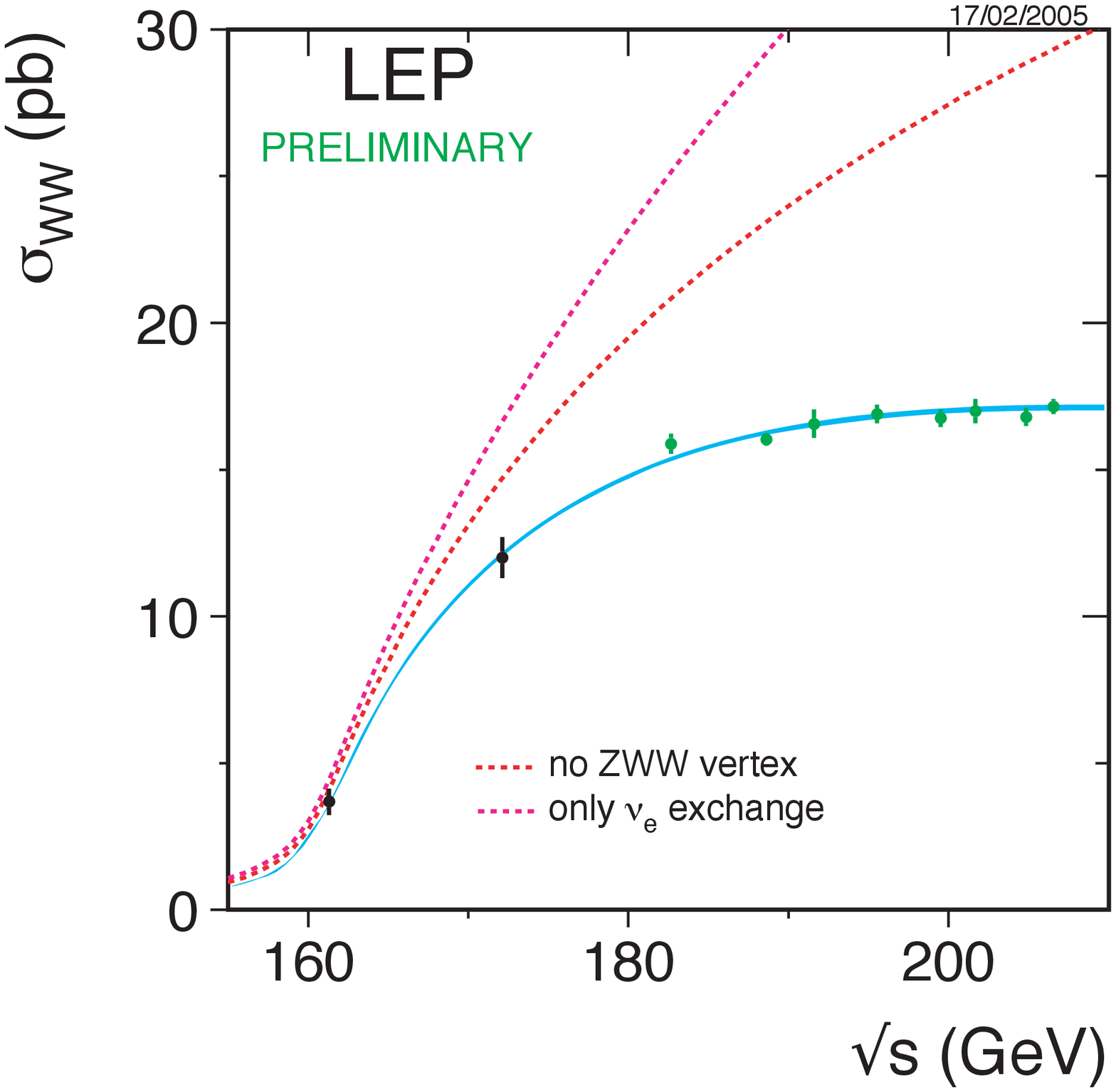}
\caption{ \small Cross section for the reaction $e+e− \rightarrow W+W−$ measured 
by the four LEP experiments, together with the full
electroweak-theory simulation and cross sections that would
result from $\nu$-exchange alone and from ($\nu +\gamma$)-exchange \cite{lepewwg}}
\label{SMgauge} 
\end{center}
\end{minipage} \hspace{7mm} 
\begin{minipage}{0.47\textwidth}
 \begin{center} 
\includegraphics[width=0.5\textwidth,angle=0,keepaspectratio]
{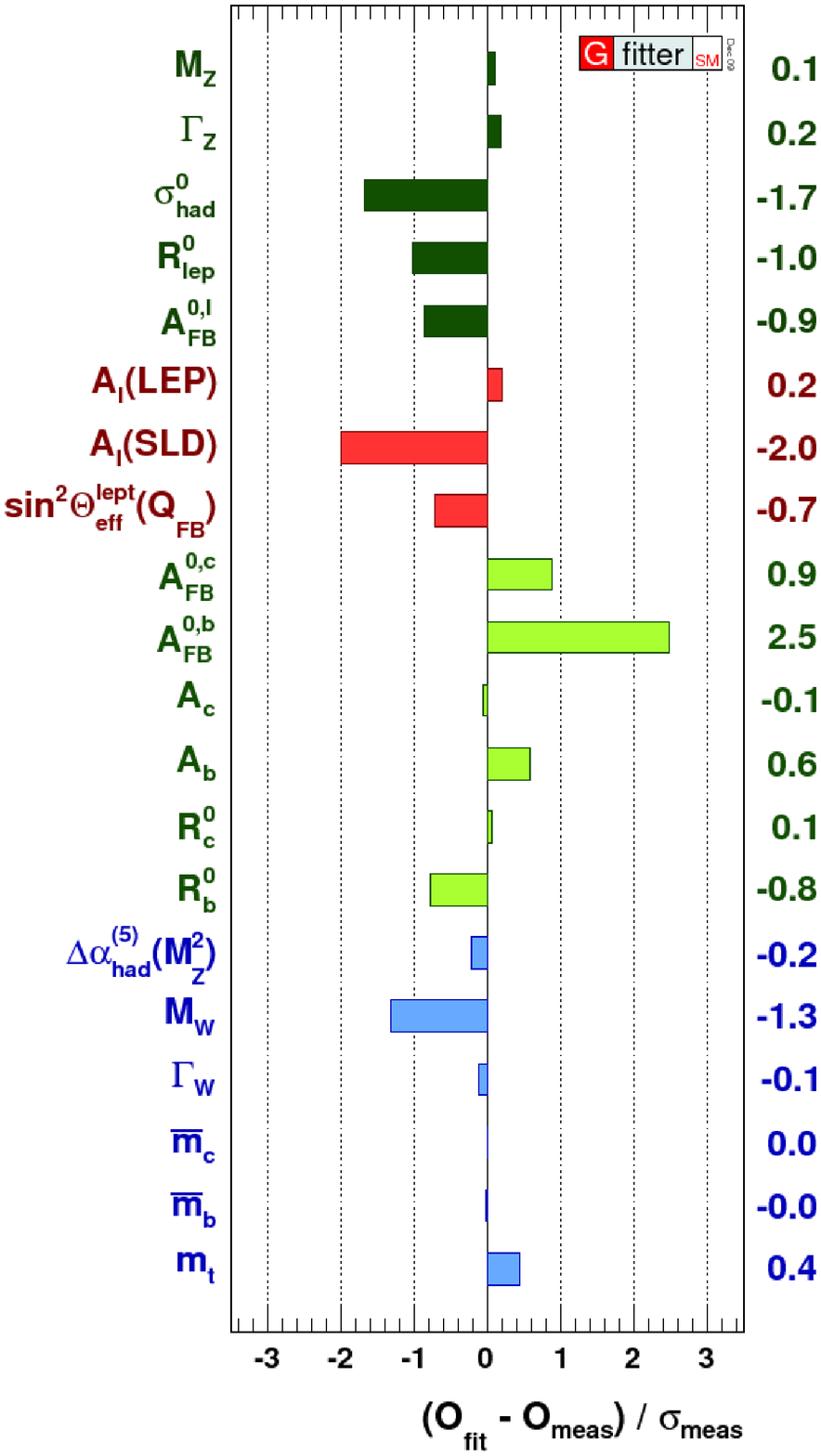} \caption{ \small TPull values
comparing \textsf{Gfitter} complete fit results with experimental
determinations~\cite{Flacher:2008zq}.}  \label{SMfit} \end{center}
\end{minipage} \end{figure}

One of the striking features of the standard model is that it has withstood
decades of increasingly intense experimental scrutiny. We briefly summerize
the present experimental status of the SM.

\textbf{Tree level:} Historically, the electroweak theory was formulated in the
context of extensive experimental information about the charged-current weak
interactions (mainly from study of $\beta$ decay).
 The Fermi Theory of the weak charged current interactions had
been developed and tested prior to the construction of the SM.  The unitarity
argument~\cite{Lee:1965js} made it clear that Fermi's four-fermion description
could not be valid above c.m. energy $\sqrt{s}\sim 600$ GeV. This necessitated
the conjecture of heavy intermediate massive charged gauge bosons.  The
smallest unitary group which provides an off-diagonal generator (corresponding
to the charged gauge boson) is SU(2). The relevant generators are $\tau^1$ and
$\tau^2$.  We further need a massless gauge boson to account for the infinite
range electromagnetic interaction. Any association of photon with the neutral
generator $\tau^3$ would lead to contradiction with respect to the charge
assignment of particles. The gauge charges of fermions coupling to $W^3$ are
$\pm {1\over 2}$, clearly different from the electric charges. Moreover, $W^3$
couples to neutrino, but photon does not. All in all, just with SU(2) gauge
theory we cannot explain both weak and electromagnetic interactions. The next
simplest construction is to avoid taking a simple group, but consider SU(2)
$\times$ U(1).  Further analysis of the reaction $\nu\bar{\nu} \to W^+ W^-$
showed that the introduction of intermediate massive vector bosons, to make
the weak interaction nonlocal, was non-renormalizable.  However, with the
advent of the Higgs mechanism, it was successfully moulded into the
renormalizable theory discussed in the previous section, which allowed the
calculation of radiative corrections.

The weak neutral current (WNC), along with the $W$ and $Z$, have been the
primary predictions of the SM.  The WNC was discovered in 1973 by the
Gargamelle collaboration at CERN and by HPW at Fermilab.  The structure of the
WNC has been tested in many processes, including (purely weak) neutrino
scattering $ \nu e \ra \nu e, \; \nu N \ra \nu N, \; \nu N \ra \nu X$;
weak-electromagnetic interference in $e^{\pm} D \ra e X$,
atomic parity violation, and recently in polarized M\"oller scattering; and in
$e^+ e^-$ scattering above and below the $Z$ pole.  The $W$ and $Z$ were
discovered at CERN by the UA1~\cite{Arnison:1985jk} and
UA2~\cite{Ansari:1987vg} groups in 1983 and the subsequent measurements of
their masses have been in excellent agreement with the SM expectations
(including the higher-order corrections~\cite{Amsler:2008zz}) discussed in the
previous section.  The cynosure of the LEP legacy is the triumphant
verification of the gauge sector of the SM which involves the spontaneous
breaking of the gauge group: $SU(2)_L \times U(1)_Y \rightarrow U(1)_Q$.
Figure~\ref{SMgauge} obtained primarily from LEP II  runs clearly
verifies the SM gauge group. On one hand it clearly shows
the existence of the $ZWW$ vertex confirming the non-abelian 
nature of the gauge group. Indirectly it also validates the
idea of spontaneous symmetry breaking. To see this, note that
the intermediate gauge bosons  have to be massive to
explain the $\beta$ decay data. However, explicit
breaking leads to non-renormalizability. But the
good behavior of the cross section with energy in~Figure~\ref{SMgauge},
indicates a renormalizable theory and thus implies spontaneous
breaking of the gauge symmetry. In summary, this plot 
clearly indicates that the charged and neutral currents
in the particle gauge interaction are in accordance with the SM
prediction. This not only confirms the $SU(2)_L \times U(1)_Y$ gauge group but
also demonstrates that it is spontaneously broken to $U(1)_Q$.

The $Z$ factories LEP and SLC allowed tests of the standard model at a
precision of $\sim 10 ^{-3}$, much greater than what had previously been
possible at high energies. In particular, the four LEP experiments ALEPH,
DELPHI, L3, and OPAL at CERN produced some $2 \times 10^{7} Z$'s at the
$Z$-pole in the reactions $e^+ e^- \rightarrow Z \rightarrow \ell^+ \ell^-/q
\bar{q}$.  The SLD experiment at SLAC had a relatively smaller number of
events, $\sim 5 \times 10^5 $, but had the significant advantage of the high
polarization ($\sim$ 75\%) of the $e^-$ beam.  The $Z$ pole observables
included the lineshape variables, $M_Z, \Gamma_Z, $ and $\sigma$; and the
branching ratios into $e^+e^-, \mu^+ \mu^-, \tau^+ \tau^-$ as well as into $q
\bar{q}, c \bar{c}, b \bar{b},$ and (less precisely) $ s \bar{s}$.  These
could be combined to obtain the stringent constraint $N_\nu = 2.9841 \pm
0.0083$ on the number of ordinary neutrinos with $m_\nu < M_Z/2$ (i.e., on the
number of families with a light neutrino).  This gave the first experimental
indication of the three generation flavor structure of the SM . At present,
all the three pairs of quarks and leptons have been directly produced at
collider experiments that give hard evidence for  the three generation
conjecture.  This also constrained other invisible $Z$ decays.

The $Z$-pole experiments also measured a number of asymmetries, including
forward-backward (FB), polarization, $\tau$ polarization, and mixed
FB-polarization, which were especially useful in determining the weak angle
$\theta_W$.  The leptonic branching ratios and asymmetries confirmed the
lepton family universality predicted by the SM. The result of fitting these
observations with the SM predictions are generally in excellent
agreement. Figure~\ref{SMfit} shows the pull of the fittings in the SM.  There
is a hint of a tension between the lepton and quark asymmetries (most apparent
in the $b$ quark forward-backward asymmetry $A_{fb}^{0,b}$ and the
polarization asymmetry $A_l$.). This may well be a statistical fluctuation,
but could possibly be suggesting new physics affecting the third family.

\begin{center} \begin{figure} \centering{
\includegraphics[width=0.5\textwidth,angle=270,keepaspectratio]
{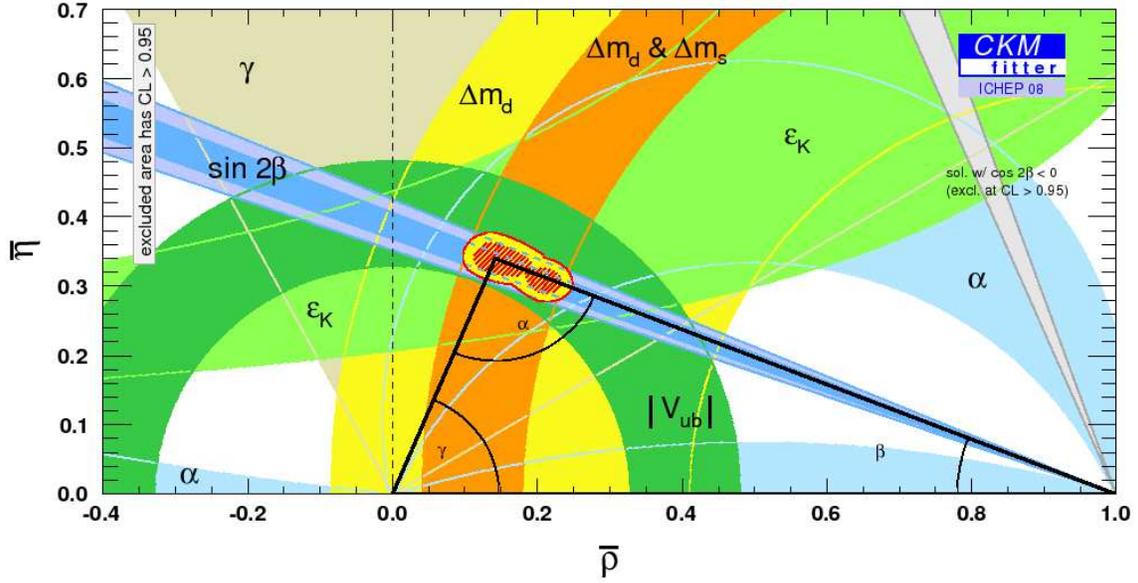}} \caption{ \small The unitarity
triangle, showing overlap regions of several CP-conserving and CP-violating
observables from the $K$ and $B$ systems. $\bar \rho$ and $\bar \eta$ are the
same as $\rho$ and $\eta$ up to higher order corrections, and
$\rho-i\eta=V_{ub}/(V_{cb}V_{us})$. Plot courtesy of the CKMfitter
group~\cite{Charles:2004jd}, {\tt http://ckmfitter.in2p3.fr}.}  \label{CKMfit}
\end{figure} \end{center}

The recent activity in charged current interaction is centered around the
study of the Cabibbo-Kobayashi-Maskawa (CKM) matrix which measures the
mismatch between the family structure of the left-handed $u$-type and $d$-type
quarks.  For 3 families, $V_{CKM}$ involves three angles and one
$CP$-violating phase after removing the unobservable $q_L$ phases as discussed
before.  There have been extensive recent studies, especially in $B$ and $K$
decays, to test the unitarity and consistency of $V_{CKM}$ as a probe of new
physics and to test the origin of CP violation. A global fit~\cite{CKMReview},
within the framework of the three-generation standard model, yields the
following \textit{magnitudes} $|{V_{ij}|}$ for the CKM matrix elements:
\small{ \be { \left( \begin{array}{ccc} 0.97419 \pm 0.00022 & 0.2257 \pm
0.0010 & 0.00359 \pm 0.00016 \\[4pt] 0.2256 \pm 0.0010 & 0.97334 \pm 0.00023 &
0.0415^{+0.0010}_{-0.0011} \\[4pt] 0.00874^{+0.00026}_{-0.00037} & 0.0407 \pm
0.0010 & 0.999133^{+0.000044}_{-0.000043} \end{array} \right).  }
\label{ckmparam} \ee } \normalsize The present experimental status of the
unitarity triangle is shown in Figure~\ref{CKMfit}.

\begin{center} \begin{figure} \centering{
\includegraphics[width=0.65\textwidth,angle=0,keepaspectratio]
{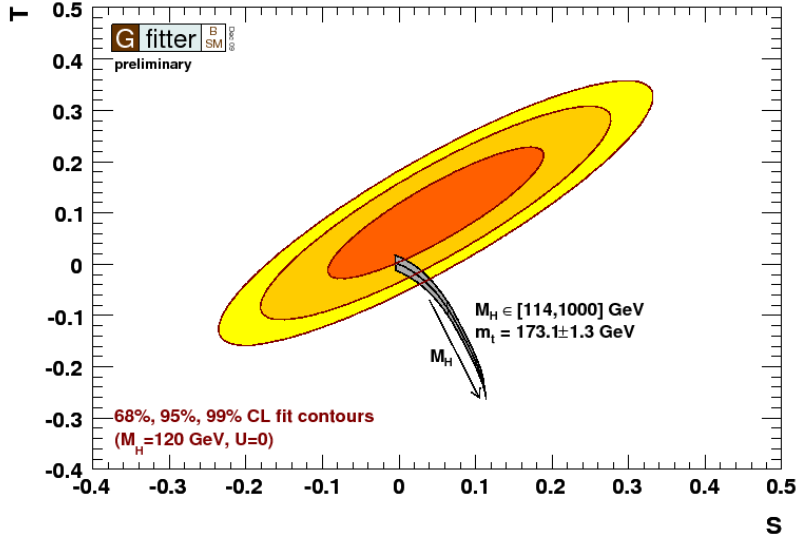}} \caption{ \small Contours of $68\%$, $95\%$, and $99\%$ CL
in the TS-plane. The gray region shows the prediction within the SM.
$M_h=120$ GeV and $m_t=173.2$ GeV defines the reference point at which all
oblique parameters vanish. Plot courtesy Gfitter
group~\cite{Flacher:2008zq}.}  \label{STplot} \end{figure} \end{center}

\textbf{Higher order:} The experimental probing of the SM has scrutinized it
beyond the tree level.  The present accuracy of experimental observations have
enabled us to probe the quantum corrections of the theory. A brief discussion
of this is in order, not only because it allows quantum verification of the
SM, but also because it puts stringent constraints on any further extension of
the theory. The discussion below closely follows the arguments laid out
in~\cite{Bhattacharyya:2009gw}.  Experimental measurements on the $Z$ pole at
LEP has verified the radiative corrections to the gauge boson propagators to
high precision. There are four two-point functions: $\Pi_{QQ}(q^2), \Pi_{Q
3}(q^2), \Pi_{33}(q^2), \Pi_{11}(q^2)$ where $Q \Rightarrow B_{\mu}$ and
$(1,2,3) \Rightarrow (W_{\mu}^1,W_{\mu}^2,W_{\mu}^3)$. Measurements have been
made at two energy scales: $q^2=0, M_Z^2$. So there are eight two-point
correlators.  Of these eight, $\Pi_{\gamma\gamma}(0)=\Pi_{\gamma Z}(0) =0$ due
to QED Ward identity\footnote{These identities ensure that the gauge
invariance of the classical Lagrangian is preserved after the quantization of
the theory.}. Three linear combinations can be absorbed in the redefinition of
the parameters: $\alpha$, $G_\mu$ and $M_Z$. The remaining three independent
combinations are called the Peskin-Takeuchi oblique electroweak parameters
($S$, $T$ and $U$). The parameters $T$ and $U$ capture the effects of
custodial symmetry and weak isospin violation, while $S$ is a measure of weak
isospin breaking alone \cite{stu}. Note that to cover all electroweak results,
one needs to expand the number of such parameters, see~\cite{Barbieri:2004qk}
for further details.  The definition of the parameters are given by,
\begin{eqnarray} \alpha T &=& \frac{\Pi_{WW}(0)}{M_W^2} -
\frac{\Pi_{ZZ}(0)}{M_Z^2} \nonumber \\ &=& \left(
\frac{e^2}{(sin({\theta_W})cos({\theta_W}))^2} \right) \frac{1}{M_Z^2}\left(
\Pi_{11}(0) - \Pi_{33}(0) \right) \label{tparam} \end{eqnarray} and \be S =
\frac{16\pi}{M_Z^2}\left( \Pi_{33}(M_Z^2) - \Pi_{33}(0) -\Pi_{3Q}(M_Z^2)
\right) \label{sparam} \ee where $\Pi_{ab}(q)$ is the vacuum polarization
amplitude with gauge bosons $a$ and $b$ in the external legs and the energy
scale associated with the amplitude is ${q}$.  A generic fermion-induced
vacuum polarization diagram with gauge bosons in the two external lines has
the following structure: \small \begin{eqnarray} &&F^{\mu\nu}
(m_1,m_2,\lambda,\lambda') = (-) \int {d^4k\over(2\pi)^4} \frac{{\rm Tr}
\left\{\g^\mu {1-\l\g_5\over2}(\slashed{q} + \slashed{k} +m_1) \g^\nu
{1-\lambda'\g_5\over2} (\slashed{k} +m_2)\right\}} {\{ (q+k)^2-m_1^2 \}(k^2
-m_2^2)} \nonumber \\ &&= \frac{i}{16\pi^2} \int_0^1dx \left[\D-\ln
\left\{\frac{-q^2 x(1-x)+m_1^2 x + m_2^2(1-x)}{\mu^2}\right\}\right]
\left[2(1+\lambda\lambda')x(1-x)(q_\mu q_\nu-q^2g_{\mu\nu}) \right. \nonumber
\\ &&\left. ~~~~+
(1+\lambda\lambda')(m_1^2x+m^2_2(1-x))g_{\mu\nu}-(1-\lambda\lambda')m_1m_2
g_{\mu\nu}\right] \, .  \end{eqnarray} \normalsize In the above equation,
$m_1$ and $m_2$ are the masses of the fermions in the loop, and $\Delta
(\equiv 2/(4-d) - \g + \ln 4\pi)$ is regularization scheme dependent divergent
quantity.  We are interested in the terms proportional to $g_{\mu\nu}$, the
$\Pi$-functions are defined as $ -i$ times these factors. By putting
$\lambda=1$ and $\lambda'=1$, we will get the left-left (LL) $\Pi$-function,
given by \small \begin{eqnarray} \Pi_{LL}(q^2,m_1^2,m_2^2)&=& -{1\over
4\pi^2}\int_0^1 dx \left[\D+\ln {\mu^2\over -q^2x(1-x)+ M^2(x)}\right]
\left[q^2x(1-x)-{1\over2}M^2(x)\right] \\ && \mbox{where,}~~~~~~
M^2(x)=m_1^2x+m_2^2(1-x) \nonumber.  \end{eqnarray} \normalsize Thus we find,
\begin{eqnarray} \Pi_{33}(q^2) & =& t^2_{3L} \Pi_{LL}(q^2,m^2,m^2) \\
\Pi_{11}(q^2) &=& \frac{1}{2} \Pi_{LL}(q^2,m_1^2,m_2^2).  \end{eqnarray}
\normalsize Now, supposing $m_1$ and $m_2$ are the masses of the two fermion
states appearing in an SU(2) doublet, it immediately follows that
\begin{eqnarray} \Pi_{33}(q^2) &=& \frac{1}{4} \left[\Pi_{LL}(q^2,m_1^2,m_1^2)
+ \Pi_{LL}(q^2,m_2^2,m_2^2)\right] \nonumber \\ \Pi_{11}(q^2) &=& \frac{1}{2}
\Pi_{LL}(q^2,m_1^2,m_2^2).  \end{eqnarray} One can now, in general, derive the
SM prediction of the oblique parameters by using the above general scheme. For
example the $T$ parameter is given by, \be T = {4\pi\over \sin^2 \t_W \cos^2
\t_W M_Z^2}\left[\Pi_{11}(0)-\Pi_{33}(0)\right].  \end{equation} The dominant
effect of isospin violation indeed comes from top-bottom mass splitting, given
by \begin{equation} T^{t-b} = {4\pi\over \sin^2 \theta_W \cos^2 \theta_W
M_Z^2} {N_c\over 32\pi^2} \left[{m_t^2+m_b^2\over2} - {m_t^2m_b^2\over
m_t^2-m_b^2} \ln {m_t^2\over m_b^2}\right] \simeq {1\over\pi} {m_t^2\over
M_Z^2} .  \end{equation} In the last step, we have assumed that $m_b^2 <
<m_t^2$. Note that in the limit $m_t = m_b$, the contribution to $T$ vanishes,
as expected.  The contribution of the Higgs boson arises from $ZZh$ and
$W^+W^-h$ interactions. It turns out that \begin{equation} \alpha T^h = -
\frac{3G_F}{8\pi^2\sqrt{2}} (M_Z^2 - M_W^2)
\ln\left(\frac{m_h^2}{M_Z^2}\right) \simeq -{\a\over2\pi}\ln{m_h\over M_Z} .
\end{equation} Figure~\ref{STplot} shows the presently allowed region in the
$S$-$T$ plane. Note that the SM contributions have been subtracted from the
parameter, i.e. $S\rightarrow S^{exp} -S^{SM}$ and $T \rightarrow T^{exp}
-T^{SM}$. The SM point on this plane would be the origin $(0,0)$ . Clearly
this is in good agreement with experimental bounds and thus puts strong
constraints on any further extension of the SM.

\subsection{Problems with the Standard Model}\label{problemssm} The SM is a
mathematically-consistent renormalizable gauge field theory which is
consistent with all experimental facts.  It successfully predicted the
existence and form of the weak neutral current, the existence and masses of
the $W$ and $Z$ bosons, and the fermion family structure, as necessitated by
the GIM mechanism.  The charged current weak interactions and quantum
electrodynamics are successfully incorporated into its folds.  The consistency
between theory and experiment indirectly tested the higher order corrections
which established the ideas of renormalization in the context of the SM.
When combined with
quantum chromodynamics for the strong interactions, the standard model is
almost certainly the approximately correct description of the elementary
particles and their interactions down to at least $10^{-16}$cm $\sim 1$ TeV. 

Despite its successes, the SM has a great deal of arbitrariness and
fine-tuning~\cite{ssm}, as is illustrated by the fact that it has 27 free
parameters (29 if we consider the Majorana neutrinos), and that is not
including electric charges. The parameters of the SM include: 3 gauge
couplings; the $Z$ and Higgs masses; the QCD $\theta$ parameter; 12 fermion
masses; 6 mixing and 2 CP phases (2 additional for Majorana $\nu$'s); and the
cosmological constant. The Planck scale (Newton constant) is not included
because only the ratios of mass parameters are observable.  It is believed
that this is a little too much for a fundamental theory of nature.  The status
of the laboratory/collider experiments in particle physics can best be
summarized as: they are in good agreement with the SM predictions but there is
still room for New Physics (NP) at the TeV or higher scale. At present there
seems to be a $3.1 \sigma$ discrepancy in the measurement of the anomalous
magnetic moment of the muon ($(g-2)_{\mu}$)\cite{Carey:2009zz}. There are some
tension in the field of b-physics as well. There is a $2 \sigma$ discrepancy
in the branching fraction of $D_s^+ \rightarrow l^+\nu$ and a $2.5 \sigma$
tension in the branching ratio of $B^+ \rightarrow \tau^+ \nu$. There are
several other unexpected observations in b-physics that hint at the existence
of NP at the TeV scale. In this regard the tension between the measured values
of $(\sin2\beta)_{\psi K_s} $ and $(\sin2\beta)_{\phi K_s} $ , the large
difference in the direct CP asymmetry $A_{CP}(B^- \rightarrow K^- \pi^0)$ and
$A_{CP}(\bar{B} ^0\rightarrow K^- \pi^+)$ etc are worth a
mention. See~\cite{Buras:2009if} for a recent review of flavor physics.

The first hint of beyond SM physics came from the observed neutrino
oscillations in solar neutrinos.  This implied a non-zero mass for the
neutrinos.  Although the original Glashow-Wienberg-Salam formulation did not
provide for massive neutrinos, they are however easily incorporated by the
addition of right-handed states $\nu_R$ (Dirac mass) or as higher-dimensional
operators, perhaps generated by an underlying seesaw (Majorana mass). The
successful explanation of light neutrino masses is considered as a major
outstanding issue with the SM. There are certain other severe deficiencies in
the SM. Some of them are enumerated below.

\begin{figure} \begin{minipage}{0.47\textwidth} \begin{center}
\includegraphics[width=\textwidth,angle=0,keepaspectratio] {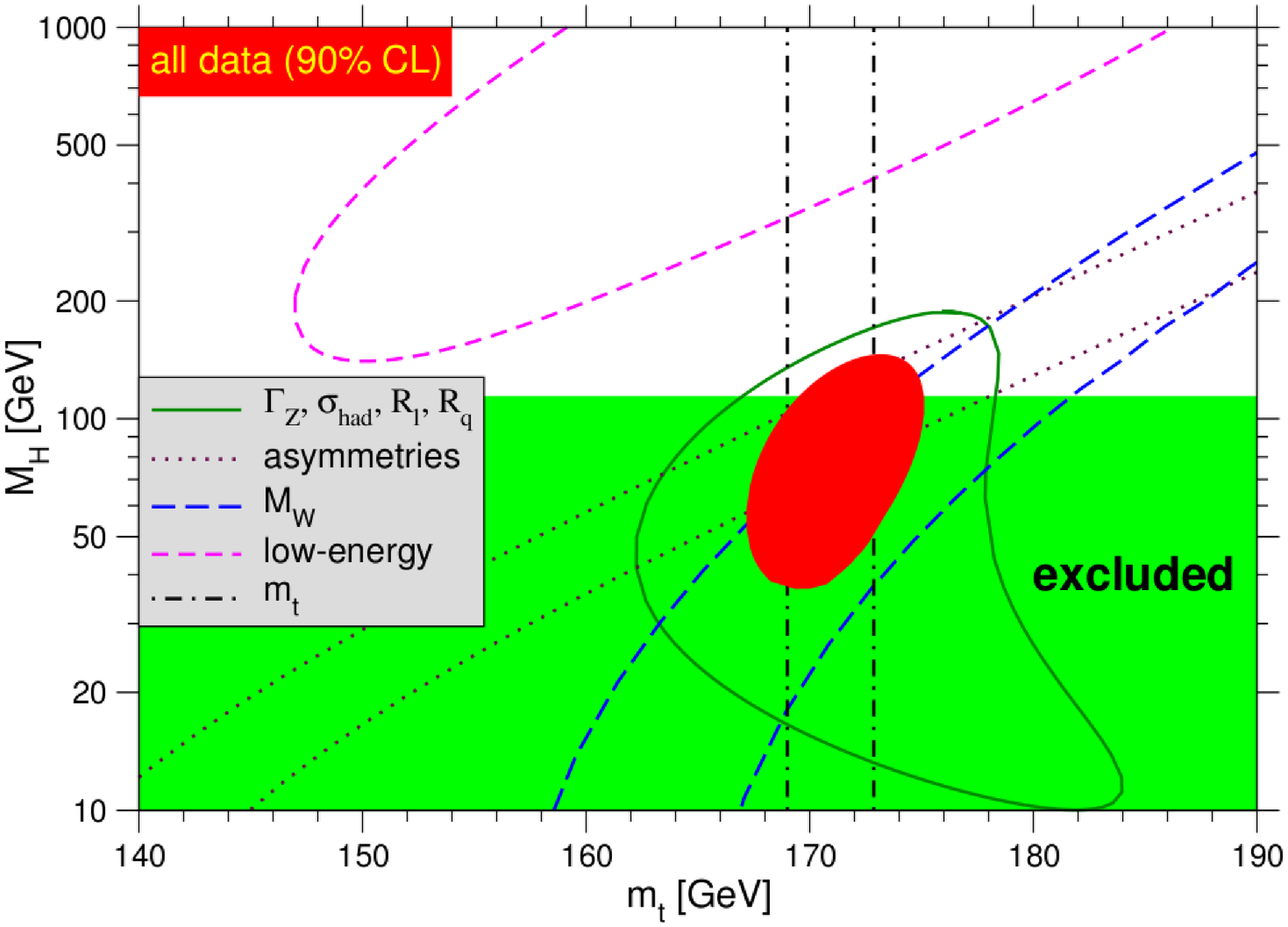}
\caption{ \small Allowed Higgs mass as a function of top mass.}
\label{hggmstm} \end{center} \end{minipage} \hspace{7mm}
\begin{minipage}{0.47\textwidth} \begin{center}
\includegraphics[width=\textwidth,angle=0,keepaspectratio]
{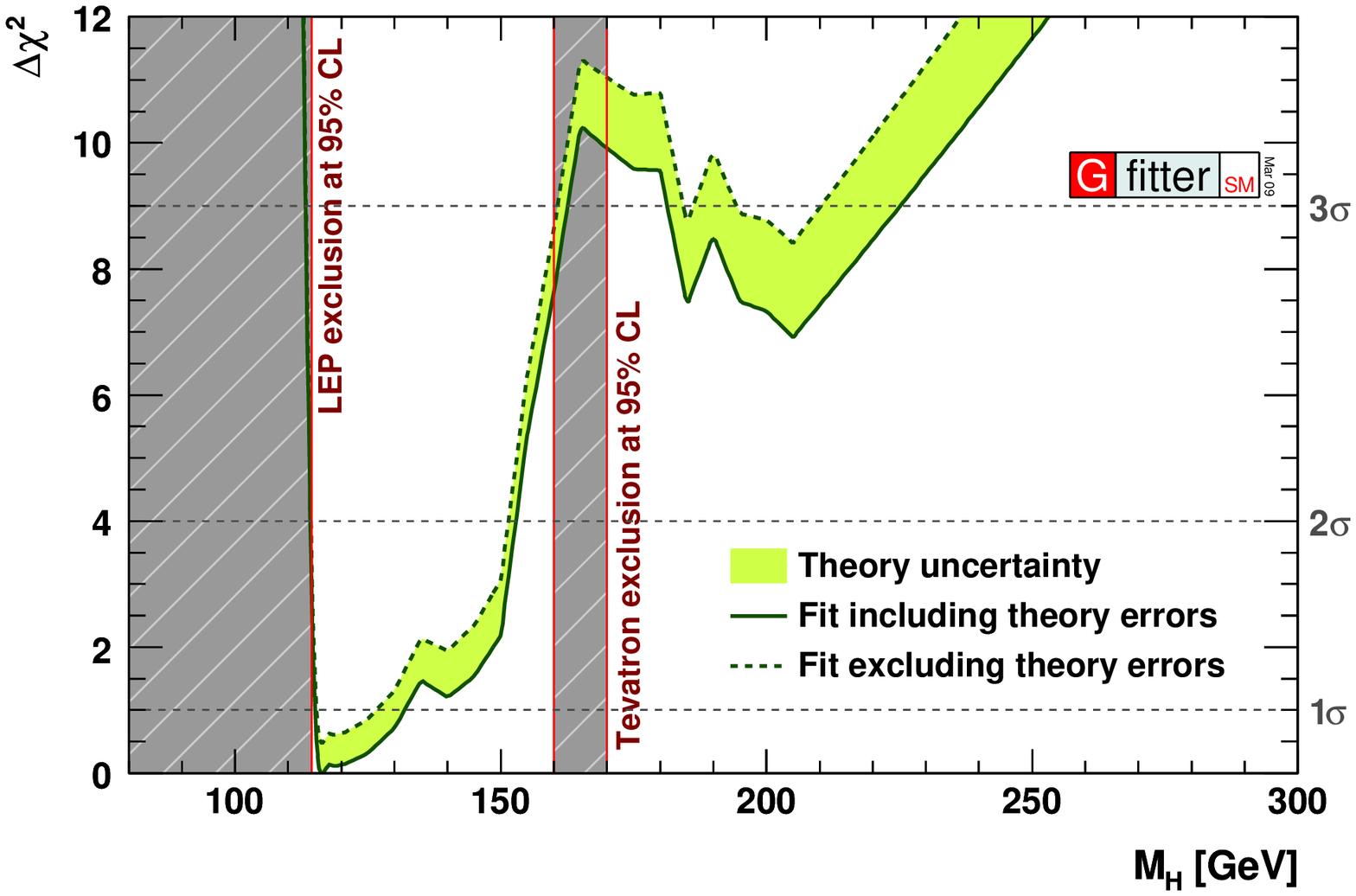} \caption{ \small $\Delta\chi^2$ as a function of the
Higgs-boson mass for the \textsf{Gfitter} complete fit, taking account of
direct searches at LEP and the Tevatron. The solid (dashed) line gives the
results when including (ignoring) theoretical errors. The minimum
$\Delta\chi^2$ of the fit including theoretical errors is used for both curves
to obtain the offset-corrected $\Delta\chi^2$~\cite{Flacher:2008zq}.}
\label{hgfit} \end{center} \end{minipage} \end{figure}
\begin{enumerate} 
\item \textbf{Cosmological consideration:} The observed
matter density of galaxies falls short of the measured matter as measured by
the rotation curves. It is theorized that the baryon matter density is $\sim 4
\%$. The rest of the universe is made up of $\sim 24\%$ dark matter and $\sim
72\%$ dark energy. In the last decade, the direct observation of gravitational
lensing and observations in galactic collision\cite{blltclst} (in the 'Bullet'
cluster) events have provided hard evidence for the existence of \textit{ Dark
Matter} (DM).  The WMAP probe has measured the dark matter density to be
between ($0.087<\Omega_{\rm DM} h^2<0.138$) \cite{Komatsu:2008hk} at $3\sigma$
range.  SM neither provides any explanation for dark energy nor does it have a
suitable dark matter candidate\footnote{Technically the QCD part of the SM
Lagrangian can have certain fields called the \textit{Axions}, theoretically
to be considered as a DM candidate.  The simplest version of this theory has
however failed to reconcile the observed dark matter density of the universe
with these axion fields.}. Similarly, the observed asymmetry between matter
and anti-matter in the universe quantified by $\eta \equiv
\frac{n_q-n_{\bar{q}}}{n_{\gamma}} \simeq (6.1^{+.3}_{-.2})\times 10^{-10}$,
cannot be explained within the framework of the SM. The minimum conditions
needed to explain this asymmetry is enshrined in the \textit{Sakharov
conditions}, not fulfilled by the SM. For example, the baryon number (B),
which should be broken to meet the Sakharov conditions, is an unbroken global
symmetry of the SM. Further, the magnitude of the CP violation generated by
the CKM picture in the SM is not sufficient to explain the baryon asymmetry in
the universe.

 \item \textbf{Gauge Hierarchy problem:} { Quantum theories involving
interacting elementary scalar fields are not natural.} This has to do with the
fact that the mass of an elementary scalar field is not associated with any
approximate symmetry. Let us consider a self-interacting theory of a real scalar
field: \be {\cal L}^{}_{scalar} =~ {\frac{1}{2}~ \partial^{\mu} \phi
\partial_{\mu} \phi ~-~ {\frac {m^2}{ 2}}~ \phi^2_{} ~-~ {\frac{\lambda}{4!}}~
\phi^4} \label{scln} \ee and consider that it is coupled to a fermion by
the following relation.  We can write the Yukawa interaction Lagrangian as \be
\mathcal{L}_Y =-h_f \phi \bar{f}_L f_R + ~\mbox{h.c.}.  \ee where $f_{L,R}$
are the left and right chiral projection of the fermion $f$.  After
spontaneous symmetry breaking, \begin{equation} \label{fermioncpl}
\mathcal{L}_Y = -\frac{h_f}{\sqrt{2}} h \bar f_L f_R -\frac{h_f}{\sqrt{2}} v
\bar{f}_L f_R + ~\mbox{h.c.}  \end{equation} The fermion mass is therefore
given by $m_f=h_f {v\over{\sqrt{2}}}$.

At the classical level, the limit mass $m \rightarrow 0$ does lead to scale
invariance; but at quantum level scale symmetry is broken. Thus smallness of
the scalar mass can not be protected against perturbative quantum
corrections. In fact such corrections appear with quadratic divergences.  Let
us compute the two-point function with the zero momentum Higgs as the two
external lines and fermions inside the loop.  The corresponding diagram is in
Figure~\ref{fig:higgscorr1}[a].  \begin{eqnarray} \label{pi_f} i
\Pi^f_{hh}(0)&=&(-)\int \frac{d^4k}{(2\pi)^4} {\rm Tr}~\left[ \left(-i
\frac{h_f}{\sqrt{2}}\right) \frac{i}{\slashed{k} - m_f}
\left(-i\frac{h_f}{\sqrt{2}}\right) \frac{i}{\slashed{k} - m_f} \right]
\nonumber\\ &=& -2h_f^2\int\frac{d^4k}{(2\pi)^4} \left[ \frac{1}{k^2 - m_f^2}
+ \frac{2m_f^2}{(k^2-m_f^2)^2}\right] .  \end{eqnarray} The correction $\D
m_h^2$ is proportional to $\Pi^f_{hh}(0)$.  The first term in the RHS is
quadratically divergent. The divergent correction to $m_h^2$ looks like
\begin{equation} \Delta m_h^2 (f) = \frac{\L^2}{16\pi^2} (-2h_f^2) \, .
\end{equation} Another divergent piece will appear from quartic Higgs vertex
(i.e., $h^4$). The corresponding diagram is similar to what is displayed in
Figure~\ref{fig:higgscorr1}[c], \begin{equation} \Delta m_h^2 (h) =
\frac{\L^2}{16\pi^2} (\l) \, .  \end{equation} We neglect the gauge boson
contributions to the scalar self energy.  Combining the above two divergent
pieces, we obtain \be \Delta m_h^2 = \frac{\L^2}{16\pi^2} \left(-2h_f^2 +
\l\right) .  \label{scradms} \ee

\begin{figure}
\begin{center} \begin{picture}(338,38)(-38,-1)
\Text(-38,0)[c]{$\Delta(m_{h}^2)\> =\, $} \SetWidth{0.7}
\DashLine(0,0)(25,0){5} \DashLine(65,0)(90,0){5} \SetWidth{1.3}
\CArc(45,0)(20,0,360) \Text(3,8)[c]{$h$} \Text(45,28)[c]{$f$}
\Text(45,-30)[c]{[a]} \SetWidth{0.7} \DashLine(120,0)(145,0){4.5}
\DashLine(185,0)(210,0){4.5} \SetWidth{1.3} \DashCArc(165,0)(20,180,360){4}
\DashCArc(165,0)(20,0,180){4} \Text(124,8)[c]{$h$} \Text(165,28)[c]{$h$}
\Text(165,-30)[c]{[b]} \SetWidth{0.7} \DashLine(240,-5)(275,-5){4.5}
\DashLine(310,-5)(275,-5){4.5} \SetWidth{1.3} \DashCArc(275,11)(16,-90,270){4}
\Text(243,3)[c]{$h$} \Text(275,35)[c]{$h$} \Text(275,-30)[c]{[c]}
\end{picture} \end{center} \vspace{1cm} \caption{\small{ One-loop quantum
corrections to the Higgs mass, due to a Dirac fermion $f$ [a], and scalars
$\tilde f_{L,R}$ ([b] \& [c]).\label{fig:higgscorr1}}} \end{figure}
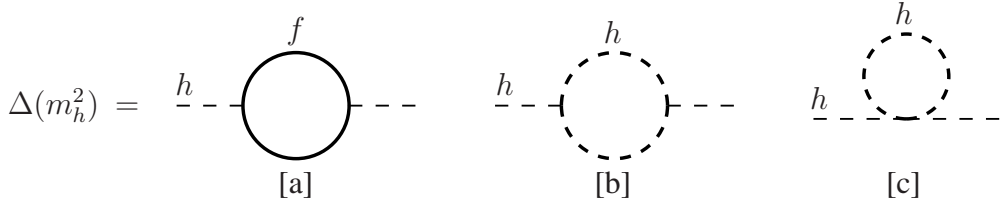
This illustrates the typical quantum correction to scalar fields generated at
one loop, that is quadratically divergent. The scalar sector of the SM faces a
similar predicament. In this regard let us note the following points:
\begin{itemize} \item In the SM, the fermion masses are protected by the
inexact chiral symmetry and the gauge boson masses are protected by the
remnant gauge symmetry after spontaneous electroweak symmetry breaking,
whereas, the Higgs field masses remain unprotected and receive quantum
corrections that are quadratically dependent on the cut off. As discussed
above, this is related to the inherent scale dependence of all fundamental
scalar theories.  \item By itself, this is not a catastrophe, as one can
envisage counter terms that will cancel such divergent quantum
corrections. Unfortunately, the cut off of the SM is believed to be of the
order of the Planck scale ($M_{pl} \sim 10^{19}$). Thus, to obtain a weak
scale Higgs mass, one needs an unnatural cancellation between two uncorrelated
numbers, i.e. the quantum correction and the counter term contribution. The
situation gets uglier when it is noted that such cancellation has to take
place order by order in the perturbation theory and there is no hope of
convergence at any finite order.  \item It is also worthwhile to know that
radiative corrections to the fermions or the gauge bosons are always
proportional to their masses. Thus one cannot generate the masses of these
fields purely from radiative contributions. This can be physically explained
by noting that there is a mismatch in the degrees of freedom of a massive and
massless gauge boson or fermion. The situation is completely different for the
case of the fundamental scalars. Here one can generate the masses radiatively
even if at tree level they are massless, as can be seen in
Eq.~\ref{scradms}. This is related to the fact that the $d. o. f.$ of a
massive scalar field is identical to that of a massless scalar field.
\end{itemize} Thus we find the lack of symmetry protecting the Higgs mass and
the large hierarchy between the weak scale and the Planck scale makes it
difficult to explain light Higgs mass within the SM. This is known as the
gauge hierarchy problem which is basically a naturalness issue with the SM.

On the other hand, the other parameter of this theory, namely the $h^4_{}$
coupling $\lambda$ is natural. This is so because, in the limit $\lambda
\rightarrow 0$, we have a free scalar theory, which indeed has higher
symmetry.

  \item \textbf{Gravity is not included:~}Gravity is not put on the same
 footing as other interactions in the SM.  The vacuum energy $ \langle V
 \rangle$ from electroweak symmetry breaking leads to an effective
 cosmological constant: $\Lambda_{\rm SSB} = 8 \pi G_N \langle V \rangle$
 which is some $ 10^{50} $ times larger than the value of the cosmological
 constant, observed from the acceleration of the universe.  Reconciliation
 with the observed value leads to extremely fine-tuned cancellation between
 the primordial value and the one generated dynamically by the electroweak
 symmetry breaking.  There is no known accepted solution to the cosmological
 constant problem, but see~\cite{landscape} for an anthropically motivated
 fine-tuning associated with the string landscape.

\end{enumerate}

Other than these severe shortcomings there are certain other criticism related
to the SM viz., \large{\textbf{(a)~}}\normalsize \textit{The strong CP
problem:} The strong CP problem~\cite{strongcp} refers to the fact that one
can add the P, T, and CP-violating term $\frac{\theta}{32 \pi^2} g_s^2 F
\tilde{F}$ to the QCD Lagrangian, where $\tilde{F}_{\mu \nu} = \epsilon_{\mu
\nu \alpha \beta} F^{\alpha \beta}/2$ is the dual field and $\theta$ is an
arbitrary dimensionless parameter.  The experimental bound on the neutron
electric dipole moment implies $\theta < 10^{-9}$. One cannot simply set
$\theta$ to zero because weak interaction corrections shift $\theta$ by
$\delta \theta|_{\rm weak} \sim 10^{-3}$, again requiring a fine-tuned
cancellation between the tree and weak
contributions. \large{\textbf{(b)~}}\normalsize \textit{The fermion mass
hierarchy problem:} Beyond the ordinarily observed matter content that can be
constituted by the following fermions $ (\nu_e, e^-, u, d)$, the first family
laboratory studies have confirmed the existence of $\geq 3$ families:
$(\nu_\mu, \mu^-, c, s)$ and $(\nu_\tau, \tau^-, t, b) $ are heavier copies of
the first family with no obvious explanation in the SM.  The SM gives no
prediction for the number of fermion generations. Furthermore, there is no
explanation or prediction of their masses, which are observed to have
hierarchical pattern spanning over 6 orders of magnitude between the top quark
and the electron.  Even more mysterious are the neutrinos, which are lighter
still by many orders of magnitude.  And \large{\textbf{(c)~}}
\normalsize\textit{The Gauge issue:} The SM gauge group is complicated: it
involves 3 distinct gauge couplings, of which only the electroweak part is
parity-violating, and charge quantization (e.g., $|q_e| = |q_p|$) is put in by
hand (anomaly cancellation by itself is not sufficient to determine all of the
hypercharge assignments). The issue of charge quantization is important
because it facilitates the electrical neutrality of atoms $(|q_p| = |q_e|)$.
The complicated gauge structure suggests that there exists underlying unity in
the interactions. This indicates the existence of
superstring~\cite{Green:110204,Polchinski:363850,Becker:1003112} or grand
unified
theory~\cite{Georgi:1974sy,Langacker:1980js,Ross:1107830,Hewett:1988xc,
Raby:2008gh}.
Charge quantization can also be explained in such classes of theories.  Charge
quantization may also be explained, at least in part, by the existence of
magnetic monopoles~\cite{Preskill:1984gd} or the absence of
anomalies\footnote{Anomaly cancellation is not sufficient to determine all of
the hypercharge assignments without additional assumptions, such as
universality of families.}.

 It is worthwhile to note that the complete
 experimental verification of the SM has to wait the discovery of the hitherto
 elusive Higgs boson. Non-observation of the Higgs fields at the LEP II
 directly excludes Higgs mass below $114.4$ GeV, whereas the precision
 electroweak observables prefer a Higgs mass below $\sim 160$ GeV
 \cite{Kielanowski:2003jg}. The major uncertainty in the electroweak fit of
 the Higgs mass comes from the uncertainty in the top quark mass. A plot of
 the values of the Higgs mass as a function of the top quark mass can be found
 in Figure~\ref{hggmstm}. The $\Delta \chi^2$ plot for the global fitting of
 the Higgs mass can be found in Figure~\ref{hgfit}. The LHC is expected to
 discover the Higgs field though accurate measurement of its mass has to wait
 for future experiments, certainly the proposed International Linear Collider
 (ILC) will be able to do a better job in this regard.  

\section{Beyond the Standard Model} \label{bsm_into}The above criticism of the
SM provides a strong motivation for advocating theoretical constructions that
extends the SM and solves some of it shortcomings. Most of the Beyond the
Standard Model (BSM) physics \cite{Bhattacharyya:2008ez} have been constructed
to solve the gauge hierarchy problem. The models that have been discussed in
the literarture may be categorized as follows:

\begin{enumerate} \item \textit{Models with no fundamental scalars:}
  Possibility to eliminate the elementary Higgs fields in favor of some
  {dynamical symmetry breaking} mechanism based on a new strong
  dynamics~\cite{Hill:2002ap}, \textit{e.g.} technicolor, higher dimensional
  Higgs-less models. In technicolor, for example, the SSB is associated with
  the expectation value of a fermion bilinear, analogous to the breaking of
  chiral symmetry in QCD. Extended technicolor, top-color, and composite Higgs
  models all fall into this class. Higher dimensional Higgs less
  models~\cite{Csaki:2003zu}) use the boundary conditions in the extra
  dimensions to break the electroweak symmetry.

  \item \textit{Models that invoke symmetry to protect Higgs mass:}
\textit{e.g.} supersymmetry, gauge-Higgs unified models, little Higgs. In
supersymmetry~\cite{Martin}, the quadratically-divergent contributions of
fermion and boson loops cancel, leaving only much smaller effects of the order
of supersymmetry-breaking.  There are also (non-supersymmetric) extended
models in which the cancellations are between bosons or between fermions. This
class includes Little Higgs
models~\cite{ArkaniHamed:2002qy,Perelstein:2005ka}, in which the Higgs is
forced to be lighter than new TeV scale dynamics because it is a
pseudo-Goldstone boson of an approximate underlying global symmetry, and
Twin-Higgs models~\cite{Chacko:2005pe}.

   \item \textit{Models that try to bridge the gap between the two scales of
the Standard Model:} \\ \textit{e.g.} ADD - large extra dimension, RS - warped
extra dimension. In these models space-time geometry is used to relate
$M_{pl}$ and a much lower fundamental scale, by providing a cutoff at the
inverse of the extra dimension scale. See ~\cite{add,Randall:1999ee} for
further details. \end{enumerate}

\section{Supersymmetry}

\begin{figure}  \begin{minipage}{0.47\textwidth}
 \begin{center}
\includegraphics[width=\textwidth,angle=0,keepaspectratio]
{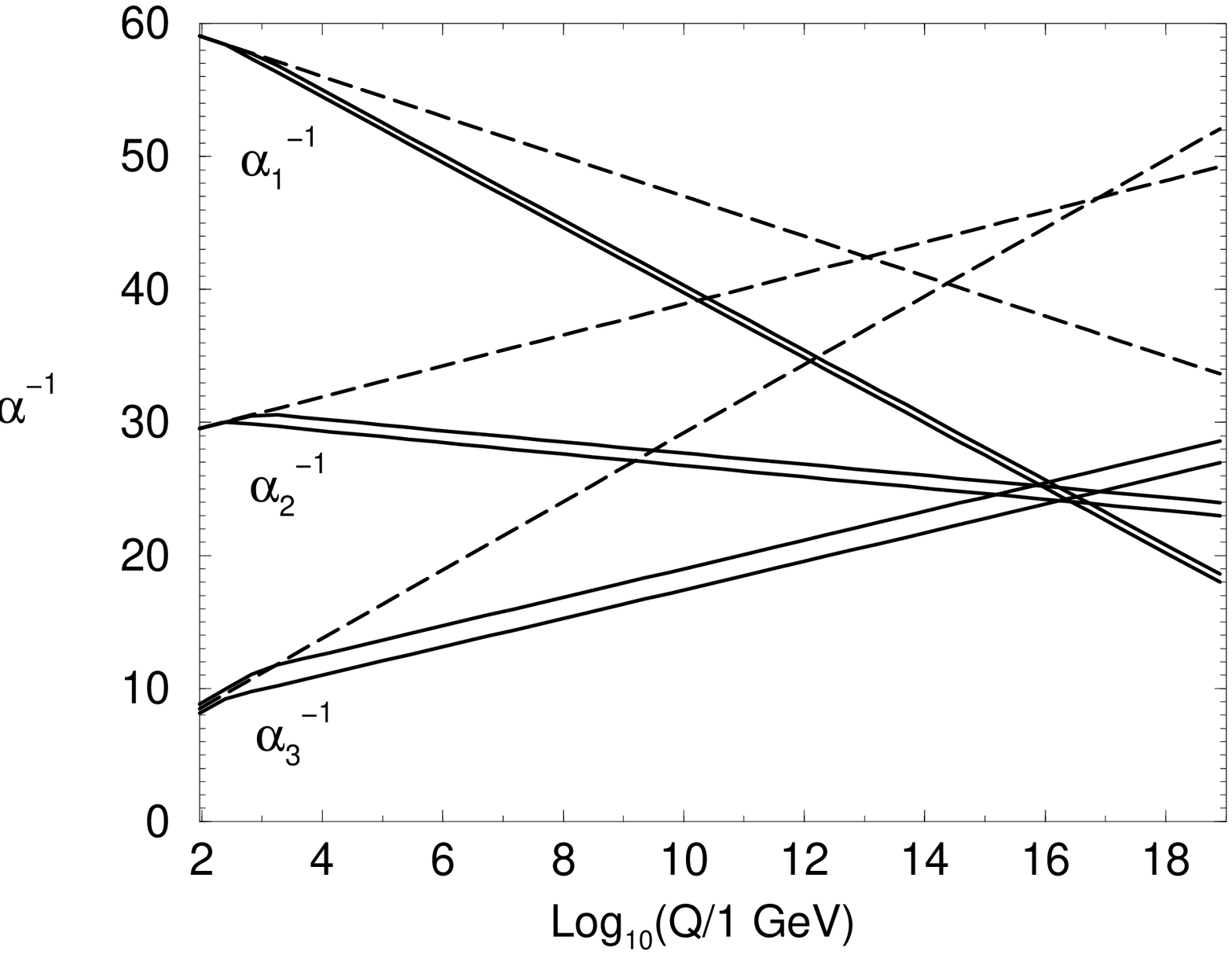}
\caption{ \small The running Gauge coupling unification within the framework 
of MSSM \cite{uni}.} 
\label{gumssm} \end{center} \end{minipage} \hspace{7mm}
\begin{minipage}{0.47\textwidth} \begin{center}
 \includegraphics[width=\textwidth,angle=0,keepaspectratio]
{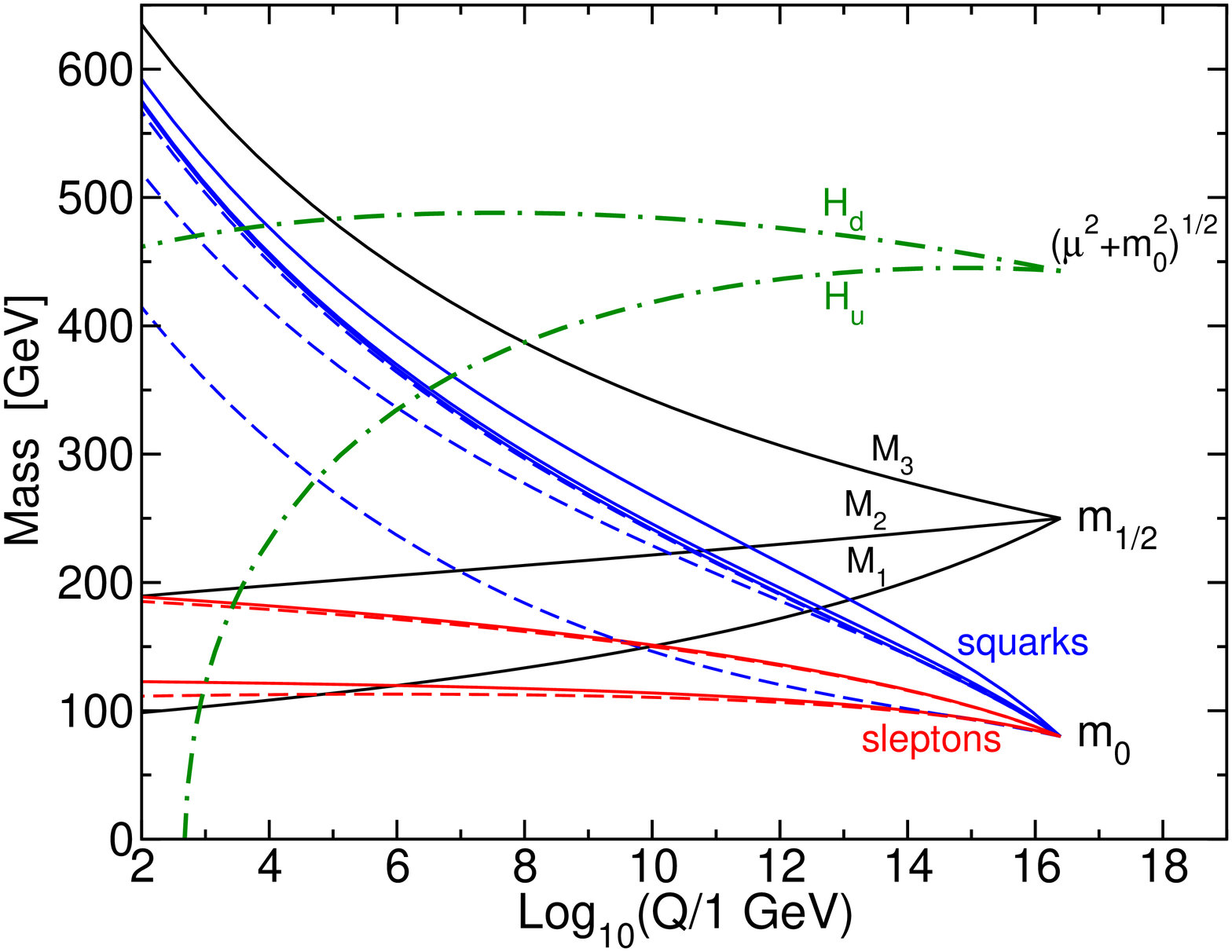} \caption{ \small The
renormalization group running of soft parameters in mSUGRA. Negative
$m_{H_u^2}$ in low energy triggers EWSB \cite{Martin}.}  \label{rewsbmssm}
\end{center} \end{minipage} \end{figure}

Let us tweak the analysis we did to reach Eq~\ref{scradms}. Consider that the
scalar inside the loop in Figure~\ref{fig:higgscorr1} [c] is not $\phi$ but
some different scalar field $\tilde{f_L}/\tilde{f_R}$, where the coupling is
$\sim \lambda \phi^2 \tilde{f}_{L/R}^2$.  Note that if there are two such
scalars ($\tilde{f}_{L}$ and $\tilde{f}_{R}$), the Eq.~\ref{scradms} becomes,
\be \Delta m_h^2 = \frac{2\L^2}{16\pi^2} \left(-h_f^2 + \l\right).  \ee

We find that the entire quadratic divergence piece in the quantum correction
to the scalar mass vanishes if, \be h_f^2 = \l.  \label{susycon} \ee There are
pairwise cancellations between fermionic contributions and the contributions
from a pair of scalars. Apriori, such relations between the coupling of two
fields are unnatural.  Supersymmetry is a space-time symmetry which relates
the bosonic degrees of freedom to the fermionic degrees of freedom
\cite{books, Martin, Ellis:2007wa}, and thus can justify relations like the
one expressed in Eq.~\ref{susycon}.

Supersymmetry (SUSY) is the most popular extension of the SM because it
provides a very aesthetic way to address the gauge hierarchy problem and
ameliorate various other shortcomings of the SM. Owing to its overwhelming
popularity in the parlays of particle physics, a brief discussion of SUSY is
now in order.  Some of the attractive features of the SUSY models are:
\begin{enumerate} \item \textbf{Supersymmetry solves the gauge hierarchy
problem :} As discussed, the quantum corrections to the Higgs mass from a
bosonic loop and a fermionic loop have {\em opposite} signs.  So if the
couplings are identical and boson is mass degenerate with the fermion, the net
contribution would cancel!  Supersymmetry fits this bill very well, as for
every particle, supersymmetry provides a mass degenerate\footnote{ The non
observation of the SUSY partners necessitates the breaking of SUSY in the real
world, as we will see later.  But if the breaking occurs through `soft' terms,
i.e., in masses and not in couplings, the condition for cancellation of
quadratic divergence given in Eq.~\ref{susycon} still remains valid. The
residual divergence is logarithmically sensitive to the supersymmetry breaking
scale.} partner differing by spin $1\over 2$ and having identical couplings.

 \item \textbf{Supersymmetry leads to unification of gauge couplings:} In the
 SM, when the gauge couplings are extrapolated to high scale from their
 measured values at the weak scale, they come close to each other but do not
 meet at a single point. In supersymmetry, the running gauge couplings do meet
 at a point\footnote{This provides motivation for construction of
 supersymmetric grand unified theories that can unify the electroweak
 interactions into a single gauge group. In many of these models
the leptons and quarks are
 incorporated into a single representation of the gauge group.}
~\cite{uni}, at the scale $M_{\rm GUT} \sim 2 \times 10^{16}$ GeV,
 provided the superparticles weigh around 1 TeV, see Figure~\ref{gumssm}.

 \item \textbf{Supersymmetry triggers EWSB:} To drive spontaneous symmetry
breaking in SM, one requires to set the scalar mass in the
Lagrangian, to a negative value by hand. In SUSY
theories, the square of one of the Higgs mass $m_{H_u}^2$, can be made
negative by radiative correction. In the Minimal Supersymmetric Standards
Model (MSSM) that we shall discuss later, one can start with a positive value
of the Higgs mass at the gauge coupling unification ($M_{\rm GUT}$) scale. The
running of the parameters drives the $m_{H_u}^2$ to a negative value at the
weak scale driving electroweak symmetry breaking, see
Figure~\ref{rewsbmssm}. In MSSM it is the heavy top quark contribution to the
radiative correction that induces the sign flip.

 \item \textbf{Supersymmetry provides a cold dark matter candidate:}
Supersymmetry with conserved $R$-parity can provide a dark matter
candidate. The lightest supersymmetric particle (LSP) cannot decay due to the
$R$-parity that forbids vertices with odd number of super-partners of the SM
fields. Thus the LSP is a stable particle and a viable cold dark matter
candidate.

 \item \textbf{Supersymmetry provides a framework to turn on gravity:} As
discussed earlier, SM do not provide a framework to unify gravity with the
other particle interactions. But SUSY does better in this regard. Space-time
transformations are naturally included in the SUSY transformations. Local
supersymmetry leads to supergravity that gives a gateway to include gravity in
a quantum field theoretic famework. Most string models invariably include
supersymmetry as an integral part.  \end{enumerate}

\subsection{SUSY algebra}

 Supersymmetry is a general space-time symmetry that is allowed by the
Poincare algebra. Unlike the Lorentz transformations supersummetric
transformations are mediated by fermionic charges. A supersymmetry
transformation turns a bosonic state into a fermionic state, and vice
versa. The operator $Q$ that generates such transformations must be an
anti-commuting spinor, generating the following transformations, \be Q |{\rm
Boson}\rangle = |{\rm Fermion }\rangle, \qquad\qquad Q |{\rm Fermion}\rangle =
|{\rm Boson }\rangle .  \ee Spinors are intrinsically complex objects, so
$Q^\dagger$ (the hermitian conjugate of $Q$) is also a symmetry generator.
Note that in general there can be arbitrary number of such generator pairs
($Q_i \& Q_i^\dagger$) that can simultaneously generate SUSY transformations.
The number of such generators are going to be represented by $N$. 
Increase in $N$ generally results in more symmetric and therefore more
constrained theories. In this
chapter we will stick to the $N=1$ version of the theory.
The possible forms for such symmetries in a quantum field theory are highly
restricted by the no go theorem put forward by Haag-Lopuszanski-Sohnius, which
is basically an extension of the Coleman-Mandula theorem \cite{HLS}. The
basic result of this theorem is, that space-time symmetry transformations by
generators of spin greater than 1 is prohibited.

 Generic
supersymmetric charges satisfy the algebra of anti-commutation and commutation
relations with the schematic form \begin{eqnarray} &&\{ Q, Q^\dagger \} =
P^\mu , \label{susyalgone} \\ &&\{ Q,Q \} = \{ Q^\dagger , Q^\dagger \} = 0 ,
\label{susyalgtwo} \\ &&[ P^\mu , Q ] = [P^\mu, Q^\dagger ] = 0
,\label{susyalgthree} \end{eqnarray} where $P^\mu$ is the four-momentum
generator of space-time translations.  Here we have suppressed the spinorial
index. Note that the appearance of $P^\mu$ on the right-hand side of
Eq.~\ref{susyalgone} is understandable, since it transforms under Lorentz boosts
and rotations as a spin-1 object while $Q$ and $Q^\dagger$ on the left-hand
side,  each transforms as a spin-1/2 object. This natural appearance of the
generator for space-time translation provides a handle to incorporate gravity
in SUSY theories.

The single-particle states of a supersymmetric theory fall into irreducible
representations of the supersymmetry algebra, called \textit{
supermultiplets}. Each supermultiplet contains both fermionic and bosonic
states, which are called \textit{ superpartners} of each other. If
$|\Omega\rangle$ and $|\Omega^\prime \rangle$ are members of the same
supermultiplet, then the $|\Omega^\prime\rangle$ can be obtained by operating
some combination of $Q$ and ${Q}^\dagger$ operators on $|\Omega\rangle $, up
to a space-time translation or rotation. The squared mass operator $ -P^2$
commutes with the operators $Q$, ${Q}^\dagger$, and with all space-time
rotation and translation operators.  It follows immediately that members of
the same supermultiplet will have equal mass eigenvalues i.e they will be mass
degenerate.  The supersymmetry generators $Q,Q^\dagger$ also commute with all
internal symmetry generators in general and the generators of gauge
transformations in particular. Therefore particles in the same supermultiplet
must also be in the same representation of the gauge group, i.e.  same
electric charges, weak isospin, color degrees of freedom etc.

Each supermultiplet contains an equal number of fermionic and bosonic degrees
of freedom. This can be demonstrated easily. Consider the operator $(-1)^{2s}$
where $s$ is the spin angular momentum. By the spin-statistics theorem, this
operator has eigenvalue $+1$ acting on a bosonic state and eigenvalue $-1$
acting on a fermionic state. Any fermionic operator will turn a bosonic state
into a fermionic state and so on. Therefore $(-1)^{2s}$ must anti-commute with
every fermionic operator in the theory, and in particular with $Q$ and
$Q^\dagger$. Now, within a given supermultiplet, consider the subspace of
states $| i \rangle$ with the same eigenvalue $p^\mu$ of the four-momentum
operator $P^\mu$. In view of Eq.~\ref{susyalgthree}, any combination of $Q$ or
$Q^\dagger$ acting on $|i\rangle$ must give another state $|i^\prime\rangle$
with the same four-momentum eigenvalue. Therefore one has a completeness
relation $\sum_i |i\rangle\langle i | = 1$ within this subspace of states. Now
one can take a trace over all such states of the operator $(-1)^{2s} P^\mu$
(including each spin helicity state separately): \begin{eqnarray} \sum_i
\langle i | (-1)^{2s} P^\mu | i \rangle &=& \sum_i \langle i | (-1)^{2s} Q
Q^\dagger|i\rangle +\sum_i\langle i | (-1)^{2s} Q^\dagger Q | i \rangle
\nonumber\\ &=& \sum_i \langle i | (-1)^{2s} Q Q^\dagger | i \rangle + \sum_i
\sum_j \langle i | (-1)^{2s} Q^\dagger |j \rangle \langle j | Q | i
\rangle\qquad{} \nonumber\\ &=& \sum_i \langle i | (-1)^{2s} Q Q^\dagger | i
\rangle + \sum_j \langle j | Q (-1)^{2s} Q^\dagger | j \rangle \nonumber\\
&=&\sum_i \langle i | (-1)^{2s} Q Q^\dagger | i \rangle - \sum_j \langle j |
(-1)^{2s} Q Q^\dagger | j \rangle \nonumber \\ &=& 0.  \end{eqnarray} The
first equality follows from the supersymmetry algebra relation
Eq.~\ref{susyalgone}; the second and third from use of the completeness
relation; and the fourth from the fact that $(-1)^{2s}$ must anti-commute with
$Q$. Now $\sum_i \langle i | (-1)^{2s} P^\mu | i \rangle = \, p^\mu$
Tr[$(-1)^{2s}$] is just proportional to the number of bosonic degrees of
freedom $n_B$ minus the number of fermionic degrees of freedom $n_F$ in the
trace, so that \be n_B= n_F \label{nbnf} \ee must hold for a given $p^\mu\neq
0$ in each supermultiplet.

The simplest possibility for a supermultiplet consistent with Eq.~\ref{nbnf}
has a single Weyl fermion (with two spin helicity states, so $n_F=2$) and two
real scalars (each with $n_B=1$). It is natural to assemble the two real
scalar degrees of freedom into a complex scalar field. This combination of a
two-component Weyl fermion and a complex scalar field is called a
\textit{chiral} supermultiplet.

Another possibility for a supermultiplet contains a spin-1 vector boson. If
the theory is to be renormalizable, this must be a gauge boson that is
massless, at least before the gauge symmetry is spontaneously broken. A
massless spin-1 boson has two helicity states, so the number of bosonic
degrees of freedom is $n_B=2$. Its superpartner is therefore a massless
spin-1/2 Weyl fermion, again with two helicity states, so $n_F=2$. Gauge
bosons transform in the adjoint representation of the gauge group, so
their fermionic superpartners, called {\it gauginos}, must also follow suit.
 Since the adjoint
representation of a gauge group is self conjugate, the gaugino fermions must
have the same gauge transformation properties for left-handed and for
right-handed components. Such a combination of spin-1/2 gauginos and spin-1
gauge bosons is called a \textit{ vector} supermultiplet.

\subsection{The generic SUSY Lagrangian}

Before zooming into the supersymmetric extension of the standard model we
review the generic features of a SUSY Lagrangian.

Consider a massless and therefore two-component Weyl fermion, $\psi$ whose
superpartner is a complex scalar $\phi.$ Both have two real degrees of
freedom.  However in the off-shell condition, the fermion is a four-component
field with four degrees of freedom, and we want supersymmetry to hold for the
full field theory.  So we introduce an additional complex scalar $F$ to match
the off-shell degrees of freedom.  $F$ is called an auxiliary field and has no
physical particle interpretation.  A complete chiral superfield will thus
contain the fields $(\psi,\phi,F)$.  The Lagrangian can be written as

\be -\mathcal{L}_{chiral}=\sum_{i}(\partial^{\mu}\phi_{i}^{\ast}\partial_{\mu}
\phi_{i}
+\bar{\psi}_{i}\gamma^{\mu}\partial_{\mu}\psi_{i}+F_{i}^{\ast}F_{i}).
\label{susy:lagchiral} \ee

The sum is over all chiral supermultiplets in the theory.  Note that the
dimensions of $F$ are $[F]=m^{2}.$ The Euler-Lagrange equations of motion for
$F$ are $F=F^{\ast}=0,$ signifying the fact that they are not physical
fields. The supersymmetry transformations defined above are so that
$\mathcal{L}_{chiral}$ is invariant.  Next we write the most general set of
renormalizable interactions, \be
\mathcal{L}_{int}=-\frac{1}{2}W^{ij}\psi_{i}\psi_{j}+W^{i}F_{i}+c.c. 
 \label{lagsup} \ee
where $W^{ij}$ and $W^{i}$ are functions of only the scalar fields
(i.e. $\phi_i$'s in our context), and $W^{ij}$ is symmetric.  If they depend
on the fermion or auxiliary fields the associated terms would have dimension
greater than four, and therefore would become non-renormalizable.

The SUSY transformations mix fermions and bosons,
$\phi\rightarrow\phi+\varepsilon\psi,\psi\rightarrow\psi+\varepsilon\phi$.
Here $\varepsilon$ must be a spinor so each term behaves the same way in spin
space, and we can take $\varepsilon$ to be a constant spinor in space-time,
and infinitesimal, which corresponds to a global SUSY transformation.  Then
the variation of the Lagrangian (which must vanish or change only by a total
derivative if the theory is invariant under the supersymmetry transformation)
contains two terms with four spinors: \be \delta
\mathcal{L}_{int}=-\frac{1}{2}\frac{\delta W^{ij}}{\delta\phi_{k}}(\varepsilon
\psi_{k})\psi_{i}\psi_{j}-\frac{1}{2}\frac{\delta W^{ij}}{\delta\phi_{k}
^{\ast}}(\varepsilon^{\dagger}\psi_{k}^{\dagger})\psi_{i}\psi_{j} +c.c.  \ee

Neither term can cancel against some other term.  For the first term there is
a Fierz identity $(\varepsilon\psi_{i})(\psi_{j}\psi_{k})+(\varepsilon\psi
_{j})(\psi_{k}\psi_{i})+(\varepsilon\psi_{k})(\psi_{i}\psi_{j})=0$, so if and
only if $\delta W^{ij}/\delta\phi_{k}$ is totally symmetric under interchange
of i, j and k, the first term vanishes identically. For the second term, the
presence of the hermitian conjugation allows no similar identity, so it must
vanish explicitly, which implies $\delta W^{ij}/\delta\phi_{k}^{\ast}=0,$and
thus $W^{ij}$ cannot depend on $\phi^{\ast}$!  $W^{ij}$ must be an analytic
function of the complex field $\phi.$ Therefore we can write \begin{eqnarray}
W^{ij}=M^{ij}+y^{ijk}\phi_{k}, \end{eqnarray} where $M^{ij}$ is a symmetric
matrix that will be the fermion mass matrix, and $y^{ijk}$ can be called
general SUSY version of the SM Yukawa couplings.  Then it is very convenient
to define

\be
W_{super}=\frac{1}{2}M^{ij}\phi_{i}\phi_{j}+\frac{1}{6}y^{ijk}\phi_{i}\phi_{j}
\phi
_{k} \label{superpot} \ee and
$W^{ij}=\delta^{2}W/\delta\phi_{i}\delta\phi_{j}.$ $W_{super}$ is the
\textit{superpotential}, an analytic function of $\phi$, and a central
function of the formulation of the theory.  $W$
is by construction, gauge invariant and Lorentz invariant, and an
analytic function of $\phi$ (i.e. it cannot depend explicitly on
$\phi^{\ast}$), so it is highly constrained\footnote{ For unbroken
supersymmetry there is a very important result, called the
\textbf{non-renormalization theorem}. In gist, the result implies that
superfields can only get a wave function renormalization in $N=1$ SUSY, so
they have the familiar log renormalization group running of couplings and
masses.  Consequently the parameters of the superpotential $W$ are not
renormalized, in any order of perturbation theory.  In particular, terms that
were allowed in $W$ by gauge invariance and Lorentz invariance are not
generated by quantum corrections if they are not present at tree level.}.  
It determines the most general
non-gauge interactions of the chiral superfields.

A similar argument for the parts of $\delta \mathcal{L}_{int}$ which contains
a spacetime derivative implies that $W^{i}$ is determined in terms of $W$ as
well, \be W^{i}=\frac{\delta
W}{\delta\phi_{i}}=M^{ij}\phi_{j}+\frac{1}{2}y^{ijk} \phi_{j}\phi_{k}.  \ee

Because of the interaction terms, the equations of motion for $F$ becomes
non-trivial, and are now modified to, \be F_{i}=-W_{i}^{\ast}.  \ee The
potential for the scalar fields of the theory is now given by, \be
V=\sum_{i}\left| F_{i}\right| ^{2}. \ee This part of the
scalar potential is called the ``F-term'' contribution, and 
is automatically bounded from below, an important feature of
SUSY theories.

Now consider massless gauge bosons, like photons, $A_{\mu}^{a},$ with gauge
index $a,$ and two degrees of freedom.  Their superpartners are two-component
spinors $\lambda^{a}.$ As stated earlier, the off shell fermion has four
degrees of freedom, while the an off shell boson has three, the two transverse
polarizations and a longitudinal polarization.  So again it is necessary to
add an auxiliary field, a real one since only one degree of freedom is needed,
called $D^{a}.$ Then the complete Lagrangian has additional pieces \be
\mathcal{L}_{gauge}=-\frac{1}{4}F_{\mu\nu}^{a}F_{a}^{\mu\nu}-i\lambda^{\dagger
a} \gamma^{\mu}D_{\mu}\lambda^{a}+\frac{1}{2}D^{a}D^{a}, \label{susy:laggauge}
\ee where, as usual, \be
F_{\mu\nu}^{a}=\partial_{\mu}A_{\nu}^{a}-\partial_{\nu}A_{\mu}^{a}
-gf^{abc}A_{\mu}^{b}A_{\nu}^{c}, \ee and the covariant derivative is \be
D_{\mu}\lambda^{a}=\partial_{\mu}\lambda^{a}-gf^{abc}A_{\mu}^{b}\lambda ^{c}.
\label{susy:covdiv} \ee It is crucial for gauge invariance that the same
coupling $g$ appears in the definition of the tensor $F_{\mu\nu}$ and in the
covariant derivative.

If we couple the chiral superfield with the vector superfields we must replace
all the derivatives in Eq.~\ref{susy:lagchiral} by the corresponding covariant
derivatives.  There are additional gauge invariant term to be added to the
Lagrangian beyond the ones discussed above given by,$
(\phi_{i}^{\ast}T^{a}\phi_{i})D^{a}$ and $\lambda ^{\dagger
a}(\psi^{\dagger}T^{a}\phi),$ and its conjugate, with an arbitrary
dimensionless coefficient.  Requiring the entire Lagrangian to be invariant
under supersymmetry transformations determines the arbitrary coefficient and
gives the final a resulting Lagrangian \begin{eqnarray}
\mathcal{L}=\mathcal{L}_{gauge}+\mathcal{L}^{cov}_{chiral}+g_{a}(\phi^{\ast}T^{a}\phi)D^{a}-\sqrt{2}g_{a}
[(\phi^{\ast}T^{a}\psi)\lambda^{a}+\lambda^{\dagger a}(\psi^{\dagger}T^{a}
\phi)] \label{susy:lagfull} \end{eqnarray} where all derivatives in earlier
forms are replaced by covariant ones. Note that the requirement of
supersymmetry requires that the couplings in the last two terms be gauge
couplings, even though they are not normal gauge interactions! The chiral part
of the Lagrangian $\mathcal{L}_{chiral}$ can be explicitly written as,
\begin{eqnarray}
\mathcal{L}_{chiral}&=&D^{\mu}\phi_{i}^{\ast}D_{\mu}\phi_{i}+\bar{\psi}_{i}\gamma^{\mu
}D_{\mu}\psi_{i} \nonumber \\
&+&(\frac{1}{2}M_{ij}\psi_{i}\psi_{j}+\frac{1}{2}y^{ijk}\phi
_{i}\psi_{j}\psi_{k}+c.c.)+F_{i}^{\ast}F_{i}.  \end{eqnarray}

The equations of motion for $D^{a}$ give $D^{a}=-g(\phi^{\ast }T^{a}\phi),$ so
the expanded scalar potential is now given by \be V=F^{\ast
i}F_{i}+\frac{1}{2}\sum_{a}D^{a}D^{a}=\left| \partial
W/\partial\phi_{i}\right| ^{2}+\frac{1}{2}\sum_{a}g_{a}^{2}(\phi^{\ast}T^{a}\phi )^{2},
\label{susyfulpot} \ee the sum is over $a=1,2,3$  for the three gauge
couplings.  The two terms are called \textit{F-terms} and
\textit{D-terms}. Note that even now the scalar potential is bounded from below
\footnote{ On one hand this is good since unbounded potentials are a
problem, but it also implies that the Higgs mechanism cannot happen for
unbroken supersymmetry since the potential will be minimized at the origin.}.

\subsection{The Minimal Supersymmetric Standard Model} \label{MSSMsection}
 \begin{table}[]
\begin{center} \begin{tabular}{|c|c|c|c|c|} \hline \multicolumn{2}{|c|}{Names}
& spin 0 & spin 1/2 & $SU(3)_C ,\, SU(2)_L ,\, U(1)_Y$ \\ \hline\hline
squarks, quarks & $Q$ & $({\stilde u}_L\>\>\>{\stilde d}_L )$&
$(u_L\>\>\>d_L)$ & $(\>{3},\>{2}\>,\>{1\over 6})$ \\ & $\overline u$
&${\stilde u}^*_R$ & $u^\dagger_R$ & $(\>{\overline 3},\> { 1},\> -{2\over
3})$ \\ & $\overline d$ &${\stilde d}^*_R$ & $d^\dagger_R$ & $(\>{\overline
3},\> {1},\> {1\over 3})$ \\ \hline sleptons, leptons & $L$ &$({\stilde
\nu}\>\>{\stilde e}_L )$& $(\nu\>\>\>e_L)$ & $(\>{1},\>{2}\>,\>-{1\over 2})$
\\ & $\overline e$ &${\stilde e}^*_R$ & $e^\dagger_R$ & $(\>{1},\> {1},\>1)$
\\ \hline Higgs, Higgsinos &$H_1$ &$(H_1^+\>\>\>H_1^0 )$& $(\stilde H_1^+
\>\>\> \stilde H_1^0)$& $(\>{1},\>{2}\>,\>+{1\over 2})$ \\ &$H_2$ & $(H_2^0
\>\>\> H_2^-)$ & $(\stilde H_2^0 \>\>\> \stilde H_2^-)$&
$(\>{1},\>{2}\>,\>-{1\over 2})$ \\ \hline & & spin 1/2 & spin 1 & \\ \hline
gluino, gluon & & $\stilde g$& $g$ & $(8,1,0)$ \\ \hline

winos, W-bosons & & $\stilde{W}^\pm$, $\stilde {W}^0$ &${W^\pm}$, ${W^0}$ &
$(1,3,0)$ \\ \hline

bino, B-boson & & $\stilde{B}^0$ & ${B^0}$ & $(1,1,0)$ \\ \hline \end{tabular}
\end{center} \caption{ \small{Supersymmetric partners with the Standard
Model members}} \label{ssp} \end{table}
The MSSM is the minimal SUSY extension of the SM. The field content includes
the SM particles and their superpartners as can be seen in
Table~\ref{ssp}. All of the quarks and leptons are put in chiral superfields
with their superpartners (\textit{squarks} and \textit{sleptons}
respectively).  In Table~\ref{ssp} all superpartners are denoted with a tilde,
and there is a superpartner for each chiral state of each SM fermion. This
enables us to treat fermions of different chirality differently.  The gauge
bosons are put in vector superfields with their fermionic superpartners (the
\textit{gauginos}).  Since $W$ is analytic in the scalar fields, we cannot
include the complex conjugate of the scalar field as in the SM to give mass to
the down quarks, so there must be a minimum of two Higgs doublets in
supersymmetric theories, and each has its own superpartner (the \textit{Higgsinos}).
The requirement that the trace anomalies vanish so that the theories stay
renormalizable, $TR(Y^{3})=TR(T_{3L}^{2}Y)=0$, also implies the existence of
even number of Higgs doublets.

The Kinetic terms of these fields are direct generalization of
Eq.~\ref{susy:lagfull}.  What remains to be specified is the
superpotential. This is given by, \be
W=\bar{u}Y_{u}QH_{u}-\bar{d}Y_{d}QH_{d}-\bar{e}Y_{e}LH_{d}+\mu H_{u} H_{d}.
\label{mssm:supot} \ee

All the fields are chiral superfields.  The bars over $u,d,e$ are in the sense,
that right chiral fields are written as left conjugates and has nothing to
do with non-analyticity.  The sign  convention is designed to generate
positive masses. The generational and fermionic indices have been
suppressed. For example the fourth term with the fermionic index would read
like $\bar
{u}_{ai}(Y_{u})_{ij}Q_{j\alpha}^{a}(H_{u})_{\beta}\varepsilon_{\alpha\beta }$

The Yukawa couplings $Y_{u}$ etc. are dimensionless 3$\times3$ family matrices
that determine the masses of quarks and leptons, and the angles and phase of
the CKM matrix after $H_{u}^{0}$ and $H_{d}^{0}$ get vevs.  They also
contribute to the squark-quark-Higgsino couplings etc. This is the most general
superpotential for the MSSM if we assume baryon and lepton number are
conserved.

\textbf{R parity:}Within the SM, B and L are accidental global symmetries of the
Lagrangian. Thus B and L violating
interactions are absent.These additional terms  could be
incorporated in $W$ keeping it
analytic, gauge invariant, and Lorentz invariant, but violating baryon
and/or lepton number conservation.  These terms are, \be
W_{R}=\lambda_{ijk}L_{i}L_{j}\bar{e}_{k}+\lambda_{ijk}^{\prime}L_{i}Q_{j}\bar{d}_{k}+
\lambda_{ijk}^{\prime\prime}\bar{u}_{i}\bar{d}_{j}\bar{d}_{k} +\mu^{\prime i}L_iH_u.
\label{MSSM:rpv} \ee The couplings
$\lambda,\lambda^{\prime},\lambda^{\prime\prime}$ are matrices in the family
space.  Combination of the second and third terms in Eq~\ref{MSSM:rpv} lead to
rapid proton decay. This requires extreme suppression of either or both terms
which again brings in the naturalness problem into the theory.  Rather, B and L
conservation consistent with observation should arise naturally from the
symmetries of the theory. This is dealt with by imposing a symmetry like the
R-parity or a variant called the matter parity, on the Lagrangian.  The R
parity is defined as, \be R=(-1)^{3(B-L)+2S} \ee where $S$ is the spin.  Then
SM particles and Higgs fields are even, superpartners odd.  This is a discrete
$\mathbb{Z}_{2}$ symmetry. Equivalently, one can use ``matter parity'', \be
P_{m}=(-1)^{3(B-L)}.  \ee It is now conjectured that a term in $W$ is only
allowed if $P_{m}=+1.$ Gauge fields and Higgs are assigned $P_{m}=+1,$ and
quark and lepton supermultiplets $P_{m}=-1.$ $P_{m}$ commutes with
supersymmetry and forbids $W_{R}.$ Matter parity could be an exact symmetry,
and such symmetries do arise in string theory.  If R-parity or matter parity
holds\footnote{R parity violating theories lead to phenomenologically rich scenarios.
However these models will not be explored further in this thesis. For a review 
see~\cite{reviews_rparity}.},
there are major phenomenological consequences, \begin{itemize} \item At
colliders, or in loops, superpartners are produced in pairs.  \item Each
superpartner decays into one other superpartner (or an odd number).  \item The
lightest superpartner (LSP) is stable.  That determines supersymmetry collider
signatures, and makes the LSP a good candidate for the cold dark matter of the
universe.  \end{itemize}

\textbf {The Soft breaking of MSSM:} Unfortunately the simple SUSY extension of
the standard model do not work. Supersymmetry predicts mass degenerate
superpartners of the SM fields, the failure to observe these in experiments
spells the doom for exact supersymmetric theory. The alternative is to break
supersymmetry in a way that will predict a mass difference between the SM
particles and their superpartners but will preserve the correlation in their
coupling that is crucial for cancellation of the quadratically divergent
quantum correction to the scalar masses. This is known as \textit{soft}
supersymmetry breaking.

Supersymmetry breaking can be driven spontaneously. To see this let us write
down the general SUSY Hamiltonian using
Eq.~\ref{susyalgone}-\ref{susyalgthree}, \be H=P^0 = {1\over 4}( Q_1
Q_{{1}}^\dagger + Q_{{1}}^\dagger Q_1 + Q_2 Q_{{2}}^\dagger + Q_{{2}}^\dagger
Q_2 ).  \label{susyham} \ee The vacuum not respecting supersymmetry translates
into the conditions: $Q |0\rangle \neq 0$ and $Q^\dagger |0\rangle
\neq0$. When these conditions are imposed on Eq.~\ref{susyham} we find that it
implies: $\langle0|H|0\rangle > 0$. In most general cases $\langle0|H|0\rangle
\equiv \langle0|V|0\rangle$.  Referring to the definition of the potential $V$
given in Eq.~\ref{susyfulpot}, the condition for spontaneous supersymmetry
breaking can be realized if either of the auxiliary fields ($F \mbox{or} D$)
develop a non-zero vev. This simple picture of spontaneous breaking of
supersymmetry cannot be implimented in the MSSM \footnote{For D fields to
develop a vev, it requires to be the auxiliary field corresponding to an
abelian gauge group. The only abelian gauge group in MSSM corresponds to
electromagnetism, association of the corresponding D fields with the required
vev would necessarily lead to breaking of electromagnetism that is
phenomenologically unacceptable. Similarly for an F term to develop a vev one
needs it to be the auxiliary field of a gauge singlet chiral
superfield. Non-existence of such gauge singlet chiral superfields makes this
mechanism inviable in the context of the MSSM.}  with the field content
defined in Table~\ref{ssp}.  Further, spontaneous symmetry breaking generally
implies certain mass sum rules that put all spontaneously broken
supersymmetric extension of the SM at variance with experimental observations.

Though it is conjectured that supersymmetry is  broken
spontaneously, possibly in some hidden sector, the pragmatic approach is to
parametrize this ignorance into certain phenomenological parameters. This
constitutes the soft breaking Lagrangian of the theory. For the MSSM we have,
\begin{eqnarray*} \hspace{-.3in}-\mathcal{L}_{soft} &
=\frac{1}{2}(M_{3}\tilde{g}\tilde{g}+M_{2}\widetilde{W}
\widetilde{W}+M_{1}\widetilde{B}\widetilde{B}+c.c.)  \end{eqnarray*}
\begin{eqnarray*}
\hspace{.8in}+\widetilde{Q}^{\dag}m_{Q}^{2}\widetilde{Q}+\widetilde{\bar{u}}^{\dag}m_{\bar{u}}^{2}\widetilde{\bar{u}}+
\widetilde{\bar{d}}^{\dag}m_{d}^{2}\widetilde{\bar{d}}+
\widetilde{L}^{\dag}m_{L}^{2}\widetilde{L}+\widetilde{\bar{e}}^{\dag}m_{\bar{e}}^{2}\widetilde{\bar{e}}
\end{eqnarray*} \begin{eqnarray*}
\hspace{.5in}+(\widetilde{\bar{u}}a_{u}\widetilde{Q}H_{u}-\widetilde{\bar{d}
}a_{d}\widetilde{Q}H_{d}-\widetilde{\bar{e}}a_{e}\widetilde
{L}H_{d}+c.c.)  \end{eqnarray*} \begin{eqnarray}
\hspace{.5in}+m_{H_{u}}^{2}H_{u}^{\ast}H_{u}+m_{H_{d}}^{2}H_{d}^{2\ast}+(bH_{u}
H_{d}+c.c.).  \end{eqnarray}

For clarity, a number of the indices are suppressed.  $M_{1,2,3}$ are the
 complex gaugino masses, e.g.  $M_{3}=\left| M_{3}\right| e^{i\phi_{3}},$ etc.
 In the second line $m_{Q}^{2},$ etc, are squark and slepton hermitian
 3$\times3$ mass matrices in family space.  The $a_{u,d,e}$ are complex
 3$\times3$ family matrices, usually called trilinear couplings.  Additional
 parameters come from $\mu_{eff}=\mu e^{i\phi_{\mu}};$ we will usually denote
 the magnitude of $\mu_{eff}$ as just $\mu$. It is
 worthwhile to note that most of the parameters of the MSSM ($>100$) actually
 come from this part of the Lagrangian.

\textbf{Physical states} In the MSSM there are 32 distinct masses
corresponding to undiscovered particles. Assuming only that the mixing of
first- and second-family squarks and sleptons is negligible, the mass
eigenstates of the MSSM are listed in Table~\ref{MSSM:physmass}
\begin{table}[tb] \begin{center} \begin{tabular}{|c|c|c|c|c|} \hline Names &
Spin & $P_R$ & Gauge Eigenstates & Mass Eigenstates \\ \hline\hline Higgs
bosons & 0 & $+1$ & $H_u^0\>\> H_d^0\>\> H_u^+ \>\> H_d^-$ & $h^0\>\> H^0\>\>
A^0 \>\> H^\pm$ \\ \hline & & &${\stilde u}_L\>\> {\stilde u}_R\>\> \stilde
d_L\>\> \stilde d_R$&(same) \\ squarks& 0&$-1$& ${\stilde s}_L\>\> {\stilde
s}_R\>\> \stilde c_L\>\> \stilde c_R$& (same) \\ & & & $\stilde t_L
\>\>\stilde t_R \>\>\stilde b_L\>\> \stilde b_R$ & ${\stilde t}_1\>\> {\stilde
t}_2\>\> \stilde b_1\>\> \stilde b_2$ \\ \hline & & &${\stilde e}_L\>\>
{\stilde e}_R \>\>\stilde \nu_e$&(same) \\ sleptons& 0&$-1$&${\stilde
\mu}_L\>\>{\stilde \mu}_R\>\>\stilde\nu_\mu$&(same) \\ & & & $\stilde
\tau_L\>\> \stilde \tau_R \>\>\stilde \nu_\tau$ & ${\stilde \tau}_1
\>\>{\stilde \tau}_2 \>\>\stilde \nu_\tau$ \\ \hline neutralinos & $1/2$&$-1$
& $\stilde B^0 \>\>\>\stilde W^0\>\>\> \stilde H_u^0\>\>\> \stilde H_d^0$ &
$\stilde N_1\>\> \stilde N_2 \>\>\stilde N_3\>\> \stilde N_4$ \\ \hline
charginos & $1/2$&$-1$ & $\stilde W^\pm\>\>\> \stilde H_u^+ \>\>\>\stilde
H_d^-$ & $\stilde C_1^\pm\>\>\>\stilde C_2^\pm $ \\ \hline gluino & $1/2$&$-1$
&$\stilde g$ &(same) \\ \hline ${\rm goldstino}\atop{\rm (gravitino)}$ &
${1/2}\atop{(3/2)}$&$-1$&$\stilde G$ &(same) \\ \hline \end{tabular}
\caption{\small The sparticles of the MSSM (sfermion mixing for the first two
generation assumed to be negligible).  \label{MSSM:physmass}} \vspace{-0.4cm}
\end{center} \end{table}
A complete set of Feynman rules for the interactions of these particles with
each other and with the Standard Model quarks, leptons, and gauge bosons can
be found in Ref.~\cite{Martin}.

\textbf{Electroweak symmetry breaking:} The MSSM has two Higgs doublets and
the combined potential term for them has three contributions, \begin{eqnarray}
V&=&\left| \mu_{eff}\right| ^{2}(\left| H_{u}\right| ^{2}+\left| H_{d}\right|
^{2})\hspace{1.6in}F^*F \\ \nonumber
&+&\frac{1}{8}(g_{1}^{2}+g_{2}^{2})(\left| H_{u}\right| ^{2}-\left|
H_{d}\right|^{2})^{2}\hspace{1.5in}D^*D \\ &+&m_{H_{u}}^{2}\left| H_{u}\right|
^{2}+m_{H_{d}}^{2}\left| H_{d}\right|
^{2}-(bH_{u}H_{d}+c.c.).\hspace{.6in}\mbox{soft breaking terms} \nonumber
\end{eqnarray} In order for the potential to be bounded from below, we need
the quadratic part of the scalar potential to be positive along the $D$-flat
directions. This requirement amounts to \be 2 b < 2 |\mu |^2 + m^2_{H_u} +
m^2_{H_d}.  \label{mssm:boundedfrombelow} \ee Now driving electroweak symmetry
breaking requires one linear combination of $H_u^0$ and $H_d^0$ to have a
negative squared mass near $H_u^0=H_d^0=0$, so that a symmetry breaking vev in
generated. This condition translates to, \be b^2 > (|\mu|^2 + m^2_{H_u}
)(|\mu|^2 + m^2_{H_d}).  \label{mssm:destabilizeorigin} \ee We write the vev's
as $\left\langle H_{u,d}\right\rangle =$v$_{u,d}.$ Requiring the Z mass be reconstructed at the weak scale,
we get, \be
\rm{v}_{u}^{2}+\rm{v}_{d}^{2}=\rm{v}^{2}=\frac{2M_{Z}^{2}}{g_{1}
^{2}+g_{2}^{2}} \approx (174{\rm GeV})^{2} \ee and it is convenient to
introduce \be \tan\beta=\rm{v}_{u}/\rm{v}_{d}.  \label{mssm:tanb} \ee Then
v$_{u}=$v$\sin\beta,$ v$_{d}=$v$\cos\beta,$ and with our conventions
$0<\beta<\pi/2.$ With these definitions the minimization conditions can be
written,
\begin{eqnarray} \left| \mu\right|
^{2}+M_{H_{d}}^{2}&=&b\tan\beta-\frac{1}{2}M_{Z}^{2} \cos2\beta \\ \nonumber
\left| \mu\right| ^{2}+M_{H_{u}}^{2}&=&b\cot\beta+\frac{1}{2}M_{Z}^{2}
\cos2\beta.  \end{eqnarray} These satisfy the EWSB conditions.

\textbf{Higgs mass:} As mentioned earlier Higgs scalar fields in the MSSM
consist of two complex $SU(2)_L$-doublet, or eight real, scalar degrees of
freedom. When the electroweak symmetry is broken, three of them, the
would-be Nambu-Goldstone bosons $G^0$, $G^\pm$, become the longitudinal
modes of the massive $Z^0$ and $W^\pm$. The remaining five Higgs
scalar mass eigenstates consist of two CP-even neutral scalars $h$ and
$H$, one CP-odd neutral scalar $A^0$, and a charge $+1$ scalar $H^+$ and its
conjugate charge $-1$ scalar $H^-$. (Here we define $G^- = G^{+*}$ and $H^- =
H^{+*}$. Also, by convention, $h$ is lighter than $H$.)

The resulting tree level masses are \begin{eqnarray}
m_{h,H}^{2}&=&\frac{m_{A}^{2}+M_{Z}^{2}}{2}\mp\frac{1}{2}\sqrt{(m_{A}^{2}
+M_{Z}^{2})^{2}-4m_{A}^{2}M_{Z}^{2}\cos^{2}2\beta}, \\ \nonumber &
&\mbox{where,} \\ \nonumber m_{A}^{2}&=&2b/\sin2\beta, \\ & &\mbox{and} \\
\nonumber m_{H^{\pm}}^{2}&=&m_{A}^{2}+M_{W^{\pm}}^{2}. \nonumber
\label{mssm:higgmass} \end{eqnarray} A little bit of algebra shows that the
lightest Higgs mass has an theoretic upper limit given by, \be
m_{h}^{tree}\leq\left| \cos2\beta\right| M_{Z}.  \ee

However, the tree-level formula for the squared mass of $h$ is subject to
quantum corrections that are relatively drastic. The largest such
contributions typically come from top and stop loops. The one loop radiative
correction is approximately given by, \be \Delta (m^2_{h}) \approx {3\over 4
\pi^2} \cos^2\!\alpha\>\, y_t^2 m_t^2 \, {\rm ln}\left (m_{\stilde t_1}
m_{\stilde t_2} / m_t^2 \right ) {}.  \label{mssm:higgradcor} \ee Including
these and other important corrections, upper bound on the
Higgs mass given by, \be m_{h} \leq 135\>{\rm GeV} \label{mssm:higgsbound}
\ee in the MSSM. This assumes that all the sparticle masses are below 1
TeV.  However by adding extra supermultiplets to the MSSM, this bound can be
stretched.  Assuming that none of the MSSM sparticles have masses exceeding 1
TeV and that all of the couplings in the theory remain perturbative up to the
unification scale, one still has, Ref.~\cite{Kane:1992kq} \be m_{h} \leq 150\>{\rm GeV}.
\label{general:higgsbound} \ee

\begin{center} \begin{figure} \centering{
\includegraphics[width=0.6\textwidth,angle=0,keepaspectratio]
{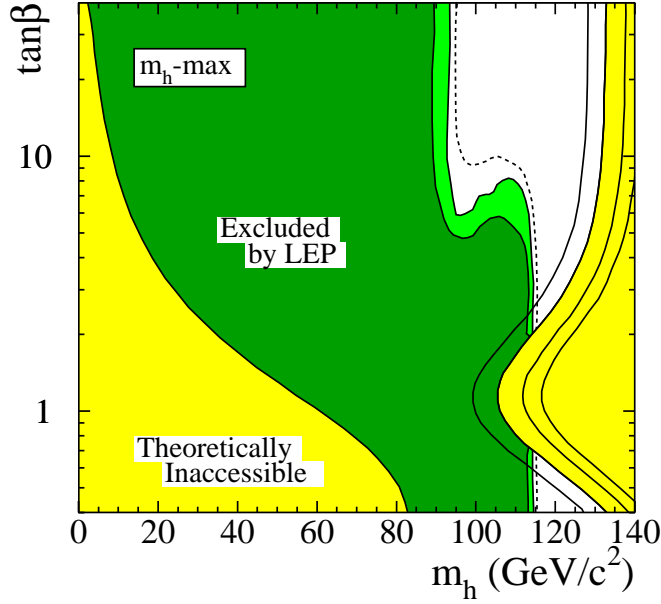}} \caption{ \small The LEP
exclusion limit on the lightest CP even neutral Higgs boson.}
\label{higg_exclusion} \end{figure} \end{center}
\subsection{The experimental status of the MSSM}
Notwithstanding the theoretic soundness and the phenomenological advantages,
discovery of supersymmetry has not yet been made, after decades of
experimentations.  No superpartners have yet been discovered at collider
experiments.  The general limits from direct experiments that could produce
superpartners are not even very strong.  They are also all model dependent,
with varying significance.  Limits from LEP on charged superpartners are near
the kinematic limits except for models having near degeneracy of the charged
sparticle and the LSP, in which case the decay products are very soft and hard
to observe, giving weaker limits.  So in most cases charginos and charged
sleptons have limits of about 94 GeV.  Gluinos and squarks have typical limits
of about 308 GeV and 379 GeV respectively, except that if one or two squarks are lighter the limits on
them are much weaker.  For stops and sbottoms the limits are about 85 GeV.

There are no clear limits on neutralinos at the LEP.  This is so because one
can easily construct models where production of LSP's are unobservable at the
LEP. There are no general relations between neutralino masses and chargino
or gluino masses, so limits on the latter do not imply limits on neutralinos.
In typical models the limits are $M_{LSP}\gtrsim 46$ GeV,
$M_{\widetilde{N}_{2}}\gtrsim 62.4$ GeV.  Superpartners get mass from both the
Higgs mechanism and from supersymmetry breaking, so one would expect them to
typically be heavier than SM particles.

The direct searches have also put constraints on the Higgs mass.  The combined
constraint on the lightest CP even neutral Higgs field is shown in
Figure~\ref{higg_exclusion}.

Theoretically if MSSM explains electroweak symmetry breaking then one needs
to reproduce Z mass in terms of soft-breaking masses, given by the relation,
\be m_Z^2 = \frac{|m^2_{H_d} - m^2_{H_u}|}{\sqrt{1 - \sin^2(2\beta)}} -
m^2_{H_u} - m^2_{H_d} -2|\mu|^2, \ee so if the soft-breaking masses are too
large, it would lead to large finetuning.  The parameters that are most
sensitive to this issue are $M_{3}$ (basically the gluino mass) and $\mu$
which strongly affects the chargino and neutralino masses.  Qualitatively one
therefore expects rather light gluino, chargino, and neutralino
masses. Argument in this direction leads to the following upper mass limits:
$M_{\tilde{g} }\lesssim 500$ GeV; $M_{\widetilde{N}_{2}},$
$M_{\widetilde{C}}\lesssim 250$ GeV; and $M_{\widetilde{N}_{1}}\lesssim 100$
GeV.  These are upper limits, seldom saturated in models.  There are no
associated limits on sfermions.

It is however expected that the LHC will finally sit on judgment for the
existence of the MSSM~\cite{LHCdiscovery}.  At the LHC, production of gluinos
and squarks by gluon-gluon and gluon-quark fusion usually dominate, unless the
gluinos and squarks are heavier than 1 TeV or so. One can also have associated
production of a chargino or neutralino together with a squark or
gluino. Slepton pair production might be observable at the
LHC~\cite{sleptonLHC} .  Cross-sections for sparticle production at hadron
colliders can be found in Refs.~\cite{gluinosquarkproduction}.

The decays of the produced sparticles result in final states with two
neutralino LSPs, which escape the detector. The LSPs carry away at least $2
m_{\tilde{N}_1}$ of missing energy, but at hadron colliders only the component
of the missing energy that is manifest in momenta transverse to the colliding
beams (denoted by $\slashed{E}_T$) is observable. So, in general the observable
signals for supersymmetry at hadron colliders are $n$ leptons + $m$ jets +
$\slashed{E}_T$. There are important Standard Model backgrounds to many of
these signals, especially from processes involving production of $W$ and $Z$
bosons that decay to neutrinos, which provide the $\slashed{E}_T$. One must
choose the $\slashed{E}_T$ cut high enough to reduce backgrounds from detector
mismeasurements of jet energies. The jets$+\slashed{E}_T$ signature is one of
the main signals currently being searched at LHC.

\section{Conclusion and Outlook} The Standard Model (SM) of elementary
particle physics provides a correct description of virtually all known
microphysical nongravitational phenomena. However, there are a number of
theoretical and phenomenological issues that the SM fails to address
adequately: the gauge hierarchy problem, triggering electroweak symmetry
breaking, gauge coupling unification, explanation of family structure and
fermion masses, cosmological challenges including the issue of dark matter
etc.

All these indicate the existence of new physics at around the 1 TeV mark,
which can be probed by collider experiments and astrophysical
observations. Low energy supersymmetry (SUSY) and compactified extra
dimensions (EDs) provide theoretically sound and phenomenologically exciting
frameworks to extend the SM and strengthen its foundations.

Supersymmetry, which is included in the most general set of symmetries of
local relativistic field theories, has the virtue of solving the gauge
hierarchy problem and is a popular choice of physics beyond the standard
model. In the simplest supersymmetric world ($N=1$), each particle has a
superpartner which differs in spin by $1/2$, and is related to the original
particle by SUSY transformations, as discussed above. Since SUSY relates the
scalar and fermionic sectors, the chiral symmetries which protect the masses
of the fermions, also protect the masses of the scalars from quadratic
divergences, leading to an elegant resolution of the hierarchy problem. We saw
that apart from this, SUSY leads to unification of gauge couplings, triggers
electroweak symmetry breaking radiatively, provides cold dark matter candidate
and provides a framework to turn on gravity.

On the other hand, theories with extra dimensions\footnote{All extra
dimensions are considered to be spatial in nature as time like EDs lead to
tachyonic fields that violate causality.}  have recently attracted enormous
attention. The study of TeV scale extra dimensions that has taken place over
the past few years has its origin in the ground breaking work of Arkani-Hamed,
Dimopoulos and Dvali (ADD) \cite{add}. Since that time, the extra dimensions
have evolved from a single idea to a new paradigm of employing EDs as a tool
to address a large number of outstanding issues that remain unanswerable in SM
context.  This in turn leads to phenomenological implications that can be
tested at colliders and elsewhere. Various variants of EDs have been used in
addressing various issues including hierarchy problem, electroweak symmetry
breaking without Higgs boson, the generation of ordinary fermion and neutrino
mass hierarchy, the CKM matrix, new sources of CP violation, grand unification
while suppressing proton decay, new dark matter candidates, new cosmological
perspectives, black hole productions at future colliders as a window on
quantum gravity, novel mechanisms of SUSY breaking etc. Technical details of
extra-dimensional theories will be given in Chapter~2.

For some time now, it is believed that string theory is a realistic attempt to
provide an unified quantum picture of all known interactions in physics.
Consistent string theories indicate the existence of supersymmetry and
compactified extra dimensions in their low energy phenomenology. Though a
rigorous connection between string theory and low energy phenomenological
models with extra dimensions has not yet been possible, it provides enough
motivation to study higher dimensional supersymmetric theories. From a purely
phenomenological point of view, such higher dimensional supersymmetric
theories have various virtues to their credit, including the explanation of
fermion mass hierarchy from a different angle, providing a cosmologically
viable dark matter candidate, interpretation of the Higgs as a quark composite
leading to a successful electroweak symmetry breaking without the necessity of
a fundamental Yukawa interaction, and lowering the unification scale down to a
few TeV. Supersymmetrization provides a natural mechanism to stabilize the
Higgs mass in extra dimensional scenarios.  It is also worthwhile to note that
all supersymmetric models in four dimensions necessarily introduce the
paradigm of further new physics that controls SUSY breaking in this class of
models. Embedding supersymmetric models in extra dimension provides various
avenues to realize soft breaking of supersymmetry.  The rest of this thesis
will focus on the phenomenology of extra dimensions and their interface with
supersymmetry.

%% file: texfiles/hggc
\section{Extra dimensions} \normalsize It is generally believed that some form of New Physics (NP)
must exist beyond the Standard Model (SM) to explain its deficiencies.
 Though there are many candidates
for NP, as discussed in Section~\ref{bsm_into}, it will be up to experiments at
future colliders, the  Large Hadron Collider (LHC) and 
the proposed International Linear Collider (ILC), to reveal its true nature.

One possibility is that extra spatial dimensions will begin to show themselves
at or near the TeV scale.  The discovery of extra dimensions (ED) would
produce a fundamental change in how we view the universe. The study of the
physics of TeV-scale EDs that has taken place over the past few years has its
origins in the ground breaking work of Arkani-Hamed, Dimopoulos and Dvali
(ADD)~\cite {add}. Since that time EDs has evolved from a single idea to a new
paradigm with various applications.  Extra dimensions have been used as a tool
to address the large number of outstanding issues that remain unanswerable in
the SM context. This in turn has lead to other phenomenological implications
which should be testable at colliders and elsewhere.  A tentative list of
some of these applications includes,

\begin{enumerate}

\item Addressing the hierarchy problem~\cite {add,Randall:1999ee}.

\item Triggering electroweak symmetry breaking without a Higgs boson~{\cite
{nohiggs}}.

\item The generation of the ordinary fermion and neutrino mass hierarchy, the
CKM matrix and new sources of CP violation~{\cite {flavor}}.

\item TeV scale grand unification or unification without SUSY while
suppressing proton decay~{\cite {unif}}.

\item New Dark Matter candidates and a new cosmological perspective~{\cite
{ued,bine}}.

\item Black hole production at future colliders as a window on quantum
gravity~{\cite {bh}}.

\end{enumerate}

An amplified discussion of all these issues is beyond the scope of the present
thesis. However it is clear from this list that EDs have found their way into
essentially every area of interest in high energy physics providing strong
motivation for exploring the phenomenology of ED in present and future
colliders.

 \textbf{The spatial Vs temporal EDs:}~Consider a massless particle moving in
5d `Cartesian' co-ordinates and assume that 5d Lorentz invariance
holds. Then the square of the 5d momentum for this particle is given by
$p^2=0=g_{MN}p^Mp^N=-p_0^2+${\bf p}$^2\pm p_5^2$ where $g_{MN}=
diag(-1,1,1,1,\pm 1)$ is the 5d metric tensor. As usual $p_0$ is the
particle energy, {\bf p}$^2$ is the square of the particle 3-momentum and
$p_5$ is its momentum along the 5th dimension. A positive sign before the
fifth component of the metric represents a space like extra dimension
whereas a negative fifth component corresponds to a time like extra dimension.
  The right hand side of the
equality is zero because of the the assumption of zero mass in 5d.  We can
re-write the equation above in a more traditional form as $-p_0^2+${\bf
p}$^2=p_\mu p^\mu =\mp p_5^2$ and we recall, for  particles 
which satisfy 4d Lorentz invariance, that $p_\mu p^\mu=-m^2$, which is
just the square of the particle mass (note $\mu$ runs from 0 to 3 where as
$M,N$ runs from 0 to 4). Notice, that if we choose a time-like extra dimension,
the sign of the square of the mass of the particle will appear to be
{\textit{negative}}, i.e., the particle is a \textit{tachyon}. Tachyons are
well known to cause severe causality problems~\cite {tach} something that is
best avoided in any theory. This implies that we should pick the space-like
solution.  Thus to avoid tachyons appearing in our ED
theories we must always choose EDs to be space-like and therefore we assume
there will always be only one time like dimension~\cite{dvali}.

\textbf{The brane world scenario:}~ED models are typically structured to have a
extra spatial dimension that is compactified with a suitable orbifolding
symmetry. The compactification enables this spatial dimension to evade all
observation of its existence at low energy. Only when the probing energy is of
the order of the compactification length scale, does one begins to see the
manifestation of the extra dimension. The end points of the compactified extra
dimension are the location of four dimensional hyper-surfaces called the
3-branes. The observed four dimensional structure of the hithertho discovered
space-time geometry corresponds to one of these 3-branes,
see~\cite{Csaki:2004ay} for further details.

In this chapter we review the \textit{Warped extra dimension} in Section~\ref{wed},
detailing the derivation of the anti-de Sitter metric that originates
naturally from the Einsteins equations with negative cosmological constant, a
review of the particle spectrum and their interaction is then made in the
context of a warped extra dimension compactified on an orbifold with
$S_1/\mathbb{Z}_2$ symmetry. In Section ~\ref{gghgaga} we demonstrate that the
loop contribution of the KK towers of quarks and gauge bosons emerging from
the compactification would have a sizable numerical impact on the rates of $gg
\to h$ and $h \to \gamma \gamma$, which are of paramount importance in the
context of Higgs search at the LHC. This happens because the Higgs coupling to
a pair of KK fermion-antifermion is not suppressed by the zero mode fermion
mass and can easily be order one . The underlying reason is simple.  Although
the zero mode wave-functions of different flavors have varying overlap at the
TeV brane depending on the zero mode masses, the KK profiles of all fermions
have a significant presence at the TeV brane where the Higgs resides. As a
result, the KK Yukawa couplings of different flavors are not only all large,
they are also roughly universal, see Section~\ref{kkyukawasec}. This large
universal Yukawa coupling in the RS scenario constitutes the corner-stone of
our study.  On the contrary, in flat Universal Extra Dimension (UED) only the
KK top Yukawa coupling is large, others being suppressed by the respective
zero mode fermion masses.  We provide comparative plots to demonstrate how the
warping in RS fares against the flatness of UED for the processes under
consideration.

\section{Warped Extra Dimension} \label{wed} Taking in account the {\it
back-reaction} of gravity to the presence of the branes themselves naturally
leads to warped extra dimensions. Careful consideration of  the back-reaction may be important,
since if one has a 4d theory with only 4d sources, it will necessarily lead
to an expanding universe with positive cosmological constant. On the other
hand, if one has 4d sources in 5d geometry, one can {\it balance} the effects
of the 4d brane sources by a 5d bulk cosmological constant thus reducing the
{\it effective} 4d cosmological constant to zero, that is the 4d universe
would still appear to be static and flat for an observer on a brane
\cite{RubShap}. Now the 5d background itself is curved, which is clear
from the fact that one had to introduce a bulk cosmological constant. In a
sense there is a transfer of the curvature from the 4d branes, which are made
flat, to the bulk which is now significantly curved~\cite{Csaki:2004ay}.  This
scenario was originally proposed by Rubakov and Shaposhnikov~\cite{RubShap}.

\subsection{The Randall-Sundrum background} With this motivation, consider a 5d
scenario with a non-vanishing 5d cosmological constant $\Lambda$ in the
bulk. We are interested in solutions where the brane itself remains static and
flat, preserving the 4d Lorentz invariance, while the extra dimension is
curved. This implies that the induced metric at every point along the fifth
dimension has to be the ordinary flat 4d Minkowski metric, and the components
of the 5d metric depend only on the fifth coordinate $y$. The ansatz for the
most general metric satisfying these properties is given by: \begin{equation}
ds^2=e^{-A(y)} dx^\mu dx^\nu \eta_{\mu\nu} +dy^2.  \end{equation} The amount
of curvature along the fifth dimension depends on the function $e^{-A(y)}$,
which is therefore called the warp-factor.  To go into the conformally flat
frame, we need to make a coordinate transformation of the form $z=z(y)$. The
coordinate transformation should not depend on the 4d coordinates $x$, which
might induce off-diagonal terms in the metric. One can ensure that the metric
be conformally flat in the new frame, if $dy$ and $dz$ are related by
\begin{equation} e^{-A(z)/2}dz=dy, \end{equation} such that the full metric in terms of the
the $z$ coordinate will be: \begin{equation} ds^2=e^{-A(z)} (dx^\mu dx^\nu
\eta_{\mu\nu} +dz^2).  \end{equation} Deriving the RS solution now reduces to
the task of finding the function $A(z)$. To do this we first note that the
above mentioned conformally flat metric leads to the following non-vanishing
components of the Einstein tensor: \begin{eqnarray} \label{e1}
G_{55}&=&-\frac{3}{2} A'^2, \nonumber \\ G_{\mu\nu}&=&-\frac{3}{2} \eta_{\mu\nu}
(-A''+\frac{1}{2}A'^2), \end{eqnarray} where
$G_{MN}=R_{MN}-\frac{1}{2}g_{MN}R$.

This should agree with the solution of the 5d Einstein-Hilbert action,
\begin{equation} S=-\int d^5x \sqrt{g} (M_*^3 R+\Lambda ).  \end{equation} One
can then use the definition of the stress-energy tensor to find the Einstein
equation: \begin{equation} \label{e2} G_{MN}=\kappa^2 T_{MN}=\frac{1}{2
M_*^3}\Lambda g_{MN}.  \end{equation} Comparing the $55$ component of the
Einstein equation will then give: \begin{equation} \frac{3}{2} A'^2
=-\frac{1}{2M_*^3} \Lambda e^{-A}.  \end{equation} The first thing that we
note is that a solution can only exist if the bulk cosmological constant is
negative $\Lambda <0$. \textit{This means that the important case for us will
be considering anti-de Sitter spaces, that is spaces with a negative
cosmological constant.} With a negative value of $\Lambda$ one can now solve
for the function $A(z)$. It is given by, \begin{equation} \label{AdS1}
e^{-A(z)}=\frac{1}{(kz+{\rm const.})^2}, \end{equation} where we have
introduced \begin{equation} k^2=-\frac{\Lambda}{12 M_*^3}.  \end{equation} To
fix the constant in Eq.~\ref{AdS1} we choose $e^{-A(0)}=1$, giving us,
\begin{equation} e^{-A(z)}=\frac{1}{(kz+1)^2}.  \end{equation} The metric in
the original $y$ coordinates can now be read off by recalling the relation
between $z$ and $y$ given by \begin{equation} e^{-A(z)/2} dz = \frac{dz}{k
z+1}=dy, \end{equation} we get that (by choosing $y=0$ to correspond to
$z=0$): \begin{equation} e^{-A(z)}=\frac{1}{(kz+1)^2}=e^{-2k y},
\end{equation} and so the RS metric in its more well-known form is finally
given by: \begin{equation} \mbox{\fbox{$ds^2=e^{-2ky} dx^\mu dx^\nu
\eta_{\mu\nu}+dy^2.$}} \end{equation}

However one still needs to check whether the 4d components $(G_{\mu\nu})$ of
  Eq.~\ref{e1} and  Eq.~\ref{e2} are in agreement. In order to
 fulfill this condition we will find that the two branes at the two ends of
 the compactified extra dimension need to have equal and opposite brane
 tension which is related to the bulk cosmological constant. If we consider
 $V_0$ and $V_1$ are the tensions at two opposite branes, then we will find that
 they are related as follows, \begin{equation} \Lambda=-\frac{V_0^2}{12M_*^3},
 \ \ V_1=-V_0.  \label{RSfinetune} \end{equation} Thus there is a static flat
 solution only, if the above {\it two} fine tuning conditions are
 satisfied. For details see for example \cite{Csaki:2004ay}.

\subsection{Compactification and KK Decomposition of Bulk Fields}

\begin{figure} \centerline{ \epsfxsize 2.5 truein \epsfbox {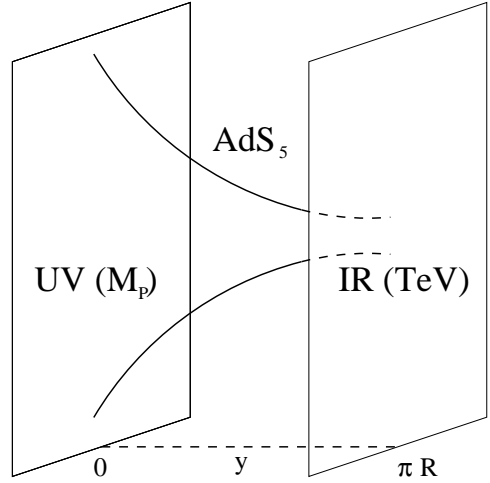}
}

\caption{\small A slice of AdS$_5$: The Randall-Sundrum scenario.}
\label{sliceAdS} \end{figure}

 We are considering a scenario \cite{Randall:1999ee,Gherghetta:2000qt}
based on a non-factorizable geometry with an extra dimension as shown in
Figure~\ref{sliceAdS}. In this scenario the fifth dimension $y$ is
compactified on a $S^1/\mathbb{Z}_2$ orbifold of radius $R$, with $-\pi R\leq
y\leq\pi R$. The orbifolding is needed to obtain chiral fermions in the zero
mode of the KK tower, in agreement with the chiral fermions of the SM.  The
orbifold fixed points at $ y = (0,\pi R)$ are also location of two
3-branes. The space-time between the two branes is simply a slice of $AdS_5$
geometry. The five dimensional metric is given by \cite{Randall:1999ee},
\begin{equation} \label{metric} ds^2=e^{-2\sigma}\eta_{\mu\nu}dx^\mu
dx^\nu+dy^2\, \end{equation} where \begin{equation} \label{sigma}
\sigma=k|y|~, \end{equation} and $1/k$ is the AdS curvature radius and
$\eta_{\mu\nu}={\rm diag}(-1,1,1,1)$. The natural mass scale associated with
the brane at $y=0$ is the Plank scale ($M_{pl}$) and the corresponding 3-brane
is called the Planck brane. The effective mass scale associated with the $y=\pi
R$ brane is $M_P e^{-\pi kR}$, which is of the TeV order for $kR\simeq
12$. The corresponding 3-brane at $y=\pi R$ is called the Weak or TeV
brane. This immediately provides a framework to address the hierarchy problem
associated with fundamental scalars.  Consider the scalar field (for example
the Higgs scalar) confined to the TeV/Weak brane. Its action would be given by
\begin{equation} S^{Higgs}=\int d^4 x \sqrt{g^{5}} [g_{\mu\nu}D^\mu hD^\nu h
-V(h)], \ \ \ V(h)=\lambda [(h^\dagger h)-v^2)^2].  \end{equation} If the size
of the extra dimension is $\pi R$, then the induced metric at the TeV brane is
given by \begin{equation} g_{\mu\nu}^{5}|_{y=\pi R}=e^{-2k\pi R}
\eta_{\mu\nu}.  \end{equation} Plugging this in for the above action we get
that the action for the Higgs is given by \begin{equation} S^{Higgs}=\int d^4x
e^{-4k\pi R}[e^{2k\pi R} \eta_{\mu\nu} \partial^\mu h\partial^\nu h -\lambda
(h^\dagger h-v^2)^2].  \end{equation} We can see that due to the non-trivial
value of the induced metric on the TeV brane the Higgs kinetic term will not
be canonically normalized. To get the action for the canonically normalized
field, one needs a field redefinition $\tilde{h} \rightarrow e^{-k\pi R} h$. In terms of
this field the action is \begin{equation} S^{Higgs}=\int d^4x [ \eta_{\mu\nu}
\partial^\mu \tilde{h}\partial^\nu \tilde{h} -\lambda [(\tilde{h}^\dagger
\tilde{h})- (e^{-k\pi R}v)^2]^2].  \end{equation} This is exactly the action
for a normal Higgs scalar, but with the vev (which sets the scale of all mass
parameters) ``warped down'' to $\tilde{v}_{Higgs}=e^{-k\pi R} v$ thus solving
the gauge hierarchy problem.

Till now we have discussed the brane bound fields (e.g Higgs field as in the
previous example). But a natural generalization of the scenario is the one
where the fields are allowed to access the 5d bulk. This requires a systematic
study of the bulk fields.  First we note that even though the space-time is 5
dimensional, we are confined to a four dimensional hyper-surface called the TeV
brane. Thus we need to derive the effective four dimensional version of the 5d
action by integrating out the fifth spatial component. The general 5
dimensional action of bulk fermions, scalars and vector bosons, is given by,
\begin{equation} \label{kin} S_5=-\int d^4x\int dy\sqrt{-g}\,
\Bigg[\frac{1}{4g^2_5}F^2_{MN}+
\left|\partial_M\phi\right|^2+i\bar{\Psi}\gamma^MD_M\Psi
+m^2_\phi|\phi|^2+im_\Psi\bar{\Psi}\Psi\Bigg]\, , \end{equation} where $g={\rm
det}(g_{MN})$, $F_{MN}=\partial_M V_N-\partial_N V_M $ and
$D_M=\partial_M+\Gamma_M $ where $\Gamma_M$ is the spin connection given by
$\Gamma_\mu= \frac{1}{2}\gamma_5\gamma_\mu\frac{d\sigma}{dy}$, $\Gamma_5=0$ for
the metric given in Eq.~\ref{metric}.\footnote{The gamma matrices,
$\gamma_M=(\gamma_\mu,\gamma_5)$ are defined in curved space as
$\gamma_M=e^\alpha_M\gamma_\alpha$, where $e^\alpha_M$ is the vierbein and
$\gamma_\alpha$ are the Dirac matrices in flat space.} The bulk masses
consistent with the orbifolding conditions are given by,
\begin{eqnarray} \label{masses} m^2_\phi&=&ak^2+b\sigma^{\prime\prime}\,
,\nonumber\\ m_\Psi&=&c\sigma^\prime\, , \end{eqnarray} where $a,b$ and $c$
are arbitrary dimensionless parameters and $\sigma$ is defined in
Eq.~\ref{sigma}.

After integrating out the compactified extra dimension the 4d Lagrangian can
be written in terms of the zero modes and their KK towers. The fermionic
\cite{Grossman:1999ra}, scalar and vector \cite{Pomarol:1999ad} fields can in
general be expanded in KK modes as follows, \begin{equation}
\label{Kaluza-Klein} \Phi(x^\mu,y)={1\over\sqrt{2\pi R}}\sum_{n=0}^\infty
\Phi^{(n)}(x^\mu)f_n(y)\, , \end{equation} with \begin{equation}
\label{solution}
f_n(y)=\frac{e^{s\sigma/2}}{N_n}\left[J_\alpha(\frac{m_n}{k}e^{\sigma})
+b_{\alpha}(m_n)\, Y_\alpha(\frac{m_n}{k}e^{\sigma})\right]\, , \end{equation}
for $\Phi=\{\phi,e^{-2\sigma}\Psi_{L,R}\,A_\mu\} $ \footnote{ The (L,R)
correspond to the left and right chiral fermions respectively.}  where
\begin{equation} \label{balpha} b_\alpha =
-\frac{(-r+\frac{s}{2})J_\alpha(\frac{m_n}{k})+\frac{m_n}{k}
J'_\alpha(\frac{m_n}{k})}{(-r+\frac{s}{2})Y_\alpha(\frac{m_n}{k})
+\frac{m_n}{k}Y'_\alpha(\frac{m_n}{k})}~, \end{equation} and \begin{equation}
N_n \simeq \frac{1}{\sqrt{\pi^2 R~m_n~e^{-\pi kR}}}~, \end{equation} with
$s=(4,1,2)$, $r=(b,\mp c,0)$ and $\alpha=(\sqrt{4+a}, c\pm \frac{1}{2}, 1)$
respectively. The Kaluza Klein\footnote{ The KK masses are basically the
quantized fifth components of the 5d momenta associated with the fields which
manifest themselves as masses in the effective 4d theory} masses are
determined by imposing boundary conditions on the solution given in
Eq.~\ref{solution}. For even (odd) fields the boundary conditions are
$\left.\left(\frac{df_n}{dy} - r \sigma^\prime f_n\right)\right|_{0,\pi R}= 0$
$\left( f_n\Big|_{0,\pi R}=0\right)$.  Thus we find that the KK masses are
given by the roots of the equation,
\begin{equation} \label{root} b_{\alpha}(m_n)=b_{\alpha}(m_ne^{\pi kR}).
\end{equation}

In the limit that $m_n \ll k$ ,$\alpha >0$ and $kR \gg 1$ the Kaluza-Klein
mass solutions are \begin{equation} \label{evenmn} \mbox{\fbox{$m_n\simeq
\left(n+\frac{1}{2}(\alpha-1)\mp\frac{1}{4}\right)\pi k~e^{-\pi kR}~$}}
\end{equation} for $n=1,2,\dots$

It is to be noted that the masses of both even and odd KK fermions are identical
and are given approximately by, \begin{equation} m_n^{\psi}\simeq
\left(n+\frac{1}{2}|(c-\frac{1}{2})| -\frac{1}{4}\right)\pi k~e^{-\pi kR}.
\end{equation} This not clear from Eq~\ref{evenmn} which is only true for
$\alpha>0$. This requires a very careful study of the fermionic sector of the
theory.  The 5 dimensional action for the fermions is given by,
\begin{equation} \label{kinf} S_5=-\int d^4x\int dy\sqrt{-g}\, \Bigg[
i\bar{\Psi}\gamma^MD_M\Psi +im_\Psi\bar{\Psi}\Psi\Bigg]\, , \end{equation}
where, \begin{eqnarray} \label{massesf} m_\Psi&=&c\sigma^\prime\, ,
\end{eqnarray} After integrating out the compactified extra dimension, the 4d
Lagrangian can be written in terms of the zero modes and their KK towers. The
fermionic \cite{Grossman:1999ra} fields can in general be expanded in KK modes
as follows, \begin{equation} \label{Kaluza-Kleinf}
\widehat{\Psi}_{L,R}(x^\mu,y)={1\over\sqrt{2\pi R}}\sum_{n=0}^\infty
\widehat{\Psi}_{L,R}^{(n)}(x^\mu)f_n(y)\, , \end{equation} The (L,R)
correspond to the $\mathbb{Z}_2$ even and $\mathbb{Z}_2$  odd fermions respectively and
$\widehat{\Psi}=e^{-2\sigma}\Psi_{L,R}$.

Where we have, \begin{equation} \label{solutionf}
f_n(y)=\frac{e^{s\sigma/2}}{N_n}\left[J_\alpha(\frac{m_n}{k}e^{\sigma})
+b_{\alpha}(m_n)\, Y_\alpha(\frac{m_n}{k}e^{\sigma})\right]\, , \end{equation}
with \begin{equation} N_n \simeq \frac{1}{\sqrt{\pi^2 R~m_n~e^{-\pi kR}}}~,
\end{equation} and $s=(1)$, $r=(\mp c)$ and $\alpha=(c\pm \frac{1}{2})$.

The Kaluza Klein masses are determined by imposing boundary conditions on the
 solution given in Eq.~\ref{solutionf}. For even (odd) fields the boundary
 conditions are $\left.\left(\frac{df_n}{dy} - r \sigma^\prime
 f_n\right)\right|_{0,\pi R}= 0$ $\left( f_n\Big|_{0,\pi R}=0\right)$.

For the even fields imposing the boundary condition gives rise to the two
equations \begin{eqnarray} \label{beven} b_\alpha &=&
-\frac{(-r+\frac{s}{2})J_\alpha(\frac{m_n}{k})+\frac{m_n}{k}
J'_\alpha(\frac{m_n}{k})}{(-r+\frac{s}{2})Y_\alpha(\frac{m_n}{k})
+\frac{m_n}{k}Y'_\alpha(\frac{m_n}{k})}~,\\
b_{\alpha}(m_n)&=&b_{\alpha}(m_ne^{\pi kR})\, .  \end{eqnarray} These two
conditions determine the values of $b_\alpha$ and $m_n$. Using the values of
$r,s,$ and $\alpha$ and the recursion relations for the Bessel functions we
get that, \begin{equation} b_{\alpha}(m_n) =
\frac{J_{\alpha-1}(\frac{m_n}{k})}{Y_{\alpha-1}(\frac{m_n}{k})} \end{equation}
The KK masses are approximately given by the roots of
$J_{\alpha-1}(\frac{m_n}{k}) = 0$, but the roots of $J_a$ and $J_{-a}$ are
identical so we can as well call it the root of $J_{|a|}$. In the limit that
$m_n \ll k$ and $kR \gg 1$ the Kaluza-Klein mass solutions for $n=1,2,\dots$
are given by \begin{eqnarray}\label{enevmass} m_n\simeq
\left(n+\frac{1}{2}|\alpha -1| -\frac{1}{4}\right)\pi k~e^{-\pi kR}\\
m_n\simeq \left(n+\frac{1}{2}|(c-\frac{1}{2})| -\frac{1}{4}\right)\pi
k~e^{-\pi kR} \end{eqnarray} since for even fields $\alpha = c+\frac{1}{2}$
thus $\alpha -1=c-\frac{1}{2}$

For the odd fields the continuity of $f_n$ at the boundaries implies that
\begin{equation} \label{boun:odd} f_n\Big|_{0,\pi R}=0\, , \end{equation} and
consequently \begin{eqnarray} \label{bodd}
b_{\alpha}(m_n)&=&-\frac{J_\alpha(\frac{m_n}{k})} {Y_\alpha(\frac{m_n}{k})}\,
,\\ b_{\alpha}(m_n)&=&b_{\alpha}(m_ne^{\pi k R})\, .  \end{eqnarray} In this
case one can check that the derivative of $f_n$ is continuous on the
boundaries and does not lead to further conditions. As in the even case, an
approximate solution for the Kaluza-Klein tower in the limit that $m_n \ll k$
and $kR \gg 1$ is \begin{equation} \label{oddmn} \mbox{\fbox{ $m_n\simeq
\left(n+\frac{1}{2}|(c-\frac{1}{2})| -\frac{1}{4}\right)\pi k~e^{-\pi kR},$}}
\end{equation} Thus the even and odd fields have degenerate KK mass
which reaches minimum at $c=\frac{1}{2}$ when it has mass equal to the gauge
boson. This is the value of c for which the fermion is in the conformal limit i.e when the bulk proffile of the fermions becone independent of the 
fifth dimension.
\normalsize
\subsection{The zero mode profile} One of the major features of warped extra
dimension is that, unlike the flat case the zero modes still have $y$ dependence
which is generally exponential. The non-trivial zero mode bulk profiles lead
to a possible explanation of the fermion mass hierarchy. Here we briefly
summerize the general features of the zero mode fields in
Table~\ref{zeromodprof}.
\begin{table} \begin{center} \label{zeromodprof} \tabcolsep=0.1cm
\begin{tabular}{|c|c|} \hline \hline Field & Profile \\ \hline scalar
$\phi^{(0)}(y)$ &$e^{(1\pm\sqrt{4+a})k|y|}$\\ \hline fermion $\psi^{(0)}_+(y)$
& $e^{\left(\frac{1}{2}-c\right)k|y|}$ \\ \hline vector $A_\mu^{(0)}(y)$ & 1
\\ \hline graviton $ h_{\mu\nu}^{(0)}(y)$ & $e^{-k|y|}$ \\ \hline
\end{tabular} \caption{\small The zero mode profiles of bulk fields.}
\end{center} \end{table}

The equation of motion for the zero mode fields can be easily derived from
Eq.~\ref{kin} and is given by \begin{equation} (\pm t \partial_t -\nu)f_0(t)
= 0 \end{equation} where, $ t =e^{\pi kR}e^{k|y|}$ and $\nu =
\frac{m_{\Phi}}{k}$. The $m_{\Phi}$'s are given in Eq.~\ref{masses}. The
solution of this equation with correct normalization is given by,
\begin{equation} f_o^{(L,R)}(t) =
\sqrt{\frac{1+2\nu}{1-\epsilon^{1+2\nu}}}\epsilon^{\pm\nu}e^{\pm\nu|k|y|}.
\label{flr} \end{equation} Only the left mode corresponding to the negative
sign in the right hand side in Eq.~\ref{flr}, gives viable solution and the
right mode does not exist as required by the orbifolding conditions.

To get the zero mode profile we need to look at the coefficient of the kinetic
term. For example the kinetic term of the fermions is given by, $
S_{kinetic}^{\psi}=-\int d^4x\int dy\sqrt{-g}\, [i\bar{\Psi}\gamma^MD_M\Psi]$
$\sim -\int d^4x\int dy\sqrt{-g}\, [i\bar{\Psi_0}
\bar{f_0}\gamma^MD_M\Psi_0f_0] $ . Now note that $\sqrt{-g}= e^{-4k|y|}$ and
$\gamma_{\mu} = E^A_a\gamma^a_{\mu}$ where $E^A_a = e^{|k|y}$. Further
$\widehat{\Psi}=e^{-2\sigma}\Psi_{L,R}$. Putting everything together we get $
S_{kinetic}^{\psi}=-\int d^4x\int dy e^{(1-2c)k|y|}
[i\bar{\Psi_0}\gamma^a_MD^M\Psi_0]$. Thus the zero mode profile is given by
$e^{(\frac{1}{2}-c)k|y|}$.

The other zero mode profiles can be derived analogously. The gauge fields are
kept in the conformal limit ($c= 1/2$) i.e. they do not have any bulk profile and thus
the zero mode is not localized at any point in the bulk. This is a direct
consequence of the 5d gauge invariance. The gauge fields (any field in the
conformal limit) couple to the two branes with equal strength whereas the
other zero mode fields are localized near one point in the bulk having
different couplings at the two branes.

\subsection{Gauge Couplings} The generic gauge coupling part of the 5d action
may be written as , \begin{equation} S_{gauge}=\int d^4 x \int dy \sqrt{-g}\,
g_5 \bar{\Psi_i}(x,y)\imath \gamma^\mu A_\mu(x,y) \Psi_i(x,y)~.
\end{equation} One can simply read off the coupling between all the zero mode
fields, which can now be written as, 
\small
\begin{equation} S_{gauge}^{(0)}=\int d^4
x \int dy \sqrt{-g}\, g_5\left( \bar{\Psi_i}(x)^{(0)}f^{(0)}_i(y)\imath
E^a_A\gamma^\mu_a A_\mu(x)^{(0)}f^A_{(0)}(y) \Psi_i(x)^{(0)}\right)\times
\left(\frac{1}{\sqrt{2\pi R}}\right)^3 \end{equation}
\normalsize
 where $E^a_A = e^{k|y|}$
is the vierbein. The 4d gauge coupling for the fermion of flavor $i$ is given
by, \begin{equation} g_{4,i}=\int dy \sqrt{-g}\,
g_5E^a_A\left(\frac{1}{\sqrt{2\pi R}}\right)^3(f^{(0)}_i(y))^2f^A_{(0)}(y),
\end{equation} remembering that all the gauge bosons are put in the conformal
limit (implies $f^A_{(0)}(y)=1 $) we can explicitly perform the y integral to
get, \large \begin{equation} \mbox{\fbox{$g_{4,i}= \frac{g_5}{\sqrt{2\pi
R}}.$}} \end{equation} \normalsize This clearly shows that the zero mode
couplings are independent of the flavor index, thus gauge invariance of the
theory is not compromised by localizing the zero modes at different location
in the bulk. Couplings of the zero mode fermions to the KK modes of the gauge
bosons may lead to four-fermion interactions and FCNC processes that are
highly constrained\footnote{Such highly constrained couplings can be evaded (atleast at the tree level) by
simply imposing KK number conservation\cite{Agashe:2007jb}.}. Using the
expression for the zero-mode fermion , the gauge coupling of a gauge boson
Kaluza-Klein mode $n$ to the zero-mode fermions is \begin{equation} g^{(n)} =
g\left(\frac{1-2c}{e^{(1-2c)\pi kR}-1}\right)\frac{k}{N_n} \int_0^{\pi R} dy\,
e^{(1-2c)\sigma} \left[ J_1(\frac{m_n}{k} e^\sigma) +b_1(m_n)Y_1(\frac{m_n}{k}
e^\sigma)\right]~.  \end{equation} 
\begin{figure} \centerline{ {
\begin{picture}(0,0) \includegraphics{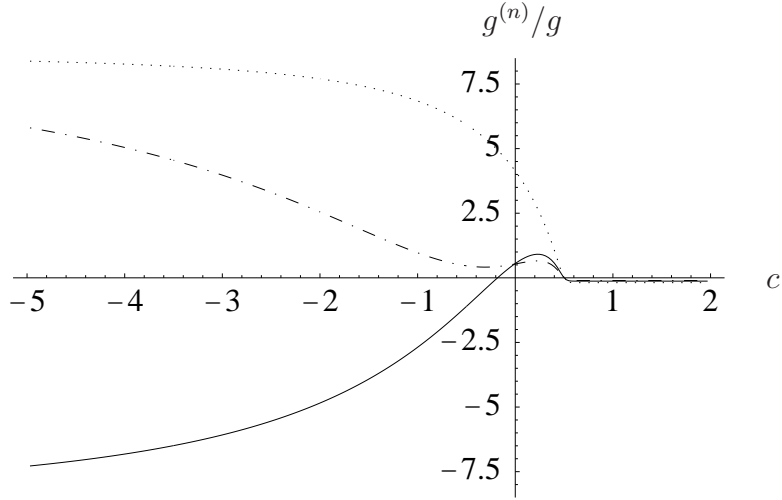} \end{picture}%
\setlength{\unitlength}{3947sp} \begin{picture}(4950,3221)(3964,-4931)
\put(7201,-2011){\makebox(0,0)[b]{$g^{(n)}/g$}}
\put(8776,-3586){\makebox(0,0)[b]{$c$}} \end{picture} }} \caption{\small The
ratio of the gauge couplings, $g^{(n)}/g$, for $n=1$ (dotted line), $n=2$
(solid line) and $n=3$ (dashed-dotted line), as a function of the
dimensionless fermion mass parameter $c$~\cite{Gherghetta:2000qt}.}
\label{fig:gc} \end{figure}

When $c$ takes a large negative value, the fermion is localized near the
TeV-brane and the ratio $g^{(1)}/g$ approaches the asymptotic limit
$g^{(1)}/g\simeq\sqrt{2\pi kR}\simeq 8.4$, which corresponds to a fermion
localized near the TeV-brane.  This leads to a restrictive lower bound on the
first excited Kaluza-Klein mass scale.  At the conformal limit $c=1/2$, the
coupling vanishes due to the conservation of the 5-momentum at this limit.
For $c> 1/2$, the coupling quickly becomes universal for all fermion
masses. Nevertheless, the FCNC and other constraints will dissappear if a KK
parity is induced into the theory. In what follows we will see that various
other considerations will also lead us to introduce a KK parity into the
theory. Thus we will restrict the rest of this discussions to models where KK parity is a good symmetry of the theory.

\subsection{Yukawa Structure}\label{kkyukawasec}
We note that the fermions and their superpartners have identical coupling to
the Higgs. Thus in what follows, we only consider the Yukawa coupling of the
fermion.  The Higgs boson is localized on the TeV brane and thus the 5d Higgs field may be written as, \begin{equation} \label{higg}
H(x,y)=H(x)\delta(y-\pi R) \end{equation}

\begin{figure}[t] \centering
\includegraphics[width=0.5\textwidth,angle=270,keepaspectratio]{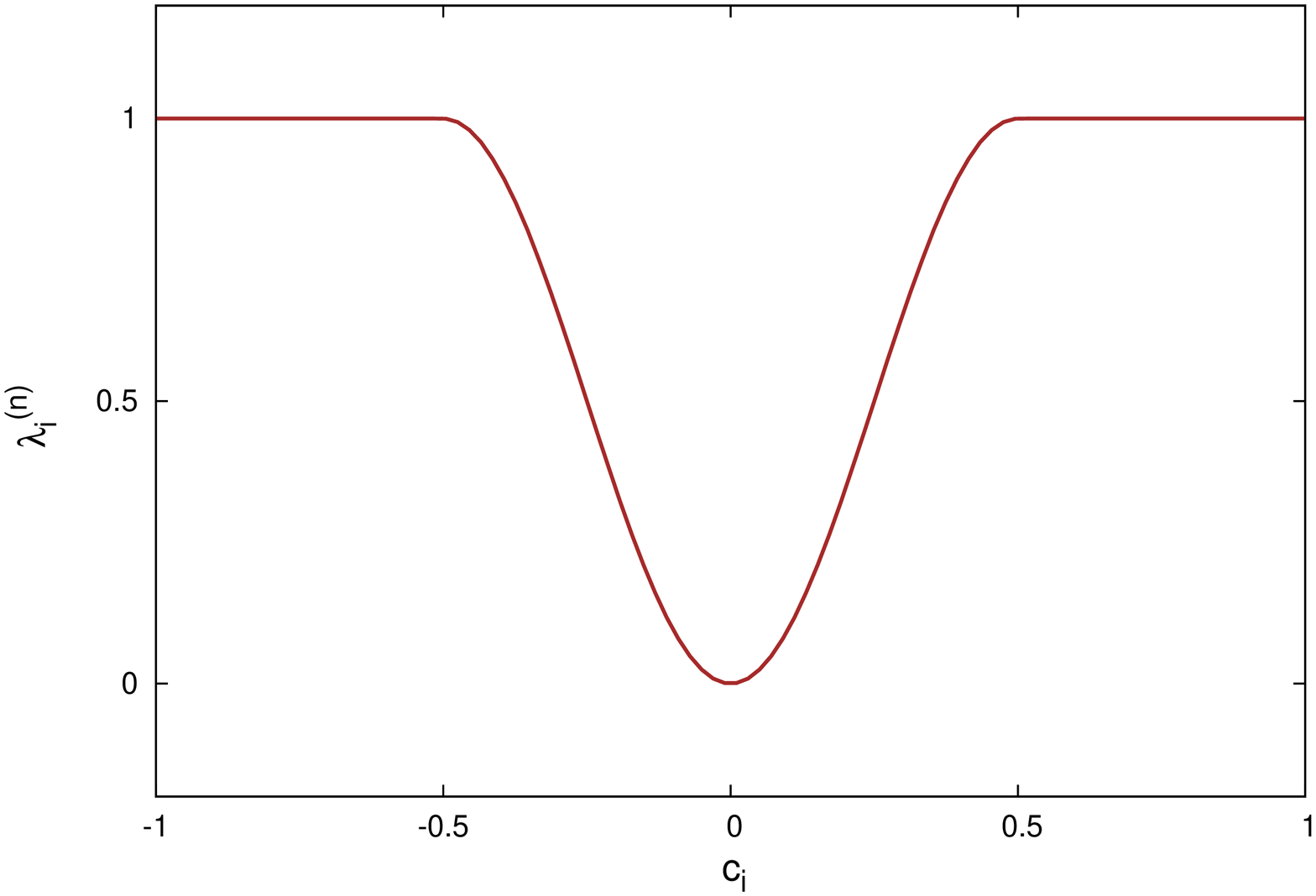}
\caption{\small The variation of the Yukawa couplings with $c_i$.}
\label{figyukawa} \end{figure}
Using Eq.~\ref{higg} and Eq.~\ref{Kaluza-Klein}, we find that the 5d Yukawa term is given by, \begin{equation} \label{5dimenyc}
S_y=\int d^4x \int dy\, \sqrt{-g}\,\,\lambda^{(5)}_{ij} H(x) \Big(
\bar\Psi_{iL}(x,y)\Psi_{jR}(x,y) + h.c.\Big) \delta(y-\pi R) \end{equation}
where, $\Psi_{i,L/R}$ is a chiral fermion with flavor index $i$ and $H(x)$ is
the Higgs boson . This can be written as,\footnote{We impose KK parity to keep
the theory calculable. It is a direct analogue of the R parity in SUSY.}
\begin{eqnarray} \label{4dimenyc} S_y&=&\int d^4x \,\,\left[ \lambda_{ij} H(x)
\Big( \bar \Psi^{(0)}_{iL}(x)\Psi^{(0)}_{jR}(x) + h.c. + \dots\Big) \right. \\
\nonumber &+& \left. \sum \lambda_{ij}^{(n)} H(x) \Big( \bar
\Psi^{(n)}_{iL}(x) \Psi^{(n)}_{jR}(x) + h.c. + \dots\Big) \right]
\end{eqnarray}

Considering a left-right symmetric model\footnote{We consider that the left
and right chiral fermions of the same flavor are identically localized in the
bulk.}, we find that the zero mode Yukawa couplings ($\lambda_{ij}$) are given
by, \large \begin{equation} \label{yc} \mbox{\fbox{$ \lambda_{ij}=
\frac{(1/2-c_{i})\lambda_{ij}^{(5)}k}{e^{(1-2c_{i})\pi kR}-1} e^{(1-2c_{i})\pi
kR},~$}} \end{equation} \normalsize If we assume $\lambda_{ij}^{(5)}k \sim 1$
for all $i,j$ we can still generate hierarchal Yukawa structure by tuning the
$c$ parameter. This is the explanation of the fermion hierarchy problem in
warped extra dimension where the warping factor is used to generate the
variation in the zero mode fermion masses.  The Yukawa couplings of the KK
modes can be read off by inserting Eq.~\ref{Kaluza-Klein} in
Eq.~\ref{5dimenyc} and comparing with Eq.~\ref{4dimenyc} and are given by,

\begin{eqnarray} {\lambda}_i^n&=&\sqrt{-g} {\lambda}_{ij}^{(5)}e^{-\pi kR}
\frac{e^{4\pi kR}}{2 \pi R}{\Big(f_n(\pi R)\Big)}^2 \nonumber \\
&=&\sqrt{-g}e^{3\pi kR}{\lambda}_{ij}^{(5)}\pi m_n {\left[
J_{\alpha}\Big(\frac{m_n}{k}e^{\pi
kR}\Big)+b_{\alpha}(m_n)Y_{\alpha}\Big(\frac{m_n}{k}e^{\pi kR}\Big) \right]
}^2 \end{eqnarray} With the approximations $m_n \ll k$, $kR \gg 1$ and
${\lambda}_{ij}^{(5)}k=1$, we can expand the Bessel function in the asymptotic
limit as, \begin{equation}\label{jalpha} J_n(x) = \sqrt{\frac{2}{\pi
x}}\cos(x-(2|n| +1)\frac{\pi}{4} \end{equation} Using this form we find that
the KK Yukawa couplings are given by \large \begin{equation} \label{kkyukawa}
\mbox{\fbox{${\lambda}_i^n \sim {\cos}^2 \Big(\left[ n+ \frac{|c-\frac{1}{2}|
- |c+\frac{1}{2}|}{2} - \frac{1}{2}\right] \pi \Big).$}} \end{equation}
\normalsize This clearly shows that all the even KK modes have Yukawa
couplings equal to unity independent of their zero mode Yukawa couplings for
$|c|>0.5$. The odd KK modes do not couple to the brane bound Higgs at all. The
important observation is that the couplings are nearly independent of the
$c_i$ parameter, the radius of compactification (R) and the KK number (n). The
actual numerical values of the Yukawa couplings are plotted as a function of
$c_i$ in Figure~\ref{figyukawa}.

\begin{center} \begin{table} \centering \begin{tabular}{|c|c|c|c|c|c|c|c|c|c|}
\hline $f_i$&$e$&$\mu$&$\tau$&$u$&$d$&$c$&$s$&$t$&$b$\\ \hline
$c_i$&0.61&0.52&0.40&0.62&0.57&0.52&0.52&-0.50&0.26\\ \hline
$m^{(1)}_i$&1073&1013&1066&1080&1047&1013&1013&1667&1160\\ \hline
$\lambda_n$&1&1&.9&1&1&1&1&1&.5\\ \hline \end{tabular} \caption{\small  The
$c_i$ parameters and $m^{(1)}_i$ (in GeV) for different flavors are shown for
$kR=12.06$ and $\tan\beta= {\langle H_u^0\rangle}/{\langle H_d^0\rangle} =
10$. For this choice, the mass gap between the consecutive KK states is
$m^{(n+1)} - m^{(n)} = 1333$ GeV, irrespective of $c_i$. The corresponding
$n=1$ KK mass for gauge boson is $1$ TeV.}  \label{table} \end{table}
\end{center}

\subsection{Radius stabilization}

The radius of the model so far has been treated as a given constant, and it
was found that radius $R$ has to be $R\sim 12/k$ in order for the hierarchy
problem to be resolved.  This raises several important issues, that needs to
be addressed: Since the radius is not {\it dynamically} fixed at the moment,
but rather just set to its desirable value, there will be a corresponding {\it
massless} scalar field in the effective theory, which corresponds to the
fluctuations of the radius of the extra dimensions, called the {\it
radion}~\cite{GW1,CGRT,GW2,CGR}.  The masslessness of this field is related to
the fact that the RS solution discussed until now did not make any reference
to the size of the extra dimension. This means that in the effective theory
this parameter is also arbitrary, and thus has no non-trivial potential.  Thus
it can have no mass. This massless radion would contribute to Newton's law and
result in violations of the equivalence principle (would cause a fifth force),
which is phenomenologically unacceptable, and therefore the radius {\it has}
to be stabilized (making the radions massive).  Even then the radius has to be
stabilized at a somewhat larger than natural value (we need $kR\sim 12$, while
one would expect $R\sim 1/k$). This reintroduces the hierarchy/finetuning
problem.  We have seen that one needed two fine tunings to obtain the static
RS solution, one of which was equivalent to the vanishing of the 4d
cosmological constant.

A mechanism for radius stabilization should address the above mentioned
issues. The solution for stabilization of the size of the extra dimension was
proposed by Goldberger and Wise~\cite{GW1}, and is known as the
Goldberger-Wise (GW) mechanism. Radius stabilization at non-trivial values of
the radius occurs dynamically, where different forces, some of which would
like to drive the extra dimension very large, and some very small, are brought
into play. Then there is a chance that these forces may balance each other at
some value and a stable non-trivial minimum for the radius could be found. A
possible way to find such a tension between large and small radii is if there
is a tension between a kinetic and a potential term of a field, one which
would want derivatives to be small (and thus large radii) and the other which
would want small radii to minimize the potential.  The Goldberger and Wise
mechanism uses exactly this scenario. A bulk scalar field is introduced into
the model, and a bulk mass term is added. This will result in a non-trivial
potential for the radius, due to the bulk mass the radius would want to be as
small as possible to minimise the potential. If there is also a non-trivial
profile (a vev that is changing with the extra dimensional coordinate) for
this scalar, then the kinetic term would want the radius to be as large as
possible, so as to minimize the kinetic energy in the 5th direction. Then
there would be a non-trivial minimum for the radius. The non-trivial profile
for the scalar is generated by adding a {\it brane potentials} for this scalar
on both fixed points, which have minima at different values from each other.
The {\it back-reaction} of the metric to the presence of the scalar field in
the bulk will be important. Simultaneous solution of the Einstein and the bulk
scalar equations are required, to have the back-reaction exactly under
control~\cite{dWGFK,CEHS,CEGH,CGK}.  Denote the scalar field in the bulk by
$\Phi$, and consider the action \begin{equation} \int d^5x\sqrt{g}
\left[-M_*^3 R++\frac{1}{2}(\nabla\Phi)^2 -V(\Phi )\right]- \int d^4x
\sqrt{g_4} \lambda_P (\Phi )- \int d^4x \sqrt{g_4} \lambda_T (\Phi ),
\end{equation} where the first term is the usual 5d Einstein-Hilbert action
and the bulk action for the scalar field, while the next two terms are the
brane induced potentials for the scalar field on the Planck and the TeV
branes. We will denote the 5d Newton constant  by
$\kappa^2=1/2M_*^3$, and look for an ansatz of the background metric again of
the generic form as in the RS case to maintain 4d Lorentz invariance:
\begin{equation} ds^2 =e^{-2 A(y)} \eta_{\mu\nu} dx^\mu dx^\nu +dy^2.
\end{equation} The Einstein equations are given by: \begin{eqnarray}
\label{bulkEinst} && 4 A'^2-A''=-\frac{2\kappa^2}{3} V(\Phi_0)
-\frac{\kappa^2}{3} \sum_{i=P,T} \lambda_i (\Phi_0) \delta (y-y_i) \nonumber
\\ && A'^2=\frac{\kappa^2}{12} \Phi_0'^2-\frac{\kappa^2}{6} V(\Phi_0).
\end{eqnarray} And the bulk scalar equation of motion in the warped space,
derived from the generic scalar equation is given by, \begin{equation}
\partial_\mu \sqrt{g} g^{\mu\nu} \partial_\nu \Phi =\frac{\partial
V}{\partial\Phi}\sqrt{g}.  \end{equation} By substituting the scalar and
metric ansatz into this equation we get \begin{equation}
\Phi_0''-4A'\Phi_0'=\frac{\partial
V}{\partial\Phi_0}+\sum_i\frac{\partial\lambda_i (\Phi_0)}{\partial\Phi}
\delta (y-y_i).  \label{bulkscalar} \end{equation} 
We can separate these
equations into the bulk equations that do not contain the delta functions, and
the boundary conditions are obtained by matching the coefficients of the delta
functions at the fixed points.  The boundary conditions derived this way are
sometimes also called the {\it jump equations}, which in our case will be
given by \begin{eqnarray} && [A']_i=\frac{\kappa^2}{3}\lambda_i(\Phi_0),
\nonumber \\ && [\Phi_0']_i=\frac{\partial\lambda_i (\Phi_0)}{\partial\Phi}.
\end{eqnarray} 
The bulk equations, Eq.~\ref{bulkEinst}-\ref{bulkscalar}, together
with these boundary conditions form the equations of the coupled
gravity-scalar system. These are coupled second order differential equations.
Let us assume that the solution to the system of equations above are given by
$A(y), \Phi_0(y)$. Where the superpotential function $W(\Phi )$ is defined via
the equations \begin{eqnarray} &&A'\equiv \frac{\kappa^2}{6} W (\Phi_0),
\nonumber \\ &&\Phi_0'\equiv \frac{1}{2} \frac{\partial W}{\partial \Phi}.
\end{eqnarray} If we use these expressions for $A'$ and $\Phi_0'$ in the
Einstein and scalar equations consistancy will demand the following relation:
\begin{equation} \label{Weq} V (\Phi )=\frac{1}{8} \left( \frac{\partial
W}{\partial \Phi}\right)^2-\frac{\kappa^2}{6} W(\Phi )^2, \end{equation} and
the corresponding jump conditions are, \begin{eqnarray} \label{jump} &&
\frac{1}{2}[W(\Phi_0)]_i=\lambda_i(\Phi_0), \nonumber \\ &&
\frac{1}{2}[\frac{\partial
W}{\partial\Phi}]_i=\frac{\partial\lambda_i(\Phi_0)}{\partial\Phi}.
\end{eqnarray} 
It is somewhat difficult to derive a superpotential for a
specific potential.  Generically the bulk potential should include a
cosmological constant term (independent of $\Phi$) and a mass term (quadratic
in $\Phi$), but for simplicity we neglect them. So we choose~\cite{dWGFK},
\begin{equation} W(\Phi )=\frac{6k}{\kappa^2}-u\Phi^2.  \end{equation} 
The
first term is just what one needs for a cosmological constant, while the
second term will provide the mass term when taking the derivative. The jump
conditions are satisfied if, \begin{equation} \lambda (\Phi )_\pm=\pm
W(\Phi_\pm)\pm W'(\Phi_\pm) (\Phi-\Phi_\pm)+ \gamma_\pm (\Phi-\Phi_\pm)^2,
\end{equation} where $\Phi_\pm$ are the values of the scalar field at the two
branes, which we will also denote by $\Phi_+=\Phi_P$ at the Planck brane, and
$\Phi_-=\Phi_T$ at the TeV brane. Then the solution is given by,
\begin{equation} \Phi_0 (y) =\Phi_P e^{-uy}.  \end{equation} 
From this, the
value of the scalar field at the TeV brane is determined to be
\begin{equation} \Phi_T =\Phi_P e^{-uR}.  \end{equation} This means that the
radius is no longer arbitrary, but given by \begin{equation} R=\frac{1}{u} \ln
\frac{\Phi_P}{\Phi_T}.  \end{equation} The value of the radius is determined
by the equations of motion, which is exactly what we were after. This is the
GW mechanism.

The metric background will then be obtained from the equation \begin{equation}
A'=\frac{\kappa^2}{6}W(\Phi_0)= k-\frac{u\kappa^2}{6} \Phi_P^2 e^{-2uy}
\end{equation} given by the solution \begin{equation} A(y)=k y
+\frac{\kappa^2\Phi_P^2}{12}e^{-2uy}.  \end{equation} The first term is the
usual RS warp factor (remember that $A$ has to be exponentiated to obtain the
metric), while the second term is the back-reaction of the metric to the
non-vanishing scalar field in the bulk.  We will assume that the back-reaction
is small, and thus that $\kappa^2\Phi_P^2, \kappa^2\Phi_T^2 \ll 1$, and that
$v>0$. The values of $\Phi_P$ and $\Phi_T$ are determined by the bulk and
brane potentials, so $\Phi_P/\Phi_T$ is a fixed value. Since we want to
generate the right hierarchy between the Planck and weak scales we need to
ensure that \begin{equation} kR\sim 12, \end{equation} from which we get that
\begin{equation} \frac{k}{u} \ln \left( \frac{\Phi_P}{\Phi_T}\right) \sim 12,
\end{equation} which implies that $u/k$ does not need to be exponentially
small. This ratio sets the hierarchy in the RS model, and we can see that
indeed one can generate this hierarchy using the GW stabilization mechanism by
a very modest tuning of the input parameters of the theory.

Once the radius is stabilized using a non-trivial potential, we know that the
radion is no longer massless. Next we find the radion mass~\cite{CGK,TM}, see
also~\cite{CGRT,GW2,KKOP}. For this, we need to find the scalar excitations of
the coupled gravity-scalar system. This can be parameterized in the following
way: \begin{eqnarray} && ds^2=e^{-2A-2F(x,y)}\eta_{\mu\nu}dx^\mu
dx^\nu+(1+G(x,y))^2 dy^2, \nonumber \\ &&\Phi (x,y)=\Phi_0(y)+\varphi (x,y).
\end{eqnarray} At this moment it looks like there would be three different
scalar fluctuations, $F,G$ and $\varphi$. However, if we plug this ansatz into
the Einstein equation the 4d off-diagonal $\mu\nu$ components are satisfied
only if \begin{equation} G=2F, \end{equation} while the $\mu 5$ components
imply the following further relation among the fluctuations: \begin{equation}
\varphi = \frac{1}{\Phi_0'} \frac{3}{\kappa^2}(F'-2A'F).  \end{equation} This
means, that in the end there is just a single independent scalar fluctuation
in the coupled equation, which we can choose to be $F$. Consistency of the
Einstein equations lead to the following equation: \begin{equation}
F''-2A'F'-4A''F-2\frac{\Phi_0''}{\Phi_0'}F'+4A'\frac{\Phi_0''}{\Phi_0'}F=e^{2A}\Box
F \label{radeq} \end{equation} in the bulk and the following boundary
condition: \begin{equation} (F'-2A'F)_i=0.  \end{equation} 
Let us first assume
that there is no stabilization mechanism, that is the background is {\it
exactly} the RS background given by $A=k|y|$, and $\Phi_0=0$. In this case
most of the terms in the above equation disappear, and we are left with
\begin{equation} F'-2kF= e^{2ky} m^2 F, \ \ (F'-2kF)_i=0.  \end{equation} The
only solution is for $m^2=0$, and the wave function of the un-stabilized
radion will be given by \begin{equation} F(y)=e^{2k|y|}.  \end{equation} 
The
metric corresponding to radion fluctuations in the unstabilized RS model
corresponds to \begin{equation} ds^2=e^{-2k|y|-2e^{k|y|}f(x)}\eta_{\mu\nu}
dx^\mu dx^\nu+(1+2e^{2k|y|}f(x))dy^2.  \end{equation} This is a single scalar
mode, that is exponentially peaked at the TeV brane, just like all the
graviton KK modes.

To find the radion mass for the case with GW stabilization, we simply need to
plug into Eq.~\ref{radeq} the full background for $A$ and $\Phi_0$ with
stabilization: \begin{equation} F''-2A'F'-4A''F+2uF'-4uA'F+m^2e^{2A}F=0.
\end{equation} To find the leading term for the radion mass we expand in terms
of the back-reaction of the metric in the parameter $l=\kappa
\Phi_P/\sqrt{2}$, and obtain the mass of the radion \begin{equation}
m_{radion}^2=\frac{4l^2(2k+u)u^2}{3k} e^{-2(u+k)r}.  \end{equation}

The radions coupling to the SM particles offers a rich phenomenology which is
beyond the scope of this discussion.

\section{Probing warped extra dimension via ${gg \to h}$ and ${h \to \gamma
      \gamma}$ at LHC } \label{gghgaga}

\textbf{This section closely follow the work published in the
paper:~G.~Bhattacharyya and T.~S.~Ray,
  ``Probing warped extra dimension via gg $\rightarrow$ h and h $\rightarrow \gamma \gamma$ at LHC,''
  Phys.\ Lett.\  B {\bf 675} (2009) 222
  [arXiv:0902.1893 [hep-ph]].}

 For an intermediate mass ($<$ 150 GeV) Higgs boson, the relevance of its
production at the CERN Large Hadron Collider (LHC) via gluon fusion ($gg \to
h$) and its subsequent decay into two photons ($h \to \gamma \gamma$) cannot
be over-emphasized. Since these are loop induced processes, a natural question
arises as how sensitive these processes are to the existence of new physics.
In this chapter, we explore such a possibility by embedding the Standard Model
(SM) in a Randall-Sundrum (RS) warped geometry \cite{Randall:1999ee}, where
the bulk is a slice of Anti-de Sitter space (AdS$_5$) accessible to some or
all SM particles \cite{bulksm,Gherghetta:2000qt}. The virtues of such a
scenario include a resolution of the gauge hierarchy problem caused by the
warp factor \cite{Randall:1999ee}, and an explanation of the hierarchy of
fermion masses by their respective localizations in the bulk keeping the Higgs
confined at the TeV brane \cite{Huber:2000ie}. Besides, the smallness of the
neutrino masses could be explained \cite{Grossman:1999ra}, and light KK states
would lead to interesting signals at LHC \cite{rslhc}. We demonstrate that the
loop contribution of the KK towers of quarks and gauge bosons emerging from
the compactification would have a sizable numerical impact on the $gg \to h$
and $h \to \gamma \gamma$ rates. This happens because the Higgs coupling to a
pair of KK fermion-antifermion is not suppressed by the zero mode fermion mass
and can easily be order one. The underlying reason
is simple.  Although the zero mode wave-functions of different flavors have
varying overlap at the TeV brane depending on the zero mode masses, the KK
profiles of all fermions have a significant presence at the TeV brane where
the Higgs resides. As a result, the KK Yukawa couplings of different flavors
are not only all large, they are also roughly universal. This large universal
Yukawa coupling in the RS scenario constitutes the corner-stone of our study.
On the contrary, in flat Universal Extra Dimension (UED) only the KK top
Yukawa coupling is large, others being suppressed by the respective zero mode
fermion masses.  We provide comparative plots to demonstrate how the warping
in RS fares against the flatness of UED for the processes under consideration.

\subsection{Contribution of KK states to ${\sigma(gg \to h)}$} The process $gg
 \rightarrow h$ proceeds through fermion triangle loops. The SM expression of
 the cross section is given by ($\tau_q \equiv 4m_q^2 / m_{H}^2$) 
 \begin{eqnarray} \label{ggh} \sigma_{gg \rightarrow h}^{\rm SM} & = &\frac{
 \alpha_{s}^2}{576 \pi v^2}\left|\sum_q A_q(\tau_q)\right|^2 \,\, , ~~{\rm
 where}~~ \left. A_q(\tau_q) \right|_{\rm SM} = 2\tau_q[1+(1-\tau_q)f(\tau_q)]
 ~,  \label{ftau} 
\end{eqnarray}
 with $f(\tau) = {\rm arcsin}^2 \left(
 \frac{1}{\sqrt{\tau}}\right) ~{\rm for}~ \tau \geq 1,$  and $ f(\tau) =
 -\frac{1}{4} \left[ {\rm ln}\left( \frac{1+\sqrt{1-\tau}} {1-\sqrt{1-\tau}}
 \right) -i\pi \right]^2 {\rm for}~ \tau < 1 ~$. 
Above,
 $\alpha_s$ is the QCD coupling at the Higgs mass scale, $v$ is the Higgs
 vacuum expectation value and $A_q$ is the loop amplitude from the $q$th
 quark.  In the SM, the dominant contribution comes from the top quark loop.
 Now, there will be additional contributions from the KK quarks. Importantly,
 due to the large universal KK Yukawa couplings, not only the KK top but also
 the KK modes of other quarks would have sizable contribution. Indeed, the
 lightest modes ($n=1$) would have dominant contributions. Setting the KK
 Yukawa couplings to unity, as suggested by Eq.~\ref{kkyukawa}, we derive
 the amplitude of the $n$th KK mediation of the $q$th flavor, with the same
 normalization of Eq.~\ref{ggh}, as \large \begin{equation}
 \mbox{\fbox{$\left. A_q(\tau_{q_n}) \right|_{\rm KK} = \frac{4v^2}{m_{h}^2}
 \left[1+(1-\tau_{q_n})f(\tau_{q_n})\right].$}} \label{FtKK} \end{equation}
 \normalsize

In 5d the sum over $n$ yields a finite result. Eq.~\ref{FtKK} is different
from the UED result \cite{Petriello:2002uu} in two ways: (i) we have set the
KK Yukawa coupling to unity irrespective of quark flavors, while in UED the KK
Yukawa coupling is controlled by zero mode masses; (ii) in UED there is an
additional factor of 2 because both $\mathbb{Z}_2$ even and odd KK modes
contribute, while in RS the odd modes do not couple to the brane-localized
Higgs.  In Figure~\ref{gghfig}, we have plotted the variation with $m_h$ of the deviation
of the production cross section $\sigma_{\rm RS} (gg \to h)$ from its SM
expectation $\sigma_{\rm SM} (gg \to h)$ normalized by the SM value. The
dominant QCD correction cancels in this normalization. We have chosen four
reference values of $m_{\rm KK}$ ($= 1.0, 1.5, 2.0$ and $3.0$ TeV), where
$m_{\rm KK}$ is the KK mass of the $n=1$ gauge bosons, which also happens to
be the lightest KK mass in the bulk (corresponding to the conformal limit,
$c=1/2$ for fermions). For $m_h$ below 150 GeV, the deviation is quite
substantial (close to 45\%) for $m_{\rm KK} = 1$ TeV. For larger $m_{\rm KK}
=$ 1.5 (3.0) TeV, the effect is still recognizable, around 18\% (5\%). In the
inset, we exhibit a comparison between RS and UED contributions to the same
observable, where the KK mass scales of the two scenarios, namely $m_{\rm KK}$
for RS and $1/R$ for UED, have been assumed to be identical ($=1$ TeV). For
$m_h <$ 150 GeV, the RS contribution is about 2.5 times larger than the UED
contribution, while the margin slightly goes down with increasing $m_h$. This
factor 2.5 can be understood in the following way: In RS, five $n=1$ KK
flavors (except the KK top) have mass around $m_{\rm KK}$ with order one
Yukawa coupling. So naively we would expect a factor of 5 enhancement relative
to UED. But in UED both $\mathbb{Z}_2$ even and odd modes contribute. This
reduces the overall enhancement factor in RS over UED to about 2.5.

\subsection{ Contribution of KK states to ${\Gamma(h \to \gamma \gamma)}$} 
\begin{figure}  \begin{minipage}[t]{0.47\textwidth}
 \begin{center}
\includegraphics[width=0.7\textwidth,angle=270,keepaspectratio]
{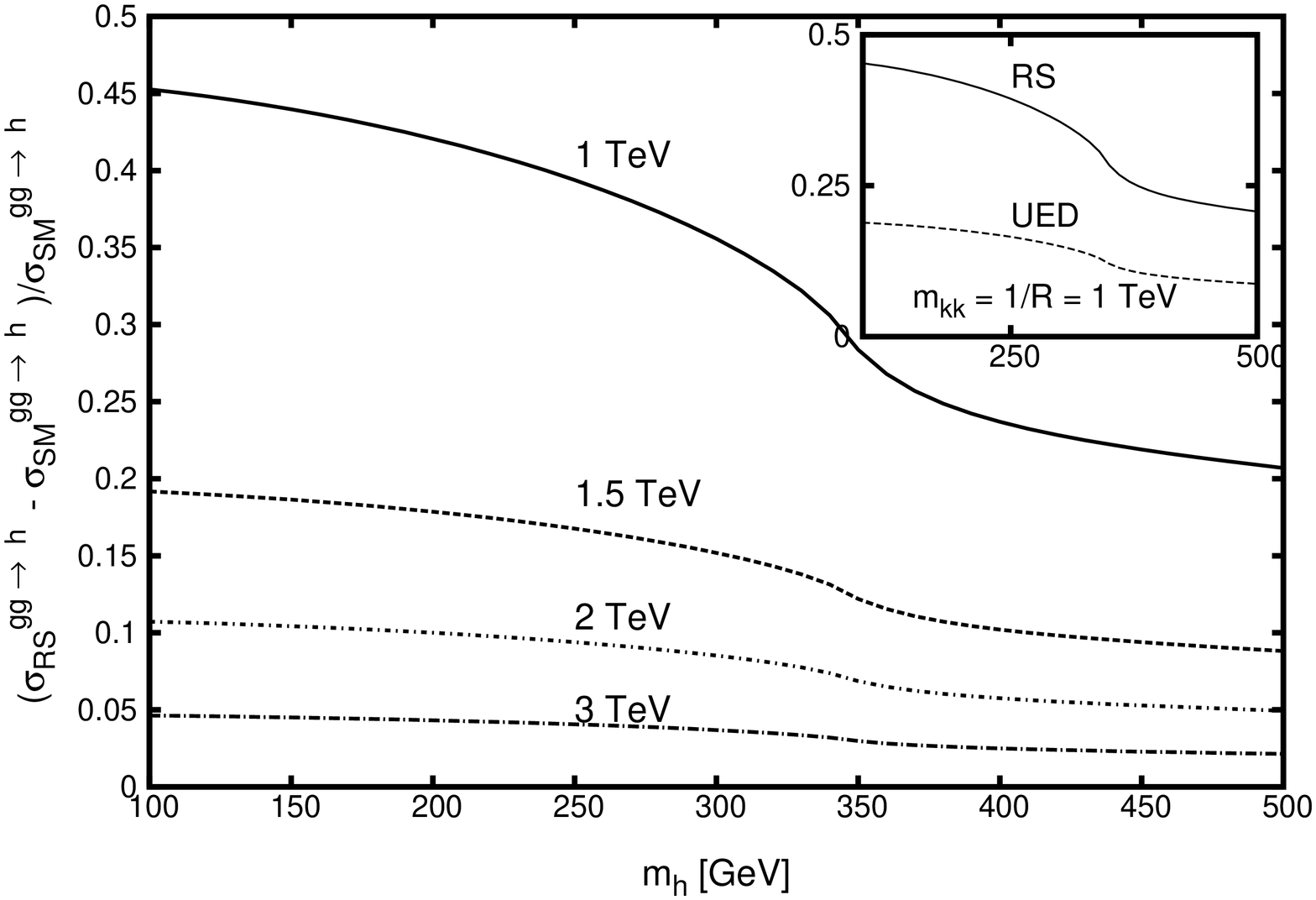} \caption{ \small The fractional deviation (from the
SM) of the $gg \rightarrow h$ production cross section in RS is plotted
against the Higgs mass.  The four curves correspond to four different choices
of $m_{\rm KK}$.  In the inset, we have compared the UED contribution for $1/R
= 1$ TeV with the RS contribution for $m_{\rm KK} = 1$ TeV.}  \label{gghfig}
\end{center} \end{minipage} \hspace{7mm} \begin{minipage}[t]{0.47\textwidth}
 \begin{center}
\includegraphics[width=0.7\textwidth,angle=270,keepaspectratio]
{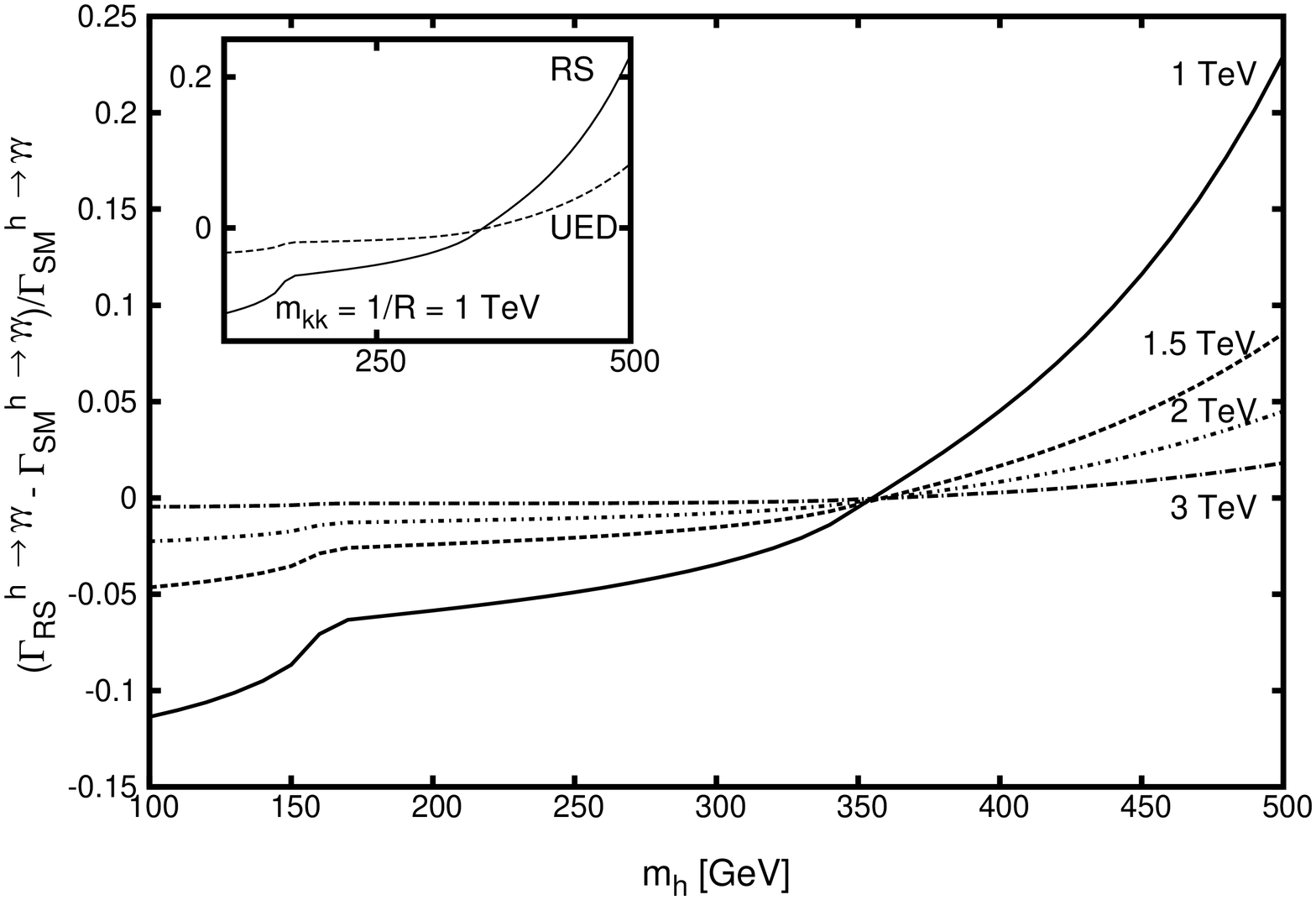}  \caption{ \small Same as in Figure~\ref{gghfig},
except that the fractional deviation in $h \to \gamma \gamma$ decay width is
plotted.}  \label{gghafig} \end{center} \end{minipage} \end{figure}

The
$h \to \gamma\gamma$ process proceeds through fermion triangles as well as via
gauge loops along with the associated ghosts.  The decay width in the SM can
be written as  \begin{equation} \Gamma_{h \rightarrow \gamma \gamma} = \frac{
\alpha m_h^3}{256 \pi^3 v^2} \left| \sum_f N_c^f Q_f^2 A_f(\tau_f) +
A_W(\tau_W) \right|^2 \, , \end{equation}  where $\alpha$ is the
electromagnetic coupling at the Higgs mass scale.  The expression for $A_f$ is
given in Eq.~\ref{ggh}, and the dominant SM contribution to $A_f$ comes from
the top quark loop.  The $W$-loop amplitude in the SM is given by 
\begin{equation} \left. A_W (\tau_W)\right|_{\rm SM} = -\left[2+3\tau_W
+3\tau_W(2-\tau_W)f(\tau_W)\right] \, .  \end{equation} 
We derive the KK
contribution of the gauge sector as \large \begin{equation} \label{awn}
\mbox{\fbox{$\left. A_W (\tau_{W_n})\right|_{\rm KK} = -\left[2+3\tau_W
+3\tau_W(2-\tau_{W_n})f(\tau_{W_n}) - 2(\tau_{W_n}
-\tau_W)f(\tau_{W_n})\right].$}} \end{equation} \normalsize 
Again, the sum
over $n$ yields finite result and in the limit of large KK mass the KK
contribution decouples. Our Eq.~\ref{awn} is very different from the
corresponding UED expression \cite{Petriello:2002uu}, primarily because the
Higgs is confined at the brane in the present scenario while it resides in the
bulk in UED.  In Figure~\ref{gghafig}, we have plotted the decay width $\Gamma(h\to
\gamma\gamma)$ in RS relative (and normalized as well) to the SM. Again, the
four choices of $m_{\rm KK}$ are 1.0, 1.5, 2.0 and 3.0 TeV. There is a partial
cancellation between quark and gauge boson loops, both in real and imaginary
parts, not only for the zero mode but also for each KK mode. The meeting of
the four curves just above the $m_h = 2 m_t$ threshold is a consequence of the
above cancellation and at the meeting point the SM contribution overwhelms the
KK contribution. Unlike in Figure~2.4, we witness both suppression and enhancement
with respect to the SM contribution. The inset carries an illustration how RS
fares against UED for identical KK masses.  

\begin{figure}
\begin{center}
\includegraphics[width=0.35\textwidth,angle=270,keepaspectratio]
{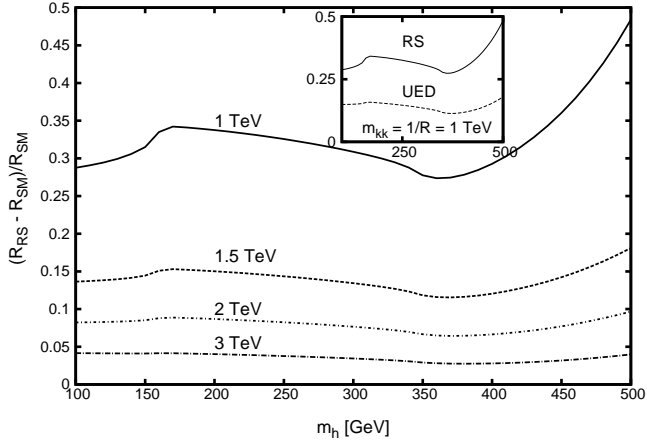}
\caption{ \small Same as in Figure~\ref{gghfig}, except that the fractional deviation
  in $R = \sigma_{gg \to h}~ \Gamma_{h \to \gamma \gamma}$ has been plotted.}
\label{hggfig}
\end{center}
\end{figure}

Next we construct a variable $R = \sigma_{gg \to h}~\Gamma_{h \to \gamma
  \gamma}$. In Figure~\ref{hggfig}, we have studied variation of $(R_{\rm RS} - R_{\rm
  SM})/R_{\rm SM}$ with $m_h$. For $m_{\rm KK} =$ 1.0, 1.5, 2.0 and 3.0 TeV,
  the fractional changes in $R$ are 30\%, 14\%, 8\% and 4\%, respectively, for
  $m_h < 150$ GeV. The comparison shown in the inset shows that RS wins over
  UED roughly by a factor of 2 for identical KK scale for $m_h < 150$ GeV.
  Incidentally, our UED plots in the insets of all the three figures are in
  complete agreement with \cite{Petriello:2002uu}. See also \cite{Rai:2005vy}
  for a numerical simulation of the Higgs signal at LHC in the UED context.

\section{Conclusions} In conclusion, we highlight the core issues: In the RS
 scenario, the brane-bound Higgs can have order one Yukawa coupling with the
 KK fermions of all flavors. Such large KK Yukawa couplings can sizably
 enhance the Higgs production via gluon fusion and alter the Higgs decay width
 into two photons, provided the KK masses are in a regime accessible to the
 LHC. Because of the proactive involvement of more flavors inside the loop, the
 effect in RS is significantly stronger (typically, by a factor of 2 to 2.5)
 than in UED for similar KK masses. Admittedly, this advantage in RS is
 somewhat offset by the fact that the lightest KK mass in UED can be as low as
 500 GeV thanks to the KK-parity, while in RS a KK mass below 1.5 TeV would be
 difficult to accommodate (see below). However, attempts have been made to
 impose KK parity in warped cases as well \cite{Agashe:2007jb}.

Electroweak precision tests put a severe lower bound on $m_{\rm KK}$ ($\sim
10$ TeV) \cite{Hewett:2002fe}. To suppress excessive contribution to $T$ and
$S$ parameters, the gauge symmetry in the bulk is extended to ${\rm SU(2)_L
\times SU(2)_R \times U(1)_{B-L}}$, and then $m_{\rm KK}$ as low as 3 TeV can
be allowed \cite{Agashe:2003zs,Bouchart:2008vp}. A further discrete symmetry
$L \to R$ helps to suppress $Zb_L\bar b_L$ vertex correction and admits an
even lower $m_{\rm KK} \sim 1.5$ TeV \cite{Carena:2007ua}.  If some other new
physics (e.g. supersymmetrization of RS) can create further room in $T$ and
$S$ by partial cancellation, $m_{\rm KK} \sim$ 1 TeV can also be accommodated.
In our analysis, values of $m_{\rm KK}$ in the range of 1-3 TeV chosen for
illustration may arise in the backdrop of such extended symmetries.
Furthermore, if the $b'$ quark, present in the case of left-right gauge
symmetry, weighs around 1 TeV, one obtains an {\em additional} $\sim$ 10\%
contribution to $\sigma(gg \to h)$ \cite{Djouadi:2007fm}.

A recent paper \cite{Cacciapaglia:2009ky} lists the relative contribution of
different scenarios (supersymmetry, flat and warped extra dimension, little
Higgs, gauge-Higgs unification, fourth generation, etc.)  to $gg\to h$ and $h
\to \gamma\gamma$ for some benchmark values.  A comparison between their work
and ours is in order.  As regards the RS scenario, the authors of
\cite{Cacciapaglia:2009ky} consider the region of parameters where the zero
mode quarks mix with their KK partners. Additionally, their choice of $c_L$ is
substantially different from $c_R$, where they observe large destructive
interference in the effective $ggh$ coupling.  On the other hand, our working
hypothesis is based on: $c \equiv c_L = c_R$ (see Eq.~\ref{yc}), and we
assume KK number conservation at the Higgs vertex.  We observe that the Higgs
coupling to KK quarks is large for any flavor (see Eq.~\ref{kkyukawa}), and
the (direct) loop effects of the KK quarks (which carry the same quantum
numbers as their zero modes) do enhance the effective $ggh$ vertex (like the
{\em enhancement} observed for the fourth family contribution
\cite{Cacciapaglia:2009ky}, or the $b'$ quark contribution
\cite{Djouadi:2007fm}, or the UED contribution
\cite{Petriello:2002uu,Cacciapaglia:2009ky}), and the magnitude is rather
insensitive to the value of $c$ as long as $|c| ~\gtap ~ 0.5$.  The authors of
\cite{ghu} also calculate the KK-induced effective $ggh$ vertex, but they rely
on the gauge-Higgs unification set-up, and hence an efficient numerical
comparison of their work with ours is not possible.

%% file: texfiles/rshg
\section{ Introduction} \label{intro_rsgh}
\normalsize
Minimal supersymmetric standard model (MSSM) with
superparticles in the 1 TeV range, primarily for its ability to settle the
gauge hierarchy problem and for providing a cold dark matter candidate, has
emerged as a leading candidate of physics beyond the standard model (SM). A
key prediction of MSSM is the existence of a light Higgs ($m_h <$ 135 GeV). If
such a light scalar exists, the CERN Large Hadron Collider (LHC) will find it
hard to miss. Moreover, if a quantum picture for all interactions including
gravity has to be woven, we have to rely on theories like the 
string theory, which invariably
includes supersymmetry (SUSY). Since string theory is fundamentally a higher
dimensional theory, a re-analysis of the four-dimensional (4d) MSSM Higgs
spectra by embedding the theory in an extra-dimensional set-up is a worthwhile
phenomenological exercise.

In Randall-Sundrum (RS) type models~\cite{Randall:1999ee} with a warped
space-time geometry, the bulk is a slice of Anti-deSitter (AdS) space in
which the SM particles were discussed in the previous
chapter. Supersymmetrization of such a scenario~\cite{bulksm,Gherghetta:2000qt}, leads to important phenomenological
consequences: (i) gauge hierarchy problem is solved thanks to the warp factor,
(ii) mass hierarchy of fermions can be explained by their relative
localizations in the bulk~\cite{Huber:2000ie}, (iii) the smallness of neutrino
masses can be explained~\cite{Grossman:1999ra}, (iv) gauge coupling
unification is achieved ~\cite{Dienes:1999sz}, (v) SUSY breaking can be
realized with a geometrical interpretation~\cite{Gherghetta:2000kr}, (vi)
light Kaluza-Klein (KK) gauge boson and fermion states can be captured at the
LHC, and some other specific signals, like top flavor-violating decays, can be
detected as well~\cite{rslhc} and (v) the so called "$\mu$" problem that
plague 4d MSSM can be ameliorated by embedding it in warped extra dimensional
scenario.

Since Higgs is the {\em most-wanted} entity at the LHC, our intention is to
 calculate how the upper limit on the lightest supersymmetric neutral Higgs
 mass changes in the warped extra-dimensional backdrop due to radiative
 corrections induced by the KK towers of fermions and sfermions. Before we
 perch on extra-dimensional details, we mention that even within the 4d set-up
 the Higgs mass receives additional contribution, beyond the MSSM limit of 135
 GeV, in the next-to-minimal MSSM~\cite{Drees:1988fc} and in the left-right
 MSSM~\cite{Zhang:2008jm}, to the tune of a few tens of a GeV in each case.

In this chapter we briefly review the salient features of the supersymmetric
warped extra dimensional scenario. The radiative correction to the Higgs
field is then discussed using the effective potential technique. Finally
we calculate the contribution of the extra dimension to the Higgs mass
quantum correction. We finally compare and contrast the results obtained with
the flat extra dimension scenario.

\section{Supersymmetric warped extradimension}\label{wedsusy}
 A 5d $N=1$ SUSY
theory becomes an $N=2$ theory when looked at from an effective 4d perspective~\cite{ArkaniHamed:2001tb}. All the fields should now arrange themselves into
valid representations of a 4d $N=2$ supersymmetric theory.
The structure of the $N=2$ supermultiplets which arises from the KK
excitations of the $N=1$ supermultiplets is well known. Here we will briefly review the multiplet
structure and  mass spectrum for an $N=2$ supersymmetric scenario.

\subsection{Supergravity multiplet} The on-shell supergravity multiplet
consists of the vierbein $e^\alpha_M$, the graviphoton $B_M$ and two
symplectic-Majorana gravitinos $\Psi^i_M$ $(i=1,2)$.  The index $i$ labels the
fundamental representation of the SU(2) automorphism group of the $N=1$
supersymmetry algebra in five dimensions.  In a slice of AdS$_5$, the
supergravity Lagrangian has extra terms proportional to the cosmological
constant: \begin{equation} \label{gravity} S_5=-\frac{1}{2}\int d^4x\int
dy\sqrt{-g}\, \Bigg[M_5^3 \Big\{{\cal
R}+i\bar\Psi^i_M\gamma^{MNR}D_N\Psi_{R}^i -i\frac{3}{2}\sigma^\prime
\bar\Psi^i_M\sigma^{MN}(\sigma_3)^{ij}
\Psi_{N}^j\Big\}+2\Lambda-\frac{\Lambda}{k^2}\sigma^{\prime\prime} \Bigg]\, ,
\end{equation} where $\gamma^{MNR}\equiv \sum_{\rm perm}
(-1)^p\gamma^M\gamma^N\gamma^R/3!$ and $\sigma^{MN}=[\gamma^M,\gamma^N]/2$.
In Eq.~\ref{gravity} we do not show the dependence on $B_M$, since in the
AdS$_5$ background we set $B_M=0$.  In order to respect supersymmetry in
AdS$_5$, the supersymmetric transformation of the gravitino must be changed in
the following way, \begin{equation} \label{sugratrans} \delta
\Psi^i_M=D_M\eta^i+\frac{\sigma^\prime}{2}\gamma_M(\sigma_3)^{ij} \eta^j\, ,
\end{equation} where $\sigma_3={\rm diag}(1,-1)$ and the symplectic-Majorana
spinor $\eta^i$ is the supersymmetric parameter.  Without loss of generality,
we have defined the $\mathbb{Z}_2$ transformation of the symplectic-Majorana
spinor as \begin{equation} \eta^i(-y)=(\sigma_3)^{ij} \gamma_5 \eta^j(y)\, .
\end{equation}

The condition that the AdS$_5$ background does not break supersymmetry is
$\delta\Psi_M^i=0$, and using Eq.~\ref{sugratrans} this leads to the Killing
spinor equation \begin{equation} \label{killing}
D_M\eta^i=-\frac{\sigma^\prime}{2}\gamma_M(\sigma_3)^{ij}\eta^j\, .
\end{equation} In a non-compact five-dimensional AdS space this condition is
always fulfilled. However in the orbifold compactification, the boundary terms
require an extra condition to be satisfied, namely \begin{equation}
\label{extrakill} \gamma_5\eta^i=(\sigma_3)^{ij}\eta^j~.  \end{equation} This
condition implies that only half of the 5d supersymmetric charges are
preserved. Therefore after compactification, one has in 4d an $N=1$
supersymmetric theory instead of $N=2$.

\subsection{Vector supermultiplet} The field content of the vector
supermultiplet is $\mathbb{V}=(V_M,\lambda^i,\Sigma)$ where $V_M$ is the gauge
field, $\lambda^i$ is a symplectic-Majorana spinor, and $\Sigma$ is a real
scalar in the adjoint representation.  For simplicity we will consider a U(1)
gauge group. The action has the form \begin{equation} \label{kinvector}
S_5=-\frac{1}{2} \int d^4x\int dy\sqrt{-g}\, \Bigg[\frac{1}{2g^2_5}F^2_{MN}+
\left(\partial_M\Sigma\right)^2+i\bar{\lambda^i}\gamma^MD_M\lambda^i
+m^2_\Sigma\Sigma^2+im_\lambda\bar{\lambda^i}(\sigma_3)^{ij}\lambda^j \Bigg]\,
.  \end{equation} Supersymmetric invariance on a slice of $AdS_5$ requires,
\begin{equation} \label{v:abc} a=-4\ , \ \ b=2\ , \ \ {\rm and}\ \
c=\frac{1}{2}\, .  \end{equation} Using Eq.~\ref{evenmn}, we find that
$\alpha=1$ for $V_\mu$ and $\lambda^1_L$, while $\alpha=0$ for $\Sigma$ and
$\lambda^2_L$.  If we assume that $V_\mu$ and $\lambda^1_L$ are even, while
$\Sigma$ and $\lambda^2_L$ are odd, then the Kaluza-Klein masses are
determined by the equation \begin{equation}
\frac{J_0(\frac{m_n}{k})}{Y_0(\frac{m_n}{k})}= \frac{J_0(\frac{m_n}{k}e^{\pi
kR})}{Y_0(\frac{m_n}{k}e^{\pi kR})}\, .  \end{equation} Thus, even though the
fields have different $\alpha$ values they still have identical Kaluza-Klein
masses.  The approximate mass of the Kaluza-Klein modes with $n=1,2,\dots$ is
given by~\cite{gauge} \begin{equation} \mbox{\fbox{$m_n\simeq
(n-\frac{1}{4})\pi k e^{-\pi kR}~,$}} \label{susy:vectormass} \end{equation}
compare this above equation with Eq.~\ref{evenmn}.

The even fields $V_\mu$ and $\lambda^1_L$ will have a massless mode while the
odd fields $\Sigma$ and $\lambda^2_L$ do not have massless modes because this
is not consistent with the orbifold condition. Therefore, the massless sector
from $V_\mu$ and $\lambda^1_L$ forms an $N=1$ supersymmetric vector multiplet.

\subsection{Hypermultiplets} The hypermultiplet consists of
$\mathbf{\Phi}=(\phi^i,\Psi)$ where $\phi^i$ are two complex scalars and
$\Psi$ is a Dirac fermion.  The action has the form \begin{equation}
\label{kinhyper} S_5=-\int d^4x\int dy\sqrt{-g}\, \Bigg[ \left|\partial_M
\phi^i \right|^2+i\bar{\Psi}\gamma^MD_M\Psi
+m^2_{\phi^i}|\phi^i|^2+im_\Psi\bar{\Psi}\Psi\Bigg]\, .  \end{equation}

Invariance under supersymmetric transformation relates the masses of the
fermions($\Psi$) with their superpartners($\phi$) by the following relation,
see Eq.~\ref{masses}, \begin{eqnarray} \label{h:susycon}
m^2_{\phi^{1,2}}&=&(c^2\pm c-\frac{15}{4})k^2 +\left(\frac{3}{2}\mp
c\right)\sigma^{\prime\prime}\, ,\nonumber\\ m_\Psi&=&c\sigma^\prime\, ,
\end{eqnarray} where $c$ is an arbitrary dimensionless parameter.

Breaking up the Dirac fermion $\Psi$ into two chiral Weyl fermions
($\psi_L,\psi_R$), and comparing Eq.~\ref{h:susycon} with Eq.~\ref{masses} we
find that all the particles of the hypermultiplet have identical KK mass given
by, \begin{equation} \label{kkhypmass} \frac{J_{| c+1/2|}(\frac{m_n}{k})}{Y_{|
c+1/2|}(\frac{m_n}{k})} =\frac{J_{| c+1/2|}(\frac{m_n}{k}e^{\pi kR})}{Y_{|
c+1/2|} (\frac{m_n}{k}e^{\pi kR})}\, .  \end{equation} approximately given by,
\begin{equation} \label{kkhypmassapp} \mbox{\fbox{$m_n\simeq (n+\frac{
c}{2}-\frac{1}{2}) \pi k e^{-k\pi R}~.$}} \end{equation}
\normalsize

\subsection{Yukawa Couplings}
The Yukawa interaction of the hypermultiplets are of importance in our calculations.
 The couplings are a
direct super-symmetrization of the non-supersymmetric ones derived in Section~\ref{kkyukawasec}.
As was done earlier, we assume that the Higgs boson
is localized at the TeV brane, i.e. $H(x,y) = H(x) \delta(y-\pi R)$ [this
immediately solves the $\mu$ problem, as $\mu \sim \cal{O}$ (TeV)].
Recall that each 5d fermion field has a bulk mass term, characterized by
$c_{iL}$ or $c_{iR}$. For simplicity, we assume that $c_i \equiv c_{iL}=
c_{iR}$. We now expand the 5d fermion fields in zero modes and higher KK
modes and obtain the corresponding 4d Yukawa couplings, see Eqs.~\ref{yc} and \ref{kkyukawa}.
For simplicity, we consider only the diagonal couplings, i.e. ignore quark mixings as their
numerical effects are negligible for our calculation. 
 This is how the
fermion mass hierarchy problem is addressed. We note here that the choice of
$c_i > 1/2$ for the first two families helps evade tight constraints ($m^{(1)}
>$ a few TeV) from FCNC processes~\cite{bulksm}. For the third generation,
FCNC constraints are not so stringent any way.
We now turn our attention to the Yukawa couplings of KK
fermions. We {\em assume} KK number conservation at the tree level Higgs
coupling with the KK fermions\footnote{Although, unlike in UED, KK-parity is
  not automatic in the warped scenario, it is still possible to implement it
  in a slice of AdS$_5$ \cite{Agashe:2007jb}. We assume this parity for
  simplicity of our analytic computation. This also helps in evading some FCNC
  constraints.}. Following arguments similar to the ones leading to  Eq.~\ref{kkyukawa}, 
we find the KK Yukawa coupling is given by,
\large
\begin{equation} 
  \mbox{\fbox{${\lambda}_{i}^{(n)} \sim {\cos}^2 \Big(\left[ n-\frac{3}{4} \mp \frac{1}{4}
  \right] \pi \Big) \, ,$}}
\end{equation}
\normalsize
where $\mp$ correspond to $\mathbb{Z}_2$ even/odd KK modes. We recall two important feature
of these couplings: (i) all KK Yukawa couplings, regardless
of their flavors (i.e. $c_i$ values) and KK numbers, are roughly equal, being
close to unity (more precisely, ${\lambda}_{i(5d)}k$), and (ii) the KK Yukawa
couplings of $\mathbb{Z}_2$ odd modes are vanishing (since the Higgs is brane-bound).

\section{Radiative Correction to the Higgs mass} 
\textbf{This section closely follow the work published in the
paper:~ G.~Bhattacharyya, S.~K.~Majee and T.~S.~Ray,
  ``Radiative correction to the lightest neutral Higgs mass in warped
  supersymmetry,''
  Phys.\ Rev.\  D {\bf 78} (2008) 071701
  [arXiv:0806.3672 [hep-ph]].}

\subsection{Tree level
relations in 4d MSSM} We briefly summerize the scalar sector of the MSSM,
discussed in Section~\ref{MSSMsection}. Within the framework of the MSSM there
are two Higgs doublets which may be represented as, \begin{eqnarray} && {H_u}
=\left(\begin{array}{c} {H_2}^+\\ {H_2}^0\end{array}\right), ~~~~~
{H_d}=\left(\begin{array}{c} {H_1}^0\\ {H_1}^- \end{array}\right)
\label{higgs} \end{eqnarray} whose SU(2)$\times$ U(1) quantum numbers are
(2,$+\frac{1}{2}$) and (2,$-\frac{1}{2}$) respectively. $H_u^0$ couples with
up-type quarks , while $H_d^0$ couples with down-type quarks and charged
leptons. Out of the eight degrees of freedom contained in the two Higgs
doublets, three are absorbed as the longitudinal modes of the massive gauge
bosons, while the remaining five modes appear as physical states. Of these five
states, two are charged $(H^\pm)$ and three are neutral ($h, H,A$). The tree
level potential involving the neutral fields is given by \begin{equation} V_0
= m_1^2|H_1^0|^2 + m_2^2|H_2^0|^2 - m_{12}^2(H_1^0 H_2^0 + {\rm h.c}) + {1
\over 8}(g^2+{g^\prime}^2)(|H_1^0|^2-|H_2^0|^2)^2 . \label{pot2}
\end{equation}

After spontaneous symmetry breaking, the minimum involves the following two
vev's: $\langle H_u^0\rangle = v_u$ and $\langle H_d^0\rangle = v_d$ where, $v
= \sqrt{v_u^2+v_d^2} = (\sqrt{2} G_F)^{-1/2} \simeq 246$ GeV, the
Fermiscale. This gives us the mass matrix for the CP-even and CP-odd neutral
Higgs bosons. One of the eigenvalues of the CP odd mass matrix is zero and
corresponds to the neutral goldstone mode that is absorbed as the longitudinal
mode of the $Z$ boson. The other eigenvalue is given by \begin{equation}
\label{ma} m_A^2 = \frac{2 m_{12}^2}{\sin 2\beta} ~,~~~{\rm where}~~ \tan\beta
= \frac{v_u}{v_d} \\ \end{equation} The $2 \times 2$ mass matrix for the
CP-even neutral Higgs is given by, \begin{eqnarray} \label{msqeven}
\left.{\cal{M}}^2_{({\rm even})}\right|_{tree} = \left(\begin{array} {cc}
M_Z^2 \cos^2 \beta + m_A^2 \sin^2 \beta & -(m_A^2 + M_Z^2) \sin\beta \cos\beta
\\ -(m_A^2 + M_Z^2) \sin\beta \cos\beta & M_Z^2 \sin^2 \beta + m_A^2 \cos^2
\beta \end{array}\right) \end{eqnarray} whose eigenvalues are given by,
\begin{equation} \label{mh} m^2_{h,H} = {1\over 2} \left[m_A^2 + M_Z^2 \mp
\sqrt{(m_A^2 + M_Z^2 )^2 - 4 m_A^2 M_Z^2 \cos^2 2\beta} \right] \end{equation}
resulting in an upper mass bound on the lightest neutral CP-even Higgs given
by the inequality, \begin{equation} m_h \leq {\rm min}~(m_A, M_Z) |\cos
2\beta| \leq {\rm min}~(m_A, M_Z) \end{equation}

\subsection{Radiative corrections from the zero mode} The zero mode of the
model considered exactly represents the 4d MSSM particle spectrum. Therefore
the correction to the lightest neutral Higgs boson mass is identical to the
correction coming from 4d MSSM. Radiative corrections to $m_h$~\cite{radcorr1,radcorr2} are dominated by the zero mode top quark Yukawa
coupling ($\lambda_t$) and the masses of the zero mode stop squarks
($\tilde{t}_1^{(0)}$, $\tilde{t}_2^{(0)}$). For large values of $\tan \beta$,
the contributions from the $b$-quark sector also assume significance. We shall
ignore loop contributions mediated by lighter zero mode quarks or the gauge
bosons. Here, we shall follow the effective potential approach as it
allows the inclusion of the new physics effects in a fairly simple way.  We
start with an RG-improved tree level potential $V_0(Q)$ which contains running
masses and gauge couplings.  The full one-loop effective potential is now
given by \begin{equation} V_1(Q) = V_0(Q) + \Delta V_1(Q) \label{epa1},
\end{equation} where, in terms of the field dependent masses $M(H)$,
\begin{equation} \Delta V_1(Q) = {1 \over {64\pi^2}}{\rm Str} M^4(H) \left
\{\ln{M^2(H)\over {Q^2}} - {3 \over 2} \right \}.  \label{epa2} \end{equation}
The scale dependence of $\Delta V_1(Q)$ cancels against that of $V_0(Q)$
making $V_1(Q)$ scale independent upto higher loop orders.  The supertrace in
Eq.~\ref{epa2}, defined through \begin{equation} {\rm Str} f(m^2) = \sum_i
(-1)^{2J_i} (2J_i+1) f(m_i^2), \end{equation} has to be taken over all members
of a supermultiplet, where $m_i^2 \equiv m_i^2(H)$ is the field-dependent
mass eigenvalue of the particle $i$ with spin $J_i$.  The contribution from
the chiral multiplet containing the up type quark (lepton) and squarks
(sleptons) is given by \begin{equation} \label{epatop} \Delta V_{u} = {c \over
{32\pi^2}} \left\{ m_{\tilde{u_1}}^4 \left(\ln{m_{\tilde
{u_1}}^2\over{Q^2}}-{3\over2}\right) + m_{\tilde {u_2}}^4 \left(\ln{m_{\tilde
{u_2}}^2\over{Q^2}}-{3\over2}\right) -2 m_{u}^4
\left(\ln{m_{u}^2\over{Q^2}}-{3\over2}\right)\right\}, \end{equation} where
$c$ is the color factor. The contribution from the down type quarks (leptons)
and squarks (sleptons) can be written analogously by replacements of up type
masses by the corresponding down type masses.

The field dependent zero mode quark (lepton) masses are given by
\begin{equation} \label{mthmbh} m_{u_i}^2(H) = \lambda_{u_i}^2 |H_u^0|^2 ~;~
m_{d_i}^2(H) = \lambda_{d_i}^2 |H_d^0|^2.  \end{equation} where $i$ is the
flavor index.  The up and down type squark (slepton) masses are given by the
eigenvalues of the corresponding mass matrix written as, \begin{eqnarray} &&
{M_{\tilde {u}}^2} (H) = \left(\begin{array}{cc} m_Q^2 +
\lambda_{u}^2|H^0_u|^2 & \lambda_{u}(A_{u} H^0_u+\mu{H_d^0}^*)\\
\lambda_{u}(A_{u}{H^0_u}^* +\mu H_d^0) & m_U^2 + \lambda_u^2|H^0_u|^2
\end{array}\right), \label{tsquark1} \end{eqnarray} and \begin{eqnarray} &&
{M_{\tilde d}^2} (H) = \left(\begin{array}{cc} m_Q^2 + \lambda_d^2|H^0_d|^2 &
\lambda_d(A_d H^0_d+\mu{H_u^0}^*)\\ \lambda_d(A_d{H^0_d}^* +\mu H_u^0) & m_D^2
+ \lambda_d^2|H^0_d|^2 \end{array}\right).  \label{bsquark1} \end{eqnarray} In
Eqs.~\ref{tsquark1} and \ref{bsquark1}, $m_Q$, $m_U$ and $m_D$ are soft
supersymmetry breaking masses, $A_u$ and $A_d$ are trilinear soft
supersymmetry breaking mass dimensional couplings, and $\mu$ is the
supersymmetry preserving mass dimensional parameter connecting $H_u$ and $H_d$
in the superpotential. We take both trilinear and the $\mu$ couplings to be
real. We have neglected the $D$-term contributions which are small, being
proportional to gauge couplings.

We shall treat the radiatively corrected $m_A$ as an input parameter. Now we
are all set to calculate the radiative corrections in the neutral CP-even mass
eigenvalues from the zero mode MSSM particles. Here only the top and the
bottom sectors are important due to the relative dominance of their Yukawa
couplings.  The one-loop corrected mass matrix square is obtained by taking
double derivatives of the full potential with respect to the scalar
excitations and is given by \begin{eqnarray} \label{msqevencor}
{\cal{M}}^2_{({\rm even})} = \left.{\cal{M}}^2_{({\rm even})}\right|_{tree} +
{3\over{4\pi^2 v^2}} \left(\begin{array}{cc} \Delta_{11} & \Delta_{12} \\
\Delta_{12} & \Delta_{22}\end{array}\right) , && \label{cpeven} \end{eqnarray}
where $\Delta_{ij} = \Delta^{\rm t}_{ij} + \Delta^{\rm b}_{ij}$ and
$\left.{\cal{M}}^2_{({\rm even})}\right|_{tree}$ is given in
Eq.~\ref{msqeven}. The individual $\Delta_{ij}$'s are explicitly written
below:\small \begin{eqnarray} \label{delta} \Delta_{11}^t &=&
{m_t^4\over{{\sin}^2\beta}}\left(\mu (A_t+\mu {\rm cot}\beta)\over{m_{\tilde
t_1}^2 - m_{\tilde t_2}^2}\right)^2g(m_{\tilde t_1}^2,m_{\tilde t_2}^2)
,\nonumber \\ \Delta_{12}^t &=& {m_t^4\over{{\sin}^2\beta}}{\mu (A_t+\mu {
\cot}\beta)\over{m_{\tilde t_1}^2 - m_{\tilde t_2}^2}}\left[{ \ln}{{m_{\tilde
t_1}^2}\over{m_{\tilde t_2}^2}}+{A_t(A_t+\mu { \cot}\beta)\over{m_{\tilde
t_1}^2 - m_{\tilde t_2}^2}} g(m_{\tilde t_1}^2,m_{\tilde t_2}^2)\right]
,\nonumber\\ \Delta_{22}^t& =& {m_t^4\over{\sin^2\beta}}\left[\ln{{m_{\tilde
t_1}^2}{m_{\tilde t_2}^2}\over{m_t^4}} + {2A_t(A_t+\mu {
\cot}\beta)\over{m_{\tilde t_1}^2 -m_{\tilde t_2}^2}} {\ln}{{m_{\tilde
t_1}^2}\over{m_{\tilde t_2}^2}} + \left(A_t(A_t+\mu
{\cot}\beta)\over{m_{\tilde t_1}^2 - m_{\tilde t_2}^2}\right)^2g(m_{\tilde
t_1}^2,m_{\tilde t_2}^2)\right] , \nonumber\\ \Delta_{11}^b& =&
{m_b^4\over{{\cos}^2\beta}}\left[{\ln}{{m_{\tilde b_1}^2}{m_{\tilde
b_2}^2}\over{m_b^4}} + {2A_b(A_b+\mu { \tan}\beta)\over{m_{\tilde b_1}^2
-m_{\tilde b_2}^2}} {\ln}{{m_{\tilde b_1}^2}\over{m_{\tilde b_2}^2}} +
\left(A_b(A_b+\mu \tan\beta)\over{m_{\tilde b_1}^2 - m_{\tilde
b_2}^2}\right)^2g(m_{\tilde b_1}^2,m_{\tilde b_2}^2)\right] , \nonumber \\
\Delta_{12}^b &=& {m_b^4\over{{\cos}^2\beta}}{\mu (A_b+\mu {
\tan}\beta)\over{m_{\tilde b_1}^2 - m_{\tilde b_2}^2}}\left[{ \ln}{{m_{\tilde
b_1}^2}\over{m_{\tilde b_2}^2}}+{A_b(A_b+\mu { \tan}\beta)\over{m_{\tilde
b_1}^2 - m_{\tilde b_2}^2}} g(m_{\tilde b_1}^2,m_{\tilde b_2}^2)\right] , \\
\Delta_{22}^b &=& {m_b^4\over{{\cos}^2\beta}}\left(\mu (A_b+\mu {
\tan}\beta)\over{m_{\tilde b_1}^2 - m_{\tilde b_2}^2}\right)^2 g(m_{\tilde
b_1}^2,m_{\tilde b_2}^2) .\nonumber \end{eqnarray} where, \begin{equation}
\label{g} g(m_1^2,m_2^2)=2 - {{m_1^2+m_2^2}\over{m_1^2-m_2^2}}\ln{m_1^2\over
m_2^2} .  \end{equation}
\normalsize
A point that deserves mention at this stage is that the tree level Higgs mass
is protected by supersymmetry. In the limit of exact supersymmetry, the entire
quantum correction vanishes. So radiative corrections to $m_h$ will be
controlled by the supersymmetry breaking scale ($M_S$).

\subsection {Radiative corrections due to extra dimensions} The KK exited
states differ from the zero mode in certain fundamental aspects.  The KK
states nearly couple universally ($\lambda_{u_i}^{(n)} \sim
\lambda_{d_i}^{(n)} \sim 1$) to the TeV brane bound Higgs, independent of its
zero mode mass. Thus the 1st and 2nd generation quarks also contribute
substantially to the corrections. We also need to incorporate the contribution
from the leptonic sector. We assume that the neutrino masses are generated by
means other than the electro-weak symmetry breaking, therefore they do not have
any coupling to the Higgs and thus do not contribute to the
correction.\footnote{The contribution from the quark sector always dominates over
the leptonic contribution due to the color factor.}

 And we also note that the field dependent masses of the KK modes for the
quarks are given by \begin{eqnarray} \label{mthmbhk}
\left(m^{(n)}_{(u,i)}\right)^2(H) = (\lambda^{(n)}_{i})^2 |H_u^0|^2 +
\left(m_{i}^{(n)}\right)^2 , \\ \nonumber \left(m^{(n)}_{(d,i)}\right)^2(H) =
(\lambda^{(n)}_{i})^2 |H_d^0|^2 + \left(m_{i}^{(n)}\right)^2 .  \end{eqnarray}
where $ \left(m_{i}^{(n)}\right)$ are the KK masses for the flavor $i$ given by
Eq.~\ref{kkhypmassapp}.  The squark masses are given by diagonalizing the mass
matrix given by Eq.~\ref{tsquark} and Eq.~\ref{bsquark} with all the Yukawa
couplings set to unity \footnote{As this is not true for the bottom quark, special
care should be taken to incorporate it. In our full numerical calculations we
have incorporated all such details.}. They can be written as, \small
\begin{eqnarray} && {M_{\tilde {{u_i}}}^2} (H) = \left(\begin{array}{cc} m_Q^2
+ |H^0_{u}|^2 & (A_{{u_i}} H^0_{u}+\mu{H^0_d}^*)\\ (A_{{u_i}}{H^0_u}^* +\mu
H^0_d) & m_U^2 +|H^0_u|^2 \end{array}\right) + \left(\begin{array}{cc}
\left(m_{i}^{(n)}\right)^2 & 0\\ 0 & \left(m_{i}^{(n)}\right)^2
\end{array}\right), \label{tsquark} \end{eqnarray} and \begin{eqnarray} &&
{M_{\tilde d_i}^2} (H) = \left(\begin{array}{cc} m_Q^2 + |H^0_d|^2 & (A_{d_i}
H^0_d+\mu{H^0_u}^*)\\ (A_{d_i}{H^0_d}^* +\mu H^0_u) & m_D^2 + |H^0_d|^2
\end{array}\right)+\left(\begin{array}{cc} \left(m_{i}^{(n)}\right)^2 & 0\\ 0
& \left(m_{i}^{(n)}\right)^2 \end{array}\right).  \label{bsquark}
\end{eqnarray} \normalsize

With this in mind we find that the contribution to the CP-even mass matrix
from a single KK mode of the MSSM may be written as , \begin{eqnarray}
\label{msqevencor1} \left.{\cal{M}}^2_{({\rm even})}\right|_{KK} =
\Sigma_i{c\over{4\pi^2 v^2}} \left(\begin{array}{cc} \Delta_{11}^i &
\Delta_{12}^i \\ \Delta_{12}^i & \Delta_{22}^i\end{array}\right) , &&
\label{cpeven1} \end{eqnarray} where $c$ is the color factor that is 3 for the
quarks and 1 for leptons and $i$ is the flavor index that runs over all the
bulk fermions in a given KK mode.

The contribution from a single up type KK fermion may be written as,\small
\begin{eqnarray} \label{deltaup1} (\Delta_{11}^u)^n &=&
{v_u^4\over{{\sqrt{2}}{\rm sin}^2\beta}}\left(\mu (A_u+\mu {\rm
cot}\beta)\over{m_{\tilde u_1^n}^2 - m_{\tilde u_2^n}^2}\right)^2g(m_{\tilde
u_1^n}^2,m_{\tilde u_2^n}^2) ,\nonumber \\ (\Delta_{12}^u)^n &=&
{v_u^4\over{{\sqrt{2}}\sin^2\beta}}{\mu (A_u+\mu \cot\beta)\over{m_{\tilde
u_1^n}^2 - m_{\tilde u_2^n}^2}}\left[\ln {{m_{\tilde u_1^n}^2}\over{m_{\tilde
u_2^n}^2}}+{A_u(A_u+\mu \cot \beta)\over{m_{\tilde u_1^n}^2 - m_{\tilde
u_2^n}^2}}g(m_{\tilde u_1^n}^2,m_{\tilde u_2^n}^2)\right] , \\
(\Delta_{22}^u)^n& =& {v_u^4\over{{\sqrt{2}}{\rm sin}^2\beta}}\left[{
\ln}{{m_{\tilde u_1^n}^2}{m_{\tilde u_2^n}^2}\over{m_{u^n}^4}} + {2A_u(A_u+\mu
\cot\beta)\over{m_{\tilde u_1^n}^2 -m_{\tilde u_2^n}^2}} \ln {{m_{\tilde
u_1^n}^2}\over{m_{\tilde u_2^n}^2}} \right.  \nonumber \\ & +&
\left. \left(A_u(A_u+\mu \cot\beta)\over{m_{\tilde u_1^n}^2 - m_{\tilde
u_2^n}^2}\right)^2g(m_{\tilde u_1^n}^2,m_{\tilde u_2^n}^2) \right] , \nonumber
\end{eqnarray} and from the down type fermion as, \begin{eqnarray}
\label{deltadn1} (\Delta_{11}^d)^n& =& {v_d^4\over{{\sqrt{2}}{\rm
cos}^2\beta}}\left[ \ln{{m_{\tilde d_1^n}^2}{m_{\tilde
d_2^n}^2}\over{m_{d^n}^4}} + {2A_d(A_d+\mu \tan\beta)\over{m_{\tilde d_1^n}^2
-m_{\tilde d_2^n}^2}} \ln{{m_{\tilde d_1^n}^2}\over{m_{\tilde d_2^n}^2}}
\right.  \nonumber \\ &+& \left. \left(A_d(A_d+\mu \tan\beta)\over{m_{\tilde
d_1^n}^2 - m_{\tilde d_2^n}^2}\right)^2g(m_{\tilde d_1^n}^2,m_{\tilde
d_2^n}^2)\right] , \nonumber \\ (\Delta_{12}^d)^n &=&
{v_d^4\over{{\sqrt{2}}\cos^2\beta}}{\mu (A_d+\mu \tan\beta)\over{m_{\tilde
d_1^n}^2 - m_{\tilde d_2^n}^2}}\left[ \ln{{m_{\tilde d_1^n}^2}\over{m_{\tilde
d_2^n}^2}}+{A_d(A_d+\mu \tan\beta)\over{m_{\tilde d_1^n}^2 - m_{\tilde
d_2^n}^2}}g(m_{\tilde d_1^n}^2,m_{\tilde d_2^n}^2)\right] ,\\
(\Delta_{22}^d)^n &=& {v_d^4\over{{\sqrt{2}}\cos^2\beta}}\left(\mu (A_d+\mu
\tan\beta)\over{m_{\tilde d_1^n}^2 - m_{\tilde d_2^n}^2}\right)^2g(m_{\tilde
d_1^n}^2,m_{\tilde d_2^n}^2) .\nonumber \end{eqnarray} \normalsize where we have made the
assumption that $\lambda_{u_i}^{(n)} \sim \lambda_{d_i}^{(n)} \sim
1$,$~v_{u/d}$ are the Higgs vev and $g(m_1^2,m_2^2)$ is given by
Eq.~\ref{g}. It is to be noted that $A_i = A_0\lambda_i$, therefore the
trilinear couplings of all the flavors are identical for a given KK mode. We
represent all the trilinear couplings for the up (down) type fermions by $A_u$
($A_d$).

A comparison with what happens in flat space supersymmetric Universal Extra
Dimension (UED) \cite{Bhattacharyya:2007te} is now in order. In UED, the KK
states are equispaced (due to space-time flatness), and the KK Yukawa
couplings are proportional to the corresponding zero mode masses. In the
warped scenario, the KK states have a sparse spectrum following the zeros of
the Bessel function, and the KK Yukawa couplings are, to a good approximation,
independent of the flavor indices and are all close to unity for a reasonable
choice of extra-dimensional parameters.  So in the warped case, only $u^{(1)},
c^{(1)}$ and $t^{(1)}$ multiplets contribute to $\Delta m_h^2$ in a
numerically significant way. The contributions from higher KK states are
negligible.  This is in sharp contrast with the SUSY UED scenario where the
first {\em few} $t^{(n)}$ (and {\em not} $u^{(n)}$ or $c^{(n)}$) chiral
multiplets provide sizable contribution to $\Delta m_h^2$. The net numerical
effects in the two cases are comparable. Recall that in UED, unlike in the
warped case, the KK spectra are not linked to fermion mass hierarchy.

\normalsize

\begin{figure}
\begin{center}
\includegraphics[width=0.4\textwidth,angle=270,keepaspectratio]{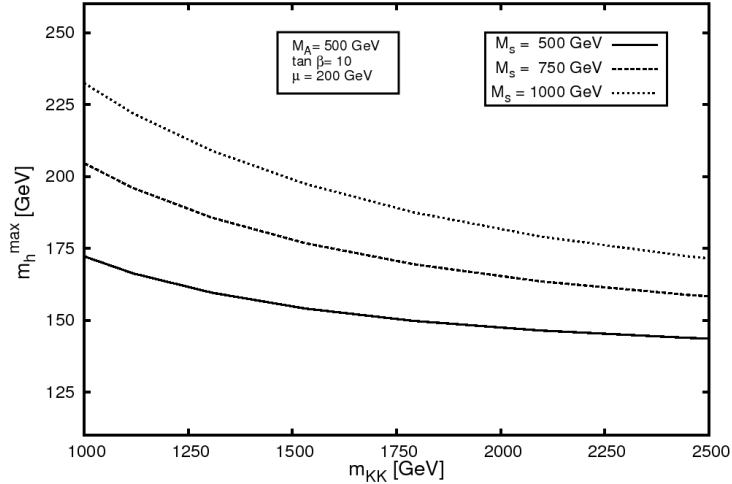}
\caption{\small The variation of $m_h^{\rm max}$ with $m_{kk}$ for
  different choices of $M_S$. We have used $A_u = A_d = \sqrt{6} M_S$ to
  maximise the radiative effect.}
\label{mhmaxr}
\end{center}
\end{figure}

\subsection{ Numerical Results} The scale of the extra dimension is best
represented by the mass of the lightest KK particle. And as discussed in
Section~\ref{wedsusy}, the lightest particles are the members of the $N=2$
vector super fields and are given by Eq.~\ref{susy:vectormass}. We denote this
by $m_{kk}$, and all our results are plotted as a function of this variable.

 In Figure~\ref{mhmaxr}, we have demonstrated that $m_h$ indeed falls with
increasing $m_{kk}$, eventually attaining its 4d value. In this plot, we have
set $A_u=A_d=\sqrt{6}M_S$, which maximizes not only the 4d MSSM radiative
correction but also the KK-induced one, which is why we have used the symbol
$m_h^{\rm max}$. The three lines correspond to three different choices of $M_S
=$ 500, 750 and 1000 GeV. All in all, $m_h$ increases by a few to several tens
of a GeV, depending on the choice of soft SUSY breaking parameters, the
radiative contribution coming primarily from all up-type multiplets.

\section{Conclusions} We have calculated one-loop correction to the lightest
 neutral Higgs boson mass in a generic MSSM embedded in a slice of
 AdS$_5$. For a reasonable choice of warped space parameters, the 4d upper
 limit of 135 GeV could be relaxed by as much as $\sim$ (50-100) GeV depending
 on $M_S$. A few other closely related highlights are the following: (i)
 matter KK spectra are controlled by the $c_i$ parameters, which, in turn, are
 determined by the zero mode fermion masses; (ii) all KK Yukawa couplings are
 close to unity to a very good approximation; (iii) the lightest KK states are
 the members of the $N=2$ vector supermultiplets; (iv) small values of
 $\tan\beta (\ltap~3)$, which are otherwise disfavored in 4d MSSM due to
 nonobservation of Higgs up to 114.5 GeV~\cite{lep}, are now resurrected
 thanks to an additional KK-induced radiative correction. Admittedly, the
 stability of the proton would require further care~\cite{Gherghetta:2000qt}. Besides, the warped models with fermions in the
 bulk, in general, pass the electroweak precision tests (EWPT) with some
 difficulty~\cite{Hewett:2002fe}, unless the KK mass is raised to tens of a
 TeV. To suppress excessive contribution to $T$ (or $\Delta \rho$), gauge
 symmetry in the bulk is enhanced to ${\rm SU(2)_L \times SU(2)_R \times
 U(1)_{B-L}}$~\cite{Agashe:2003zs}, while, to keep the contributions to
 $Zb_L\bar b_L$ vertex and other loop corrections under control, a further
 discrete $L \leftrightarrow R$ symmetry has been employed~\cite{Carena:2007ua}. This allows us to consider $m_{kk}$ as light as 1.5 TeV
 (i.e.  KK gauge boson is of that order). Furthermore, $\tan\beta$ can be
 tuned to reduce the contribution to $T$.  Since our primary intention here
 has been to develop a simple analytic framework (for the first time) to compute KK-induced radiative corrections to $m_h$ in a supersymmetric warped
 space, we pared the scenario down to its bare minimum. The further details
 necessary to overcome the above constraints are unlikely to alter the
 essential qualitative and quantitative features.

%% file: texfiles/rg5d
\section{Introduction}
The CERN Large Hadron
Collider (LHC) is all set to search out the yet elusive Higgs boson. But, LHC is also expected to reveal a new ruler of the
tera-electron-volt (TeV) territories. The standard model (SM) has so far been
remarkably successful in explaining physics up to a few hundred GeV energy
scale. But theoretical inconsistencies of the SM (like, gauge hierarchy
problem) and experimental requirements (like, a candidate to account for the
dark matter of the universe) suggest that there are good reasons to believe
that new physics beyond the SM is just around the corner crying out for
verification. Among the different possibilities, supersymmetry and extra
dimension stand out as the two leading candidates for dictating terms in the
TeV regime. These two apparently distinct classes of scenarios cover a wide
variety of more specific models. The usual practice from a bottom-up approach
is to attach an `either/or' tag on supersymmetry and extra dimension, as if
the presence of one excludes the other. A more careful thought would reveal
that the relationship between these two is {\em not necessarily} mutually
exclusive.  In fact, the presence of higher dimensions is a common feature of
any fundamental theory valid at high scale.  We will get back to this issue a
little later. For the moment, to put things into perspective, we recapitulate
the chronological evolution of the extra dimensional scenarios without
invoking supersymmetry {\em a priori}.  We restrict our discussion to the flat
space scenarios, as we are not pursuing the warped path in this chapter.

Flat extra dimensions were first studied \cite{add} in a scenario where
gravity propagates in a millimeter (mm) size compact space dimension, with the
SM particles confined to a 4d brane. The motive was to bring down the
fundamental Planck scale to about a TeV. Subsequently, it was conceived that
the brane where the SM particles live may actually have a very small size,
like $10^{-16}$ cm $\sim {\rm TeV}^{-1}$, leading to the concept of a `fat
brane' \cite{Antoniadis:1990ew}. In the context of the present chapter, we stick
to the fat brane scenario. {\em What are the experimental bounds on the
  fatness of such a brane, more precisely, on the radius of compactification
  ($R$)}? For universal extra dimension (UED) models \cite{acd}, in which {\em
  all} the SM particles access the extra dimensional bulk, a safe estimate is
$R^{-1} ~\gtap~ 500$ GeV.  More specifically, the $g-2$ of the muon
\cite{nath}, flavor changing neutral currents \cite{chk,buras,desh}, $Z \to
b\bar{b}$ decay \cite{santa}, the $\rho$ parameter \cite{acd,appel-yee}, and
hadron collider studies \cite{collued} reveal that $R^{-1}~\gtap~300$
GeV. Consideration of $b \to s \gamma$, however, implies a somewhat tighter
bound ($R^{-1}~\gtap~600$ GeV \cite{bsg}).  Methods to decipher its signals
from the LHC data have recently been discussed too
\cite{Bhattacharyya:2009br}.  On the other hand, in the non-universal scenario
where both the SM gauge bosons and the Higgs boson propagate in the bulk but
the fermions are confined to a 4d brane \cite{nued}, $R^{-1}$ cannot be below
$(1-2)$ TeV \cite{nued-bounds}. The reason behind the difference in
constraints is the following.  The KK parity, defined by $(-1)^n$ for the
$n$th KK label, is conserved in UED, while it is not a good symmetry in the
non-universal scenario. As a result, while in the non-universal models KK
states can mediate many processes at tree level yielding strong constraints,
in the UED model, thanks to the KK parity, KK states appear only inside a loop
leading to milder constraints.  {\em In any case, in the presence of
  supersymmetry, all those analyses need to be modified with more parameters,
  which would expectedly lead to a set of more relaxed bounds on $R^{-1}$}.

The motivation of studying a TeV scale (or, a fat brane)
  extra dimension scenario has been
investigated from the perspective of string theory, phenomenology,
cosmology/astrophysics and high energy experiments.  Such models provide a
cosmologically stable dark matter candidate \cite{Servant:2002aq}, trigger
successful electroweak symmetry breaking successfully through a composite
Higgs \cite{Arkani-Hamed:2000hv}, address the fermion mass hierarchy problem
from a different point of view \cite{Arkani-Hamed:1999dc}\footnote{Generation of
  non-universality in fermion localization imposes $R^{-1} > 5000$ TeV due to
  large flavor-changing neutral currents and CP violation
  \cite{Delgado:1999sv}.}, and stimulate power law renormalization group (RG)
running yielding a lower (few tens of a TeV) gauge coupling (near-)unification
scale \cite{Dienes:1998vg,dienes2,blitz}\footnote{The power law loop
  corrections are admittedly ultraviolet (UV) cutoff dominated. It has been
  argued that if the higher dimensional theory contains a larger gauge
  symmetry which is perturbatively broken, then the difference of gauge
  couplings of the unbroken subgroups is a calculable quantity independent of
  UV completion \cite{Hebecker:2002vm}.}. Besides, the running of neutrino
mixing angles generated from effective Majorana mass operator in a 5d set-up
has been studied in both non-supersymmetric \cite{Bhattacharyya:2002nc} and
supersymmetric \cite{Deandrea:2006mh} contexts.

We argue that
supersymmetry and extra dimension need not always be seen as {\em two} new
physics considered simultaneously. In fact, they may nicely complement each
other in {\em some situations} through mutual requirements\footnote{
For a tentative list of advantages of supersymmetrizing  extra dimensional scenarios
 see Section~\ref{intro_rsgh}.}. 
From a top-down approach, string theory provides a rationale behind
  linking supersymmetry and extra dimension. The string models are
  intrinsically extra dimensional, and more often than not contain
  supersymmetry as an integral part.  That said, we must also admit that
  establishing a rigorous connection between a {\em realistic} low energy
  supersymmetric model with string theory is still a long shot, though a lot
  of efforts have already been put in that direction \cite{Acharya:2007rc}.
Even after embedding the SM in an extra dimensional set-up, the scalar
  potential remains unstable under quantum correction. Supersymmetrization
  stabilizes it and ameliorates the hierarchy problem.  It is interesting to
  note that by admitting chiral fermions and their scalar partners in the same
  multiplet tacitly provides a rationale behind treating the Higgs boson as an
  elementary object. An elementary Higgs can be perfectly accommodated in a
  flat extra dimensional set-up. As a corollary, the upper limit on the
  lightest supersymmetric neutral Higgs is relaxed beyond the 4d upper limit
  of 135 GeV due to the presence of the KK towers of top/stop chiral
  multiplets, and the {\em hitherto} disfavored low $\tan\beta$ region can be
  resurrected \cite{Bhattacharyya:2007te}.
Finally, each 4d supersymmetric scenario has its own supersymmetry breaking
  mechanism. The origin of this mechanism may be linked to the existence of
  extra dimension. In fact, one of the earliest motivations of a TeV scale fat
  brane was to relate the scale of 4d supersymmetry breaking with $R^{-1}$
  \cite{Antoniadis:1990ew,Antoniadis:1992fh}.

Keeping these in mind, we outline the formalism of a 5d supersymmetric model
in an $S^1/Z_2$ orbifold which contains the 4d supersymmetric states as zero
modes. In section \ref{5dmssm}, we state our assumptions leading to the
construction of the 5d model and comment on supersymmetry
breaking. Furthermore, we explicitly write down the particle content and their
5d Lagrangian and illustrate the KK decompositions of the different 5d
fields. In section \ref{beta}, we derive the beta functions of the gauge and
Yukawa couplings as well as those of the different soft supersymmetry breaking
parameters {\em diagram by diagram}, pointing out how they are all modified
from their 4d values due to the presence of KK states. In section
\ref{numerical}, we discuss the numerical effects of RG running and highlight
the reason behind the differences between the 4d and 5d scenarios. We also
point out under what conditions we can ensure electroweak symmetry
breaking. In section \ref{m0mhalf}, we exhibit the numerical impact of RG
running through plots showing constraints in the $m_0$--$M_{1/2}$ plane. We
standardize our numerical codes by reproducing the known 4d plots before
encoding the necessary alterations for producing the new plots pertaining to
5d supersymmetry.  This also enables us to compare and contrast the 4d and 5d
allowed regions. Finally, in section \ref{concl}, we showcase the essential features
we have learnt from this analysis.

\textbf{This rest of this chapter closely follows the work published in the
paper: ~G.~Bhattacharyya and T.~S.~Ray,
  ``A phenomenological study of 5d supersymmetry,''JHEP 1005 (2010) 040
  [arXiv:1003.1276 [hep-ph]].}

\section{5d supersymmetry}
\label{5dmssm} 
\subsection{A brief summary of our model}
We highlight the salient features of supersymmetry in higher dimension and
outline below the various assumptions that lead to a calculable
phenomenological framework.

 We consider a 5d flat space time metric. The 5th dimension is
  compactified on a $S_1/Z_2$ orbifold. Orbifolding is necessary to reproduce
  chiral zero mode fermions as a 5d theory is vector-like. 
We embed the minimal supersymmetric standard model (MSSM) in this higher
  dimensional set-up (several consequences of such embedding, mainly the
  effects on gauge and Yukawa couplings' evolution, have been studied in
  \cite{Dienes:1998vg}). From a 4d point of view, this leads to a tower of KK
  states. The massless sector corresponds to the 4d MSSM states.
  Since in 5d bulk the fermion representation is vectorial, the
  two-component spinor $Q$ that generates 4d supersymmetry will in 5d be
  accompanied by its chiral conjugate mirror $Q^c$. Thus a $N=1$ supersymmetry
  in 5d corresponds to two different $N=1$ supersymmetry, or equivalently, a
  $N=2$ supersymmetry from a 4d perspective. In fact, all the KK modes of a
  given level must fall into a valid representation of $N=2$ supersymmetry.
  In fact, each 4d supermultiplet is augmented by new chiral conjugate
  states and together they form a hypermultiplet.
Here we are talking about a massive representation of supersymmetry,
  where the supersymmetry preserving Dirac mass plays the r\^ole of central
  charge for $N=2$ supersymmetry. This charge is not renormalized, as a
  consequence of which the bulk hypermultiplets do not receive any
  wave-function renormalization \cite{Dienes:1998vg,
    Barbieri:1982nz}\footnote{In other words, for $N=2$ supersymmetry, it turns
    out that $m_R= m_B$, which is analogous to $g_R = g_B$ for $N=4$
    supersymmetry. Here $m$ is the Dirac mass (central charge) and $g$ is
    gauge coupling, while $R$ and $B$ are labels for renormalized and bare
    quantities. Since the Dirac mass of $N=2$ hypermultiplets appears on the
    right-hand side of the anti-commutation relation of the conserved
    supersymmetry charges, this mass cannot be renormalized. This is
    intertwined with the observation that only those terms are renormalized
    which can be written as integrals over all superspace volume. The kinetic
    term of $N=2$ hypermultiplets cannot be written as any such integral (see
    discussions and related earlier references in \cite{Barbieri:1982nz}.}. We
  observe that this $N=2$ non-renormalization has serious numerical
  consequences in RG evolution of parameters. The most notable effect is the
  blowing up of the Yukawa couplings into the non-perturbative regime around
  18 TeV, which we will take to be the cutoff of our theory. This is below the
  scale of perturbative gauge coupling unification, which is around 30
  TeV. Recall that in 5d we encounter power law running which results in early
  (compared to 4d) unification.

We allow the gauge and the Higgs multiplets access the 5d bulk. Thus far
  what we said is nothing but a supersymmetrization of UED. Only the matter
  multiplets make the difference. In the UED framework, {\em all} SM particles
  access the bulk, and thus even though there are two fixed points, there is
  no brane. One could as well have kept some or all of the fermion generations
  in a brane at a fixed point; the difference would be that the scenario would
  cease to be universal. In the present supersymmetric context too we have the
  freedom of keeping some or all of the matter multiplets at an orbifold fixed
  point. We note that unless we confine at least two generations of matter
  multiplets on a brane, the requirement of {\em perturbative} gauge coupling
  unification leads to a constraint $R^{-1} > 10^{10}$ GeV \cite{blitz},
  spoiling its relevance for LHC. On the other hand, unless we keep the third
  family of matter multiplet in the bulk we cannot ensure electroweak
  breaking. In view of the above, we let the third generation matter multiplet
  access the bulk, but fix the first two generations at $y=0$.
$N=2$ supersymmetry forbids Yukawa interaction in the 5d bulk as this
  interaction involves odd (three) number of {\em chiral}
  multiplets. Therefore, we localize Yukawa interaction at the orbifold fixed
  point where the supersymmetry corresponds to $N=1$.

Now we come to the important question as how we break the residual $N=1$
  supersymmetry. Different ideas have been advanced for its realization. One
  way is to break it by the Scherk-Schwarz mechanism \cite{ss} in which
  fermions and bosons satisfy different periodic conditions over the
  compactified dimension. Explicit realizations towards this using a TeV-scale
  orbifold can be found in \cite{pomarol}. Another interesting approach was to
  break the residual supersymmetry by a second compactification on an orbifold
  with two reflection symmetries, viz.~$S^1/(Z_2 \times Z_2^\prime)$
  \cite{Barbieri:2000vh}. This can be viewed as a discrete version of the
  Scherk-Schwarz mechanism. Both these scenarios yield soft masses which are
  UV insensitive due to the non-local nature of supersymmetry breaking. From a
  completely different viewpoint, supersymmetry breaking may be infused from
  the brane-bulk interface \cite{Mirabelli:1997aj}, or transmitted from a
  distant brane \cite{Kaplan:1999ac}, or arisen from a gaugino mediation
  set-up \cite{Chacko:1999mi} (possibly with a much lower cutoff than
  $10^{16}$ GeV), or triggered by some completely unknown brane dynamics, for
  example, by a spurion $F$-term vacuum expectation value (vev).
  In the context of the present analysis, we keep the exact mechanism of the
  $N=1$ brane supersymmetry breaking {\em unspecified}.  We assume that the
  supersymmetry breaking scale is of the order of the inverse of radius of
  compactification, for example $c/R$, where $c$ is an ${\cal {O}}(1)$
  dimensionless parameter.  

 {\em Our main goal is the following}: Just like in the conventional but
  constrained version of 4d supersymmetry one starts with a common scalar and
  a common gaugino mass at high scale (e.g. the GUT scale) and then run them
  down using the MSSM beta functions to find the weak scale spectrum, we do
  exactly the same here by assuming a common scalar mass ($m_0$) and a common
  gaugino mass ($M_{1/2}$) at low cutoff scale (18 TeV) and follow the running
  using the KK beta functions through successive KK thresholds to obtain the
  weak scale parameters. By adopting a phenomenological approach, we scan
  $m_0$ and $M_{1/2}$ over a set of values $c/R$, with $c$ varying in the
  range $0.1$ to $1$ and $R^{-1}$ fixed at $1$ TeV.

\subsection{Multiplet Structures}
As mentioned in the introduction, from a 4d perspective, the KK towers of
matter and gauge fields rearrange in the form of $N=2$ hypermultiplets. A
judicious choice of $Z_2$ parity of the 5d fields allows us to break the $N=2$
supersymmetry to $N=1$ supersymmetry.  We briefly review below the multiplet
structures of the fields following the prescription suggested in
\cite{ArkaniHamed:2001tb}.

\subsubsection{Vector hypermultiplet}
The 5d super Yang-Mills theory contains a 5-vector gauge field, a 4-component
Dirac gaugino and a real scalar. When dimensionally reduced to 4d, the gauge
field splits into a 4-vector and a scalar, the gaugino splits into 2 Majorana
gauginos, and we still have the real scalar previously mentioned. All these
fit into a vector multiplet and a chiral multiplet in $N=1$ language.  If we
represent the $N=2$ vector supermultiplet by $V$, the 4-vector gauge field by
$A_{\mu}$, the gauginos by $\lambda$ and $\psi$, and define a complex scalar
field $\phi\equiv \frac{1}{2}(\Sigma + iA_5)$, where $\Sigma$ is the 5d real
scalar and $A_5$ is the 5th component of the 5-vector field, then one can
schematically represent the 5d vector supermultiplet as
\begin{equation}
\label{5dV}
  V \equiv \pmatrix{ A_{\mu} & \phi\cr \lambda & \psi \cr} \, . 
\label{Tvec}
\end{equation}
From a 4d perspective (where the compactified 5th coordinate $y$ is just a
label), and in the $N=1$ language, one can visualize the vector hypermultiplet
by a vector multiplet $\mathcal{V}$ (first column) and a chiral multiplet in
the adjoint representation by $\Phi$ (second column):
\begin{eqnarray}
{\mathcal{V}}(x,y) &=& -\theta {\sigma}^{\mu} \overline{\theta} A_{\mu}{(x,y)}
+ i ~{\overline{\theta}}^2 \theta {\lambda}{(x,y)} - i~ {\theta}^2
\overline{\theta}~{\overline{\lambda}}{(x,y)} + \frac{1}{2}
 {\overline{\theta}}^2 {\theta}^2 D_V(x,y) \, , \nonumber \\ 
{\Phi}{(x,y)} &=& {\phi}{(x,y)} + \sqrt{2}~ \theta {\psi}{(x,y)} + 
{\theta}^2 F_{\Phi}(x,y)  \, .
\end{eqnarray}
The $Z_2$ parity of $V$ is so chosen that the $\mathcal{V}$ contains a zero
mode, but $\Phi$ does not have any zero mode.

The gauge invariant action may be written as ($\int d^5x \equiv \int d^4x \int
dy$)
\begin{equation}
 S^5_{\rm gauge} = \int d^5 x \left[ \frac{1}{4 g^2} \int d^2 \theta
   ~W^{\alpha} W_{\alpha} + {\rm h.c.} + \int d^4 \theta ~\frac{1}{g^2}
   {\left( {\partial}_5 \mathcal{V} - \frac{1}{\sqrt{2}} \left( \Phi +
     \overline{\Phi} \right) \right)}^2 \right],
\end{equation} 
where the $~W^{\alpha}(x,y)$ is the field strength superfield
corresponding to ${\mathcal{V}}(x,y)$.

\subsubsection{Higgs hypermultiplets}
From the $N=1$ perspective, the $N=2$ hypermultiplet splits into two chiral
multiplets. Thus we have a $H_u$ hypermultiplet and a $H_d$ hypermultiplet.
We can represent them as (the tilde symbol represents Higgsino) 
\begin{equation}
\label{huhd-hyper}
{H_{(u,d)}} \equiv \pmatrix{ {H_{L(u,d)}} & {H_{R(u,d)}}\cr {\tilde{H}_{L(u/d)}}
  & {\tilde{H}_{R(u/d)}} \cr} \, . 
\end{equation} 
If we denote the two chiral multiplets inside the hypermultiplet $H(x,y)$ as
$h(x,y)$ in left column and $h^c(x,y)$ in right column, then one can expand
the chiral superfield as
\begin{eqnarray} {h/h^c} = {H_{L/R}} + \sqrt{2}~ \theta {\tilde{H}_{L,R}} +
  {\theta}^2 F_{h/h^c} \, .
\end{eqnarray}
We assign even $Z_2$ parity to $h$ so that it has a zero mode, and odd $Z_2$
parity to $h^c$ which does not have zero mode. 
The free action of the hypermultiplets 
can be written as
\begin{equation}
S^5_{\rm Higgs} = \int d^5 x \left[ \int d^4 \theta \left(
  {\overline{h}}^{c} h^{c} + {\overline{h}} h \right)
  + \left( \int d^2 \theta ~h^{c} \left( {\partial}_5 + m \right)h
  + {\rm h.c.}\right) \right] \, . 
\label{higgact}
\end{equation}

\subsubsection{Matter hypermultiplets}
Matters have hypermultiplet structures similar to Higgs:
\begin{equation}
  \Psi \equiv \pmatrix{
     {{\phi}_L} & {{\phi}_R}\cr
    {{\psi}_L} & {{\psi}_R} \cr}~,
\end{equation}
where, ${{\mathcal{F}}_L} \equiv \left( {{\phi}_L} , {{\psi}_L} \right)$
($Z_2$ even) and ${{\mathcal{F}}_R} \equiv \left( {{\phi}_R} , {{\psi}_R}
\right)$ ($Z_2$ odd) represent the two $N=1$ chiral multiplets. 
The free matter hypermultiplet action will be similar to Eq.~\ref{higgact}.
There are five matter representations, two SU(2) doublets $Q$ and $L$ and
three singlets $u, d, e$, where the symbols have their standard meaning.

\subsubsection{Gauge interactions}  
When the hypermultiplets are charged under gauge symmetry, their free action 
can be promoted to take care of the interaction in the following way: 
\small
\begin{eqnarray}
S^5_{\rm int} &=& \int d^5x \left[\int d^4 \theta \left(
  {{\mathcal{F}}_L} e^{\mathcal{V}}
  {{\overline{\mathcal{F}}}_L} + {{\mathcal{F}}_R}
  e^{-{\mathcal{V}}} {{\overline{\mathcal{F}}}_R} \right)
+ \left\{ \int d^2\theta {\mathcal{F}}_L
  \left( m + {\partial}_5 - \frac{1}{\sqrt{2}} {\Phi}
  \right) {{\mathcal{F}}_R} + {\rm h.c.} \right\} \right]
\end{eqnarray}
\normalsize
where, ${\mathcal{V}} = {\mathcal{V}}^{a} T^a$ and ${\Phi} =
{\Phi}^{a} T^a$ are Lie-algebra-valued gauge and matter superfields. 

\subsubsection{Yukawa Interactions}
Since Yukawa interaction involves three (i.e. odd number) chiral superfields,
it is not possible to write a bulk Yukawa interaction maintaining $N=2$
supersymmetry. For this reason, we confine Yukawa interaction at the branes.
We denote the Yukawa part of the superpotential by $W_Y$, which contains the
usual chiral superfield combinations $QH_uu$, $QH_dd$ and $LH_de$. Then the
action can be written as
\begin{equation}
S^5_{\rm Yuk}= \int d^5x\left( \int d^2 \theta ~
W_Y \right)\left[ \delta(y)+\delta(y-{\pi}R)\right] \, . 
\end{equation} 
As the $Z_2$ odd fields vanish at the fixed points, they do not contribute to
Yukawa interactions.

\subsection{KK decomposition of fields}
In order to obtain the action in terms of 4d component fields, we need to
write down the KK decomposition of the 5d fields in terms of zero modes and
higher KK modes \cite{acd}. Each 5d field is either $Z_2$ even or $Z_2$
odd. Only the even fields have zero modes. The decomposition of the Higgs
fields will be exactly like the matter fields. 
\begin{eqnarray}
\label{fourier}
\mathcal{V}(x,y)&=&\frac{\sqrt{2}}{\sqrt{2\pi
R}}\mathcal{V}^{(0)}(x)+\frac{2}{\sqrt{2\pi
R}}\sum^{\infty}_{n=1}\mathcal{V}^{(n)}(x)\cos\frac{ny}{R} \, , \nonumber\\
\Phi(x,y) &=& \frac{2}{\sqrt{2\pi
R}}\sum^{\infty}_{n=1}\Phi^{(n)}(x)\sin\frac{ny}{R} \, , \\
{\mathcal{F}}_L(x,y)&=&\frac{\sqrt{2}}{\sqrt{2\pi
R}}{\mathcal{F}}_L^{(0)}(x)+\frac{2}{\sqrt{2\pi
R}}\sum^{\infty}_{n=1}{\mathcal{F}}_L^{(n)}(x)\cos\frac{ny}{R} \,, \nonumber \\
{\mathcal{F}}_R(x,y)&=&\frac{2}{\sqrt{2\pi
R}}\sum^{\infty}_{n=1}{\mathcal{F}}_R^{(n)}(x)\sin\frac{ny}{R} \, . \nonumber
\end{eqnarray}

\section{RG evolution and derivation of the beta functions}
\label{beta}
The technical meaning of RG evolution in a higher dimensional context has been
amply clarified in \cite{Dienes:1998vg}, and we merely reiterate it in the
present context. The multiplicity of KK states renders any such higher
dimensional scenario non-renormalizable. So `running' of couplings or
parameters with the energy scale does not make much of a sense. Rather, one
can estimate the finite quantum corrections that these couplings/parameters
receive whose size depends on some explicit cutoff $\Lambda$. The contribution
comes from $\Lambda R$ number of KK states which lie between the scale of the
first KK state, which is $1/R$, and the cutoff $\Lambda$. With this
interpretation of RG running, we compute the one loop beta functions of the
gauge and Yukawa couplings and various soft supersymmetry breaking masses.  We
make the following observations:
\begin{enumerate}
\item The contribution to the beta function from a given KK mode does not
  depend on its KK label.
\item When we consider different KK thresholds we neglect their zero mode
  masses, i.e. we assume that the $n$th level KK state is kicked into life
  when we cross the energy scale $n/R$.
\item As we cross different KK thresholds, the beta functions also change.
  The beta function of the quantity $X$ at an energy scale $Q$, where 
  $n~\ltap~QR < (n+1)$, can be written as ($t = \ln (Q/Q_0)$, where $Q_0$ is a
  reference scale, e.g. the electroweak scale)
\begin{equation}
\label{master}
\frac{\partial{X}}{\partial{t}} = \beta_X \, ,~~{\rm where}~~
\beta_X = \beta_{0X} + n \tilde{\beta}_X \, . 
\end{equation}
Here $\beta_{0X}$ is the contribution induced by the zero mode (i.e. ordinary
4d MSSM) states (which may be found, for example, in the review
\cite{Martin}) and $\tilde{\beta}_X$ arises from a single KK
mode. Eq.~\ref{master} is our master equation, using which we perform a
diagram by diagram calculation for the estimation of $\tilde{\beta}_X$ for
various couplings and parameters.
\end{enumerate}

\subsection{Gauge couplings and gaugino masses}
The running of the gauge couplings ($g_i$) and gaugino masses ($M_i$) are
controlled by
\begin{eqnarray}
\label{gaugerg}
\beta_{g_i} = \frac{g_i^3}{16\pi^2} \left[b_i^0 + n \tilde{b}_i\right] \, ,~~  
\beta_{M_i} = \frac{g_i^2 M_i}{16\pi^2} \left[b_i^0 + n \tilde{b}_i \right] \, .
\end{eqnarray}
For the gauge groups U(1) (which corresponds to $g_1 = \sqrt{5/3} g'$, which
unifies), SU(2) and SU(3), $b_i^0 = (33/5,1,-3)$, and $\tilde{b}_i =
(26/5,2,-2)$ respectively.

\subsection{Yukawa and scalar trilinear couplings}
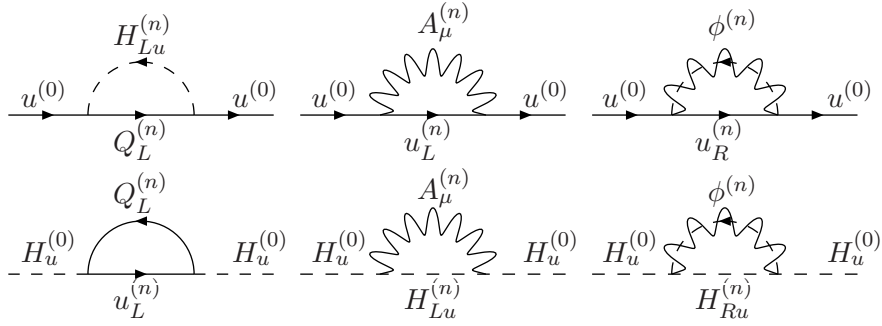
\begin{figure}
\begin{center}
\fcolorbox{white}{white}{
  \begin{picture}(350,92) (105,-89)
    \SetWidth{0.5}
    \Text(110,-25)[lb]{\small{\textbf{$u^{(0)}$ }}}
    \Text(190,-25)[lb]{\small{\textbf{$u^{(0)}$ }}}
    \Text(145,-45)[lb]{\small{\textbf{$Q_{L}^{(n)}$ }}}
    \Text(145,-5)[lb]{\small{\textbf{$H_{Lu}^{(n)}$ }}}
    \Text(220,-25)[lb]{\small{\textbf{$u^{(0)}$ }}}
    \Text(300,-25)[lb]{\small{\textbf{$u^{(0)}$ }}}
    \Text(255,-45)[lb]{\small{\textbf{$u_{L}^{(n)}$ }}}
    \Text(260,0)[lb]{\small{\textbf{$A_{\mu}^{(n)}$ }}}
    \Text(330,-25)[lb]{\small{\textbf{$u^{(0)}$ }}}
    \Text(415,-25)[lb]{\small{\textbf{$u^{(0)}$ }}}
    \Text(365,-45)[lb]{\small{\textbf{$u_{R}^{(n)}$ }}}
    \Text(370,0)[lb]{\small{\textbf{$\phi^{(n)}$ }}}

    \Text(110,-85)[lb]{\small{\textbf{$H_u^{(0)}$ }}}
    \Text(190,-85)[lb]{\small{\textbf{$H_u^{(0)}$ }}}
    \Text(145,-105)[lb]{\small{\textbf{$u_{L}^{(n)}$ }}}
    \Text(145,-65)[lb]{\small{\textbf{$Q_{L}^{(n)}$ }}}
    \Text(220,-85)[lb]{\small{\textbf{$H_u^{(0)}$ }}}
    \Text(300,-85)[lb]{\small{\textbf{$H_u^{(0)}$ }}}
    \Text(255,-105)[lb]{\small{\textbf{$H_{Lu}^{(n)}$ }}}
    \Text(260,-63)[lb]{\small{\textbf{$A_{\mu}^{(n)}$ }}}
    \Text(330,-85)[lb]{\small{\textbf{$H_u^{(0)}$ }}}
    \Text(415,-85)[lb]{\small{\textbf{$H_u^{(0)}$ }}}
    \Text(365,-105)[lb]{\small{\textbf{$H_{Ru}^{(n)}$ }}}
    \Text(370,-63)[lb]{\small{\textbf{$\phi^{(n)}$ }}}
    \ArrowLine(105,-28)(135,-28) \ArrowLine(135,-28)(175,-28) 
    \ArrowLine(175,-28)(205,-28) \DashArrowArc(155,-28)(20,0,180){5}
    \DashLine(105,-88)(135,-88){5} \ArrowLine(135,-88)(175,-88)
    \DashLine(175,-88)(205,-88){5} \ArrowArc(155,-88)(20,0,180)
    \DashLine(215,-88)(315,-88){5} \ArrowLine(215,-28)(245,-28)
    \ArrowLine(245,-28)(285,-28) \ArrowLine(285,-28)(315,-28)
    \PhotonArc(265,-28)(20,0,180){-5}{7.5} \ArrowLine(325,-28)(355,-28)
    \ArrowLine(355,-28)(395,-28) \ArrowLine(395,-28)(425,-28)
    \PhotonArc(375,-28)(20,0,180){-5}{5.5} \DashArrowArc(375,-28)(20,0,180){5}
    \PhotonArc(265,-88)(20,0,180){-5}{7.5} \DashLine(325,-88)(425,-88){5}
    \PhotonArc(375,-88)(20,0,180){-5}{5.5} \DashArrowArc(375,-88)(20,0,180){5}
  \end{picture}
}
\end{center}
\caption{\small{ Feynman diagrams showing the KK contributions to the
    wave-function renormalizations of the zero mode $u_3$ and $H_u$.  Similar
    diagrams for the other fermions and scalars may be drawn analogously. Here
    $A_{\mu}$ is a generic gauge field and $\phi$ is an adjoint scalar, both
    arising from a vector hypermultiplet.}  }\label{selfenergies}
\end{figure}

We recall that $N=1$ non-renormalization relates the beta functions of the
Yukawa couplings ($y_{ijk}$) to the anomalous dimension matrices
($\gamma^i_j$) of the superfields.  This theorem implies that logarithmically
divergent contributions can always be written in terms of wave-function
renormalizations.  Generically, $y_{ijk}$ may be written as
\begin{equation}
\beta_{y^{ijk}} = \gamma^i_n y^{njk} + \gamma^j_n y^{ink} + \gamma^k_n y^{ijn}
\, . 
\label{nonrenorm}
\end{equation}
The Feynman diagrams showing the KK contributions to the wave-function
renormalizations of the scalars and fermions are displayed in
Figure~\ref{selfenergies}. The contribution from the gauge sector cancels
exactly as a consequence of the $N=2$ non-renormalization theorem mentioned in
section \ref{5dmssm}.  Diagrammatically, the origin of this cancellation may
be traced to a relative sign between the $A_\mu$- and $\phi$-propagators - see
Eq.~\ref{5dV}.  Only the brane localized Yukawa interactions contribute to
the Yukawa evolution. We also keep track of the fact that the $\mathbb{Z}_2$ odd fields
have vanishing wave-functions at the two branes, leaving the even fields alone
to contribute to the diagrams in Figure~\ref{selfenergies}. Here we have made
a tacit assumption that although the Yukawa interaction is brane localized,
only one KK level ($n$) states float inside the loop at a time. This is a
technical assumption to ensure calculability by avoiding KK divergence which
would have arisen while summing more than one KK index in a loop calculation.

\normalsize

To appreciate the numerical impact of the bulk $N=2$ non-renormalization, we
first write down the conventional 4d MSSM beta functions (i.e. those coming
from zero mode states in the 5d context) which contribute to the evolution of
the third generation Yukawa couplings \cite{Martin}:
\begin{eqnarray}
 \beta_{t}^0 &=& {y_t \over 16 \pi^2} \Bigl [ 6 y_t^* y_t + y_b^* y_b -
   {16\over 3} g_3^2 - 3 g_2^2 - {13\over 15} g_1^2 \Bigr ],\nonumber
 \\ \beta_{b}^0 &=& {y_b \over 16 \pi^2} \Bigl [ 6 y_b^* y_b + y_t^* y_t +
   y_\tau^* y_\tau - {16\over 3} g_3^2 - 3 g_2^2 - {7\over 15} g_1^2 \Bigr] 
\, , \\ \beta_{\tau}^0 &=& {y_\tau \over 16 \pi^2} \Bigl [ 4 y_\tau^*
   y_\tau + 3 y_b^* y_b - 3 g_2^2 - {9\over 5} g_1^2 \Bigr ] \, . \nonumber 
\end{eqnarray}
The corresponding KK contributions are given by 
\begin{equation} 
\label{betaf}
\tilde{\beta}_f = \beta_f^0 (g_i \to 0) ~~ (f \equiv t,b,\tau) \, , 
\end{equation} 
where the vanishing gauge contributions are a direct consequence of the bulk
$N=2$ non-renormalization.

The effects of the above non-renormalization can also be felt in the evolution
of the trilinear scalar couplings.  The relevant Feynman diagrams are
displayed in Figure~\ref{trilinear}.  Again, for illustration, we first write
down the contributions to the beta functions from the zero mode (i.e. 4d MSSM)
states \cite{Martin}:
\begin{eqnarray}
\label{atrge}
{\beta}_{a_t}^0 \!&=&\! \frac{1}{16\pi^2} \Bigl [a_t \left( 18 y_t^* y_t +
  y_b^* y_b - {16\over 3} g_3^2 - 3 g_2^2 - {13\over 15} g_1^2 \right) + 2 a_b
  y_b^* y_t\Bigr .  \nonumber\\ && \!+ ~y_t \Bigl . \left( {32\over 3} g_3^2
  M_3 + 6 g_2^2 M_2 + {26\over 15} g_1^2 M_1 \right) \Bigr ] \, , \nonumber
\\ {\beta}_{a_b}^0 \!&=&\!\frac{1}{16\pi^2}\Bigl [ a_b \left( 18 y_b^* y_b +
  y_t^* y_t + y_\tau^* y_\tau - {16\over 3} g_3^2 - 3 g_2^2 - {7\over 15}
  g_1^2 \right) + 2 a_t y_t^* y_b + 2 a_\tau y_\tau^* y_b \Bigr .
  \phantom{xxxx} \nonumber \\&& \!+ y_b \Bigl. \left( {32\over 3} g_3^2 M_3 +
  6 g_2^2 M_2 + {14 \over 15} g_1^2 M_1 \right) \Bigr ] \, ,\qquad{} 
\\ \beta_{a_{\tau}}^0 \!&=&\! \frac{1}{16\pi^2}\Bigl [a_\tau \left( 12
  y_\tau^* y_\tau + 3 y_b^* y_b - 3 g_2^2 - {9\over 5} g_1^2 \right) + 6 a_b
  y_b^* y_\tau + y_\tau \left( 6 g_2^2 M_2 + {18\over 5} g_1^2 M_1 \right)
  \Bigr ] \, . \nonumber
\end{eqnarray}
As expected, the beta functions of the {\em soft supersymmetry breaking
  parameters} are proportional {\em not only} to those parameters but to
others as well, since any non-renormalization theorem ceases to work when
supersymmetry is broken. For the computation of $\tilde{\beta}_{a_f}$, we need
to keep in mind the essence of Eq.~\ref{betaf}, i.e. the {\em absence of
  gauge contributions} in $\tilde{\beta}_{f}$, while solving the coupled
differential equations. However, that part of the gauge contributions
(proportional to $g_i^2$) to the trilinear scalar couplings which multiply the
gaugino masses ($M_i$) in Eq.~\ref{atrge} would still remain while computing
the KK contribution.  All in all,
\begin{equation} 
\label{betaaf}
\tilde{\beta}_{a_f} = \beta_{a_f}^0 (a_f g_i \to 0) \, . 
\end{equation} 
\begin{center}
 \begin{figure}[t]
\begin{center}
\begin{center}
\fcolorbox{white}{white}{
  \begin{picture}(318,255) (60,-75)
    \SetWidth{0.5}
\Text(65,125)[lb]{\small{\textbf{$H_u^{(0)}$ }}}
\Text(118,134)[lb]{\small{\textbf{$\tilde{u}^{(0)}$ }}}
\Text(170,47)[lb]{\small{\textbf{$\tilde{Q}^{(0)}$ }}}
\Text(150,134)[lb]{\small{\textbf{${Q}_L^{(n)}$ }}}
\Text(128,165)[lb]{\small{\textbf{$\tilde{H}_{Lu}^{(n)}$ }}}
\Text(170,180)[lb]{\small{\textbf{$\tilde{u}^{(0)}$ }}}

\Text(258,125)[lb]{\small{\textbf{$H_u^{(0)}$ }}}
\Text(311,134)[lb]{\small{\textbf{$\tilde{u}^{(0)}$ }}}
\Text(343,65)[lb]{\small{\textbf{$\tilde{Q}^{(0)}$ }}}
\Text(343,134)[lb]{\small{\textbf{$\tilde{u}_L^{(n)}/\tilde{u}_R^{(n)}$ }}}
\Text(290,165)[lb]{\small{\textbf{${A}_{\mu}^{(n)}/{\phi}_{\mu}^{(n)}$ }}}
\Text(363,180)[lb]{\small{\textbf{$\tilde{u}^{(0)}$ }}}

\Text(65,-10)[lb]{\small{\textbf{$H_u^{(0)}$ }}}
\Text(145,20)[lb]{\small{\textbf{$\tilde{Q}_L^{(n)}$ }}}
\Text(145,-65)[lb]{\small{\textbf{$\tilde{u}_L^{(n)}$ }}}
\Text(195,30)[lb]{\small{\textbf{$\tilde{Q}^{(0)}$ }}}
\Text(195,-70)[lb]{\small{\textbf{$\tilde{u}^{(0)}$ }}}

\Text(258,-10)[lb]{\small{\textbf{$H_u^{(0)}$ }}}
\Text(310,-0)[lb]{\small{\textbf{${Q}_L^{(n)}$ }}}
\Text(310,-40)[lb]{\small{\textbf{${u}_L^{(n)}$ }}}
\Text(355,-10)[lb]{\small{\textbf{$\lambda^{(n)}$ }}}
\Text(378,30)[lb]{\small{\textbf{$\tilde{Q}^{(0)}$ }}}
\Text(378,-70)[lb]{\small{\textbf{$\tilde{u}^{(0)}$ }}}
\DashLine(60,120)(120,120){5} \DashLine(120,120)(142,141){5}
\ArrowLine(142,141)(162,160) \DashLine(162,160)(182,180){5}
\ArrowArc(152,150.5)(15,42.52,223.29)
\DashLine(120,120)(181,60){5} \DashLine(60,-15)(120,-15){5}
\DashArrowArc(150,-15)(30,0,360){5} \DashLine(180,-15)(225,45){5}
\DashLine(180,-15)(225,-75){5} \DashLine(253,120)(315,120){5}
\DashLine(315,120)(375,180){5} \DashLine(315,120)(378,59){5}
\PhotonArc(345.31,151.61)(15,42.52,228.29){-5}{8.5}
\DashLine(253,-15)(315,-15){5} \ArrowLine(315,-15)(346,15)
\ArrowLine(346,-45)(315,-15) \ArrowLine(346,15)(347,-45)
\DashLine(346,15)(375,45){5} \DashLine(346,-45)(375,-75){5}
\Gluon(346,15)(346,-45){5}{6}
  \end{picture}
}
\end{center}
\end{center}
\caption{ \small{  Feynman diagrams showing the KK loop contribution to the
  evolution of a trilinear scalar coupling.  The diagrams contributing to
  other trilinear couplings may be drawn analogously.}}
 \label{trilinear}
 \end{figure}
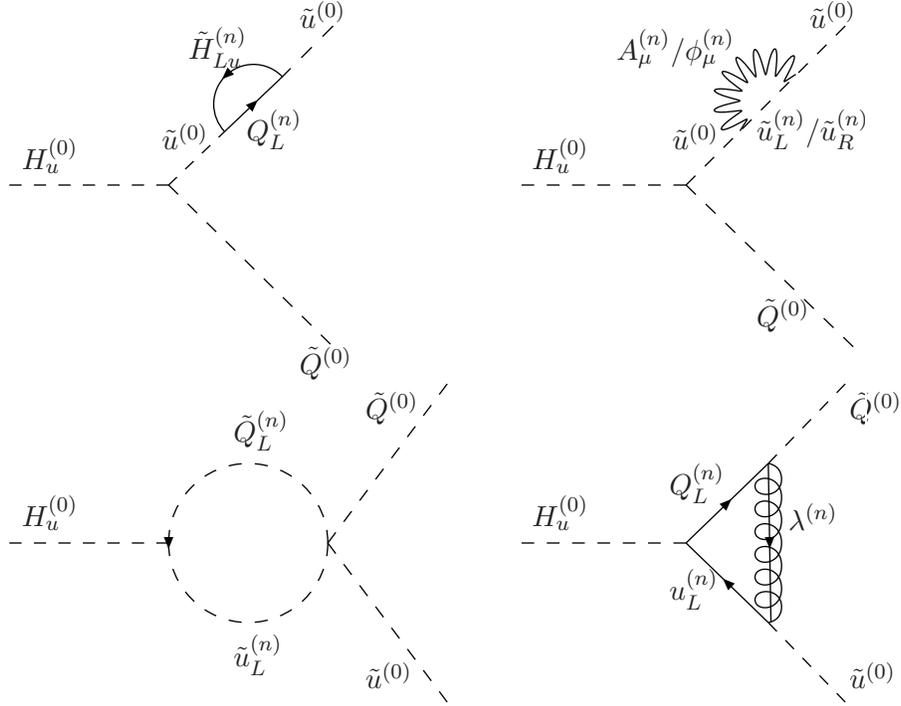
 \end{center}
\normalsize

\subsection{Scalar masses}
We make three observations regarding the KK contributions to the evolution of
scalar masses (see Figure~\ref{3higg}):  
\begin{enumerate}

\item The two diagrams in the lower row of Figure~\ref{3higg} depend on the
  Yukawa couplings. Hence, they are important only for the third generation
  matter fields. 

\item Recall that in the evolution of the Yukawa couplings the KK
  contributions from the gauge field $A_\mu$ and the complex scalar $\phi$
  exactly cancelled thanks to the bulk $N=2$ non-renormalization. However,
  their fermionic superpartners contribute to the scalar mass evolution and
  those contributions add up instead of canceling out. This happens because
  these contributions yield gaugino masses which are $N=1$ supersymmetry
  breaking parameters and hence the non-renormalization theorem ceases to be
  applicable.

\item Each KK state in the two diagrams in the top row of Figure~\ref{3higg}
  contributes twice that of the SM because of the doubling of the fermions
  (this factor of 2 is highlighted in bold-face in Eqs.~\ref{scalarbeta} and
  \ref{higgsbeta} below). However, each KK state at the lower row diagrams
  contributes the same as in the SM because the odd fermion modes vanish at
  the brane where Yukawa interaction is confined.

\end{enumerate}

Below we write down the beta functions of the third generation scalars:
\begin{eqnarray}
\label{scalarbeta}
\tilde{\beta}_{u_3} &=& \frac{1}{16\pi^2} \left[ 2 \left( 2y_t^2 \left(
  m_{Hu}^2 + m_{\tilde{Q}_3}^2 + m_{\tilde{u}_3}^2 \right) + 2a_t^2 \right)
  \right.- \left. {\bf 2} \left( \frac{32}{3}g_3^2|M_3|^2 +
  \frac{32}{15}g_1^2|M_1|^2 + \frac{4}{5}g_1^2 S \right) \right], \nonumber
\\ \tilde{\beta}_{d_3} &=& \frac{1}{16\pi^2} \left[ 2\left( 2y_b^2 \left(
  m_{Hd}^2 + m_{\tilde{Q}_3}^2 + m_{\tilde{d}_3}^2 \right) + 2a_b^2 \right)
  \right. - \left. {\bf 2} \left( \frac{32}{3}g_3^2|M_3|^2 +
  \frac{8}{15}g_1^2|M_1|^2 - \frac{2}{5}g_1^2 S \right) \right], \nonumber
\\ \tilde{\beta}_{Q_3} &=& \frac{1}{16\pi^2} \left[ \left(2 y_t^2 \left(
  m_{Hu}^2 + m_{\tilde{Q}_3}^2 + m_{\tilde{u}_3}^2 \right) + 2a_t^2 \right)
  \right. + \left. \left( 2y_b^2 \left( m_{Hd}^2 + m_{\tilde{Q}_3}^2 +
  m_{\tilde{d}_3}^2 \right) + 2a_b^2 \right)\right. \nonumber\\ & &
  \left. - {\bf 2} \left( \frac{32}{3}g_3^2|M_3|^2 + 6g_2^2|M_2|^2 +
  \frac{2}{15}g_1^2|M_1|^2 - \frac{1}{5}g_1^2 S \right) \right],
\\ \tilde{\beta}_{L_3} &=& \frac{1}{16\pi^2} \left[ \left( 2y_{\tau}^2 \left(
  m_{Hd}^2 + m_{\tilde{L}_3}^2 + m_{\tilde{e}_3}^2 \right) + 2a_{\tau}^2
  \right) - {\bf 2} \left( \frac{6}{5}g_1^2|M_1|^2 + \frac{3}{5}g_1^2 S \right)
  \right],\nonumber \\ \tilde{\beta}_{e_3}
&=& \frac{1}{16\pi^2} \left[ 2\left( 2y_{\tau}^2 \left( m_{Hd}^2 +
  m_{\tilde{L}_3}^2 + m_{\tilde{e}_3}^2 \right) + 2a_{\tau}^2 \right) - {\bf 2}
  \left( \frac{24}{5}g_1^2|M_1|^2 - \frac{6}{5}g_1^2 S \right) \right] \, .
\nonumber
\end{eqnarray}

The beta functions for the Higgs scalars are given by 
\begin{eqnarray}
\label{higgsbeta}
{\beta}_{H_u} &=& \frac{1}{16\pi^2} \left[ 3 \left( 2y_t^2 \left( m_{Hu}^2 +
  m_{\tilde{Q}_3}^2 + m_{\tilde{u}_3}^2 \right) + 2a_t^2 \right) \right.-
  \left. {\bf 2} \left( 6g_2^2|M_2|^2 + \frac{6}{5}g_1^2|M_1|^2 -
  \frac{3}{5}g_1^2 S \right) \right] \, , \nonumber \\ 
{\beta}_{H_d} &=& \frac{1}{16\pi^2}
\left[ 3 \left( 2y_b^2 \left( m_{Hd}^2 + m_{\tilde{Q}_3}^2 + m_{\tilde{d}_3}^2
  \right) + 2a_b^2 \right) \right.+ \left( 2y_{\tau}^2 \left( m_{Hd}^2 +
  m_{\tilde{L}_3}^2 + m_{\tilde{e}_3}^2 \right) + 2a_{\tau}^2 \right)
  \\ & &- \left. {\bf 2} \left( 6g_2^2|M_2|^2 +
  \frac{6}{5}g_1^2|M_1|^2 + \frac{3}{5}g_1^2 S \right) \right] \nonumber \, .
\label{higgsrg}
\end{eqnarray}

\begin{center}
\begin{figure}
\fcolorbox{white}{white}{
  \begin{picture}(469,257) (0,0)
    \SetWidth{0.5}

\Text(85,112)[lb]{\small{\textbf{$H_u^{(0)}$ }}}
\Text(238,112)[lb]{\small{\textbf{$H_u^{(0)}$ }}}
\Text(135,85)[lb]{\small{\textbf{$\tilde{H}_{Lu}^{(n)},
\tilde{H}_{Ru}^{(n)}$}}}
\Text(150,150)[lb]{\small{\textbf{$\lambda^{(n)}, \psi^{(n)}$ }}}
\DashLine(74,107)(136,107){5} \ArrowLine(136,107)(198,107)
\Text(210,130)[lb]{\small{\textbf{$\times~2$ }}}
\DashLine(198,107)(258,107){5} \ArrowArc(167,107)(30,0,180)
\GlueArc(167,107)(30,0,180){5}{6.87}
\Text(300,112)[lb]{\small{\textbf{$H_u^{(0)}$ }}}
\Text(445,112)[lb]{\small{\textbf{$H_u^{(0)}$ }}}
\Text(420,150)[lb]{\small{\textbf{$\times~2$ }}}
\Text(310,170)[lb]{\small{\textbf{$\phi_L^{(n)},\phi_R^{(n)}$ }}}
\DashLine(285,107)(468,107){5} \Photon(376,107)(376,153){7.5}{2}
\DashArrowArc(376,173)(20,0,360){5} \Text(85,7)[lb]{\small{\textbf{$H_u^{(0)}$ }}}
\Text(238,7)[lb]{\small{\textbf{$H_u^{(0)}$ }}}
\Text(110,50)[lb]{\small{\textbf{$\tilde{Q}_L^{(n)},\tilde{u}_L^{(n)}$ }}}
\DashLine(75,2)(255,2){5} \DashArrowArc(155,30)(30,-68,68){5}
\DashArrowArc(175,30)(30,112,248){5}
\Text(300,7)[lb]{\small{\textbf{$H_u^{(0)}$ }}}
\Text(445,7)[lb]{\small{\textbf{$H_u^{(0)}$ }}}
\Text(373,-13)[lb]{\small{\textbf{${Q}_L^{(n)}$ }}}
\Text(373,40)[lb]{\small{\textbf{${u}_L^{(n)}$ }}}
\DashLine(285,2)(347,2){5} \ArrowLine(347,2)(407,2)
\DashLine(407,2)(469,2){5} \ArrowArc(377,2)(30,0,180)
  \end{picture}
}
\caption{\small { Feynman diagrams showing the KK contributions to the
  running of the up-type Higgs mass. Diagrams contributing to the evolution of
  the other soft scalar masses may be drawn analogously.}}
\label{3higg}
\end{figure}
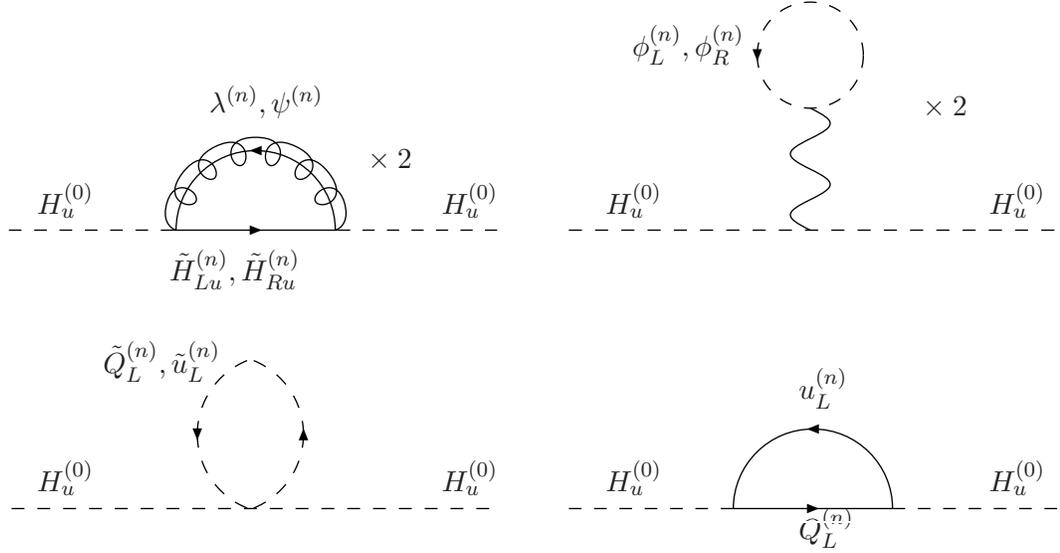
\end{center}
\normalsize

\section{Special numerical features of  RG running in  5d scenario}
\label{numerical}
In this section, we highlight the special features of RG evolution in the 5d
scenario. We also compare and contrast them with the 4d features.  For all our
numerical estimates we have fixed $1/R = 1$ TeV.

\subsection{The Gauge and Yukawa couplings}
The power law running of the gauge and Yukawa couplings has been discussed in
\cite{Dienes:1998vg,blitz} for the non-supersymmetric scenario and in
\cite{Dienes:1998vg} for the supersymmetric case. As far as the Higgs
multiplets are concerned, there is a crucial difference between our model and
that considered in \cite{Dienes:1998vg}.  In our scenario there are separate
up- and down-type Higgs hypermultiplets - see Eq.~\ref{huhd-hyper}.  Inside
each hypermultiplet only the left column with label ($L$) is $\mathbb{Z}_2$ even and
its scalar zero mode receives a vev, whereas the right column with label ($R$)
is projected out.  In other words, the hypermultiplet $H_u$ contains the vev
$v_u$ and, similarly, $H_d$ contains $v_d$.  On the other hand,
\cite{Dienes:1998vg} contains a single hypermultiplet, each column of which
has a zero mode, one to be identified with the up-type chiral multiplet which
contains the vev $v_u$, and the other to be identified with the down-type
containing $v_d$. While our approach constitutes a straightforward
generalization of $H_u$ and $H_d$ from chiral multiplets to hypermultiplets,
the choice made in \cite{Dienes:1998vg} requires non-trivial boundary
conditions.  These two different assumptions lead to significant numerical
differences.  In our approach, the gauge couplings converge to one another but
actually do not meet at a single point, while in \cite{Dienes:1998vg} the
gauge couplings do meet at a point. The difference in the number of KK scalar
excitations makes the difference between the two approaches. 

Indeed, both gauge and Yukawa couplings exhibit power law running due to
summation over the KK states as one crosses the energy thresholds.  As we have
mentioned in section \ref{5dmssm}, keeping the first two matter generations
confined at the brane ensures that the couplings remain perturbative even with
$R^{-1}$ as low as 1 TeV.  Starting from their LEP-measured values at the weak
scale, as we extrapolate the gauge couplings using the KK beta functions we
observe that the three couplings approach very close to one another near 32
TeV, but they do not actually meet at any point, as mentioned in the previous
paragraph.

A crucial point of immense numerical significance is that on account of the
special $N=2$ bulk non-renormalization, the third generation matter
hypermultiplet kept in the bulk does not receive any wave-function
renormalization from the gauge hypermultiplet, which we have illustrated below
Eq.~\ref{nonrenorm}.  As an important consequence of this, the Yukawa
couplings blow up to large (non-perturbative) values around $\Lambda \sim$ 18
TeV, which we therefore take to be the effective cutoff of our theory.

\subsection{The gaugino and scalar masses}
We assume that at the highest scale $\Lambda = 18$ TeV of our theory,
i.e. just before the Yukawa couplings blow up, all scalar masses unify to
$m_0$ and all gaugino masses to $M_{1/2}$. Our high scale parameters are then
$m_0, M_{1/2}, \mbox{sgn}(\mu)$ and $\tan\beta \equiv v_u/v_d$.

\begin{center}
  \begin{figure}[t]
\centering \includegraphics[width=.44\textwidth,angle
  =270,keepaspectratio]{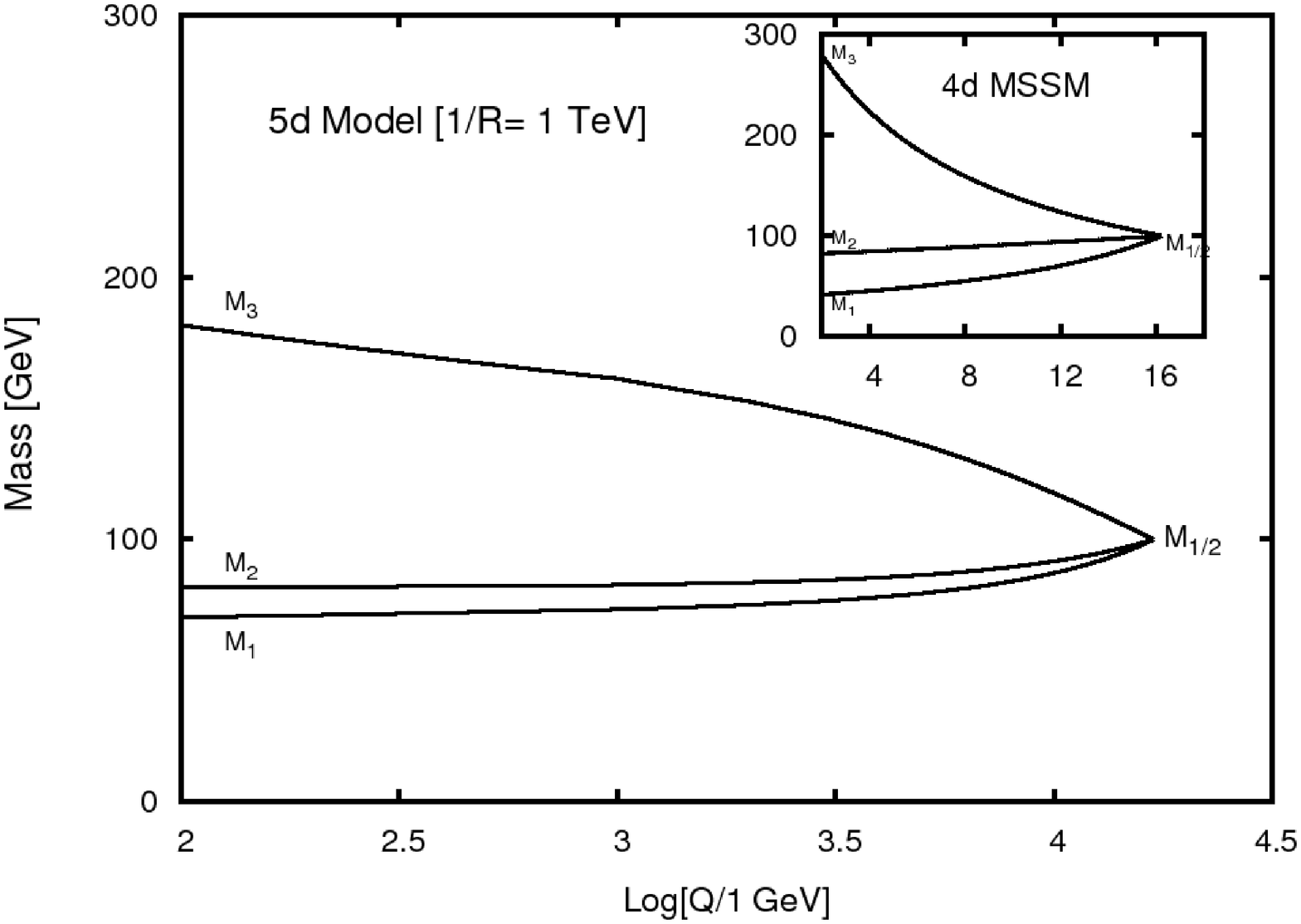}

\caption {\small RG running of the gaugino masses.}
\label{gauge}
\end{figure}
\end{center}

The gaugino mass running is governed by the evolution of gauge
couplings. Since gauge couplings {\em nearly} meet around 32 TeV, the gaugino
masses tend to converge also at that scale.  But in the present context, as
mentioned before, we forced the gaugino masses to unify at 18 TeV.  Recall
that in 5d the running is short but fast (power law), but in 4d it is long and
slow (logarithmic). This leads to a general expectation that, starting from a
given high scale value, the low scale predictions would be similar in 4d and
5d. But since we forcibly unified the gaugino masses in our set-up, earlier
than otherwise expected, we obtain a somewhat different set of low scale
values. The gaugino mass scaling in 5d is shown in Figure~\ref{gauge}, while in
the inset, the 4d running is displayed.  A rough comparison of the weak scale
ratios of the three gaugino masses is presented below:
\begin{eqnarray}
 M_1,M_2,M_3 &\sim& (0.4,0.8,3.0)\times M_{1/2}~ \mbox{(in 4d)}\, , \nonumber
  \\ M_1,M_2,M_3 &\sim& (0.7,0.8,2.0)\times M_{1/2}~
  \mbox{(in 5d)} \, .
\label{gauginom}
 \end{eqnarray}
 If $R$-parity remains conserved, the lightest neutralino remains the lightest
 supersymmetric particle (LSP), only that its mass is heavier than what is
 expected in the standard 4d scenario - see Eq.~\ref{gauginom}.

\begin{center}
 \begin{figure}[h]
\centering
 \includegraphics[width=.44\textwidth,angle =270,keepaspectratio]{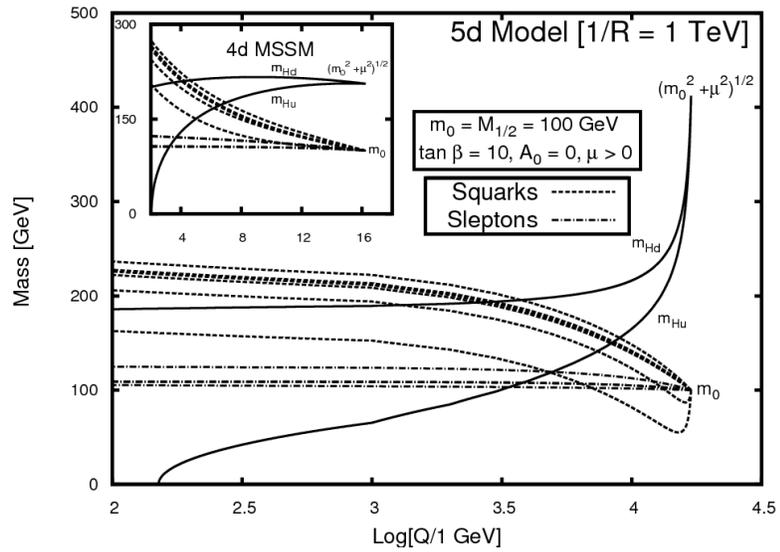}
 \caption{\small RG running of the scalar masses and radiative electroweak
   symmetry breaking.}
\label{rewsb}
\end{figure}
\end{center}

 Figure \ref{rewsb} shows the running of the soft scalar masses.  The large
 top quark Yukawa coupling continues to play a crucial r\^ole as in 4d. A
 rough comparison of the weak scale predictions in 4d and 5d is:
\begin{eqnarray}
m_{\tilde{Q}_3}^2&\sim& m_0^2 + 5.5 M_{{1}/{2}}^2~ \mbox{(in 4d)} \, , 
\nonumber\\  m_{\tilde{Q}_3}^2&\sim& m_0^2 +
3.5 M_{{1}/{2}}^2~ \mbox{(in 5d).}
\label{scalarm}
\end{eqnarray}
Even for the brane localized scalars, the 5d model predicts slightly higher
weak scale masses compared to 4d. 

During power law running we ensure that radiative breaking of electroweak
symmetry does happen at the desired scale\footnote{Radiative electroweak
  symmetry breaking has been discussed in the context of some specific
  realization of supersymmetry breaking in an orbifold
  \cite{Barbieri:2000vh,ArkaniHamed:2001mi}.}. Just like in 4d, only
$m_{H_u}^2$ turns negative while all other scalars remain positive. Again, the
large top quark Yukawa coupling drives this phase transition. A point to note
is that {\em unless} we keep the third generation matter in the bulk,
electroweak symmetry would never break radiatively in our class of models.

\section{The $m_0 -M_{1/2}$ parameter space}
\label{m0mhalf}

\subsection{Numerical procedure} 
For our numerical estimates we go through the following steps:
\begin{enumerate}
\item We scan $m_0$ and $M_{1/2}$ over a range $[0.1-1.0]/R$. We choose
  $\tan\beta = 10$ and take both positive and negative values of $\mu$.  We
  use one loop RG equations as displayed in section \ref{beta}.
\item For each input combination, we perform a consistency check to ensure
  correct electroweak symmetry breaking, and accept only those inputs which
  admit this phenomenon.
\item We then feed the weak scale spectrum into the code {\sf micrOMEGAS}
  \cite{Belanger:2001fz}, and using this software package calculate the dark
  matter density ($\Omega_{\rm DM}$), ${\rm Br}~(b \rightarrow s \gamma)$,
  $\Delta a_\mu = (g-2)_\mu/2$, and $\Delta \rho$. Since we consider $1/R = 1$
  TeV, which is a bit too high compared to the lighter section of the zero
  mode superparticle spectrum, we neglect the direct loop contributions of the
  KK particles. In other words, the KK effects feed into the calculation of
  low energy spectra {\em via} power law running, but after that we rely on
  the standard 4d computations encoded in {\sf micrOMEGAS}. This approximation
  is good enough for our purpose.
\item We compare the predictions of the above observables with their
  experimental values/constraints, and translate the information into the
  inclusion/exclusion plots, given in Figures \ref{momh1} and \ref{momh2} in the $m_0
  -M_{1/2}$ plane. The 4d plots have been reproduced to serve as a guide to
  the eyes for capturing the 5d subtleties. We note that our 4d plots are in
  agreement with the ones in the existing literature, e.g. with
  \cite{Djouadi:2006be}.
\end{enumerate}

\subsection{Comparison between 4d and 5d models}
We highlight only the major differences between the 4d and 5d models that
appear in the $m_0$--$M_{1/2}$ plane.

\begin{center}
\begin{figure}
\begin{center}
\includegraphics[width=0.5\textwidth,angle=270,keepaspectratio]
{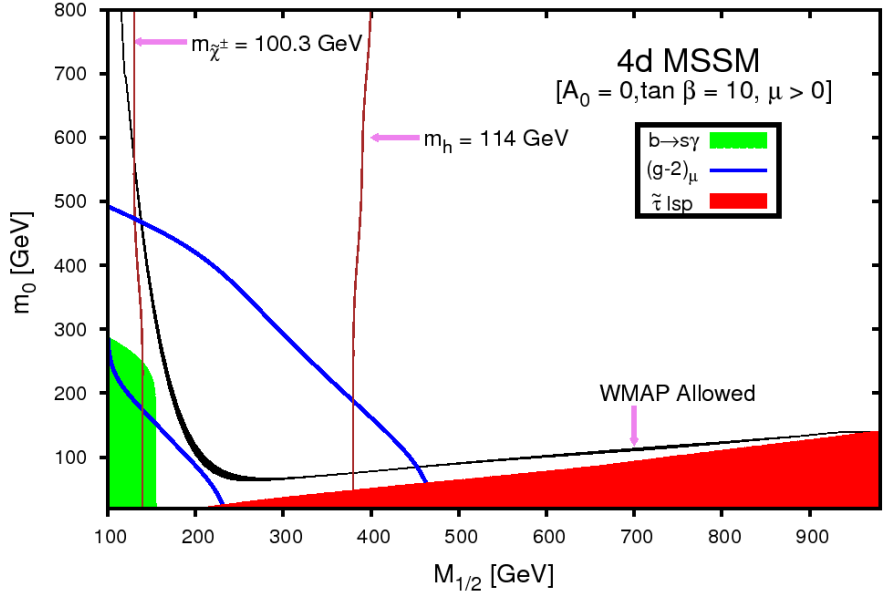}
\centering{\hspace{16mm} (a)}
\includegraphics[width=0.5\textwidth,angle=270,keepaspectratio]
{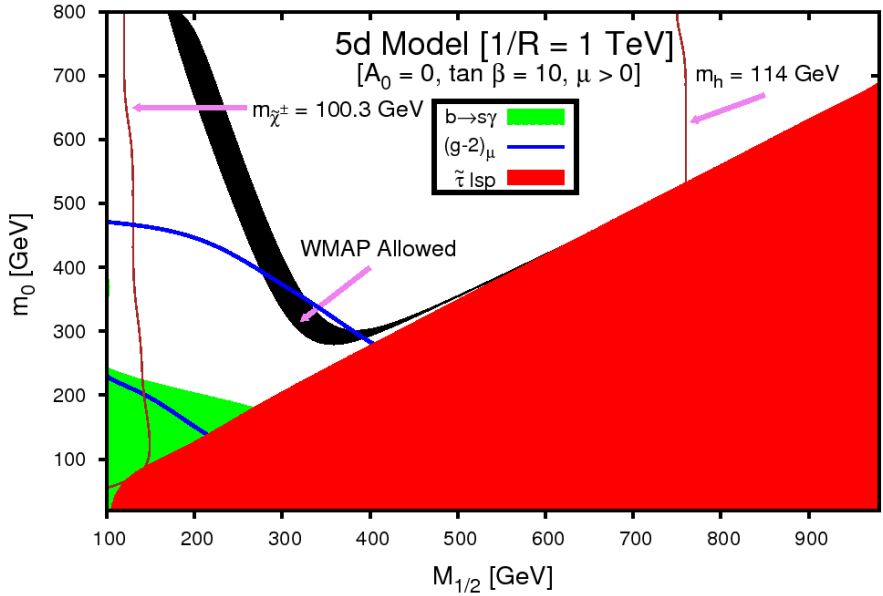}
\centering{\hspace{16mm} (b)}
\end{center}
\caption{ \small The $m_0-M_{1/2}$ parameter space for $\mu>0$ . We keep
  $tan \beta =10$ for all the plots. The region ruled out by $B(b \rightarrow
  s \gamma)$ is shaded in light green (lightest shade) ,the $\tilde{\tau}$ LSP
  region is shaded in red (darker shade) and the region favored by
  $(g-2)_{\mu}$ is the region between the two blue (darkest shade) lines. The
  WMAP allowed region where $.087<\Omega_{DM} h^2<.138$ is shaded in black. We
  also show the contours for $m_h =114 GeV$ and $m_{\tilde{\chi}^{\pm}} =103.3
  GeV$, the region to the left of these lines are ruled out by the lep
  exclusion limits. For the 5d models, the Higgs contour shown does not
  include the KK contribution.}
\label{momh1}
\end{figure}
\end{center}

\begin{figure}
\hspace{7mm}
\begin{center}
\includegraphics[width=0.5\textwidth,angle=270,keepaspectratio]
{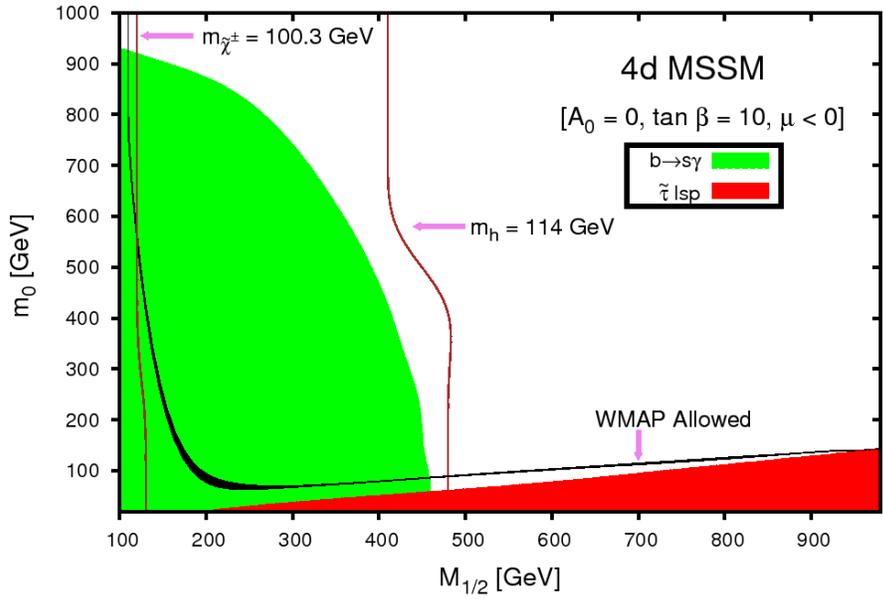}
\centering{\hspace{16mm} (a)}
\includegraphics[width=0.5\textwidth,angle=270,keepaspectratio]
{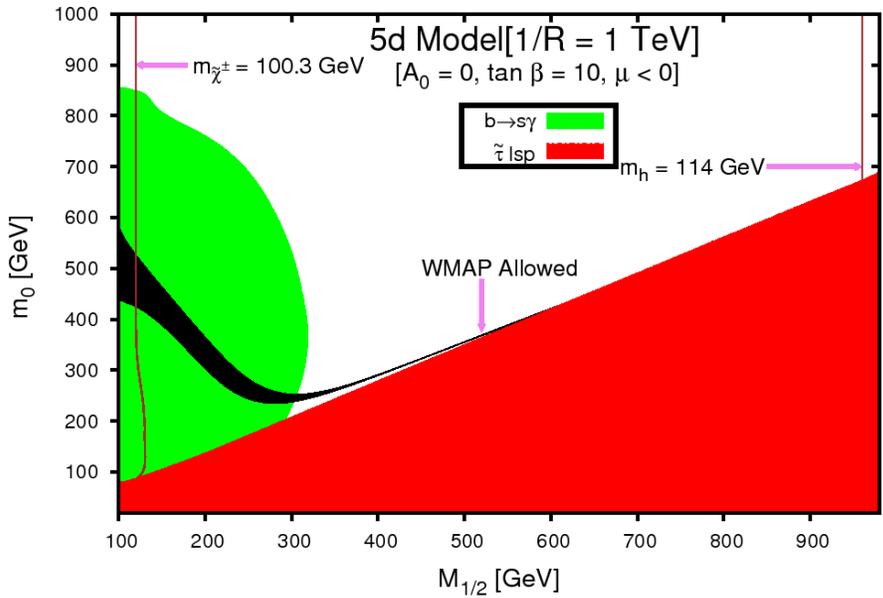}
\centering{\hspace{16mm} (b)}

\caption{ \small Same as Figure \ref{momh1} for $\mu<0$. The
          entire region is now disfavored by the $(g-2)_{\mu}$. }
\label{momh2}
\end{center}
\end{figure}

\begin{enumerate} 
\item We assume that $R$-parity is conserved.  In the 4d scenario the lightest
  neutralino is the most likely candidate for an LSP.  In the 5d model the
  situation is somewhat tricky. Indeed, the 4d LSP is still an LSP here which
  is the zero mode neutralino. Besides, {\em if} the KK parity remains
  conserved, then the $n=1$ mode of photon tower, namely $\gamma_1$, and its
  supersymmetric partner $\tilde{\gamma}_1$ are also stable dark matter
  candidates. However, the KK parity is unlikely to be respected by the
  brane-bulk interaction. In our numerical analysis, we have treated the zero
  mode LSP as the dark matter candidate.  

\item We have taken a $3 \sigma$ range of the five year average of WMAP dark
  matter density ($0.087<\Omega_{\rm DM} h^2<0.138$) \cite{Komatsu:2008hk}. We
  raise a caution here that {\em if} KK parity remains conserved and we have
  two {\em more} dark matter candidates, as mentioned above, then the edge of
  the allowed band arising from the lower limit of $\Omega_{\rm DM}$ would be
  further stretched. Note further that in the 5d case there is a slight
  broadening of the WMAP allowed strip compared to 4d. This happens because of
  a combined effect of Eqs.~\ref{gauginom} and \ref{scalarm} leading to a
  reduced sensitivity to $M_{1/2}$ variation. 

\item As a consequence of Eq.~\ref{gauginom}, to arrive at a given value of
  $M_1$, one needs to start from a {\em smaller} $M_{1/2}$ in 5d compared to
  4d.  For this reason, the region where the lightest neutralino satisfies the
  dark matter constraints extends to a {\em lower} value of $M_{1/2}$ in 5d
  compared to 4d.

\item We have taken $2.65 \times 10^{-4} \lesssim {\rm Br}~(b \to s \gamma)
  \lesssim 4.45 \times 10^{-4}$ \cite{Barberio:2008fa}, and $10.6\times
  10^{-10}\lesssim \Delta a_{\mu}^{\rm new} = (g-2)_{\mu}/2 \lesssim
  43.6\times 10^{-10}$ \cite{Carey:2009zz}. There is nothing much to
  distinguish between 4d and 5d from these two observables.

\item We have {\em not} included the {\em direct} loop effects of the virtual
  KK states for any of the weak scale observables. For $R^{-1} = 1$ TeV or
  more, for processes like muon anomalous magnetic moment or $b\to s \gamma$,
  such effects are numerically negligible, but {\em only } for the Higgs mass
  it makes a difference. In Figures.~\ref{momh1} and \ref{momh2}, the entire
  region to the left of the line marked with $m_h = 114$ GeV is disfavored
  from the non-observation of the Higgs boson. However, if we include the KK
  loop correction to the Higgs mass \cite{Bhattacharyya:2007te}, the entire
  {\em hitherto} disfavored region is resurrected.

\item To ensure correct electroweak symmetry breaking, we had to take a factor
  2 to 3 larger (than 4d) value of $\mu$ in 5d at the cutoff scale.  Otherwise
  $m_{H_u}^2$ would become negative at a scale higher than required, thanks
  again to the bulk $N=2$ non-renormalization.
\end{enumerate}

\section{Conclusions and Outlook}
\label{concl} 

We reiterate that the presence of extra dimensions is an essential part of any
high scale fundamental theory, and supersymmetry is quite often an integral
component of such theories. Furthermore, extra dimension may trigger
supersymmetry breaking.  Be it a Scherk-Schwarz mechanism, or a breaking
triggered by a spurion $F$-term vev, or due to compactification on the
orbifold $S^1/(\mathbb{Z}_2 \times \mathbb{Z}_2^\prime)$, or for that matter any top-down
scenario that contains supersymmetry, would find a common ground in our
phenomenological model where we varied $m_0$ and $M_{1/2}$ in a reasonable
range $[0.1-1.0]R^{-1}$.

The logarithmic running in 4d from 100 GeV to $10^{16}$ GeV is replaced in 5d
by fast power law running on a shorter interval from 100 GeV to about 30 TeV
in 5d thanks to the KK states. This has nothing to do with supersymmetry. What
is special about 5d supersymmetry is a special $N=2$ non-renormalization that
forces us to consider an early cutoff ($\sim$ 18 TeV).

The constraints in the $m_0$--$M_{1/2}$ plane have been placed for the {\em
  first time} in this work. The ratio $M_1/M_{1/2}$ is higher in 5d compared
to 4d. For this reason the {\em allowed} region in the 5d plot extends to {\em
  lower} values of $M_{1/2}$ compared to the 4d plot. 

Two issues require further studies that is beyond the scope of this thesis: (i)
Besides the lightest neutralino (the usual 4d LSP), there are two other
candidates of dark matter in this model. One is $\gamma_1$, the $n=1$ level
photon, and the other is its superpartner $\tilde{\gamma}_1$.  Both are stable
dark matter candidates if KK parity remains conserved. The {\em combined
  effects} of all three candidates need to be investigated. It will also be
interesting to revisit the lower limit on $R^{-1}$ in a supersymmetric
scenario, which we suspect would be relaxed.

\newpage

\newpage

\newpage

%% file: texfiles/con
We have entered the LHC era with confidence in our belief in the existence of
a theory beyond the standard model.  We already have experimental evidences in
its favour from the electro-weak sector, \emph{like} the exixtance neutrino
mass.  Expectations are mounting that the LHC will discover some of these new
physics, turning decades of speculation into experimental realities. Certainly
it will pave the way for future research in this field.

 New physics of different incarnations, especially supersymmetry and/or extra
dimensions, are crying out for validation. And the LHC is expected to sit on
judgement on them.  At this crucial juncture in the development of particle
physics, we consider it relevant to study the phenomenology of these scenarios
that are testable at the collider experiments already underway (like the LHC)
or is at the planning stage (like the ILC). In this thesis we have studied the
formal and phenomenological aspects of supersymmetry, extra dimension and
their interface.

In the second chapter of this thesis we discuss the impact of warped extra
dimension on the processes $gg \to h$ and $h \to \gamma \gamma$, that are of
paramount importance in the context of Higgs search at the LHC.  These
processes are loop driven and hence could be sensitive to the presence of any
new colored fermion states having a large coupling with the Higgs. We confine
Higgs field to the TeV brane and the hierarchy of fermion masses is addressed
by localizing them at different positions in the bulk. We show that the Yukawa
coupling of the Higgs with the fermion Kaluza-Klein (KK) states can be order
one irrespective of their zero mode masses. We observe that the $gg \to h$ and
$h \to \gamma \gamma$ rates are substantially altered if the KK states lie
within the reach of LHC.  We found that inspite of completely different
reasons, the numerical impact of the RS model is comparable to the UED
scenario.

In chapter three we compute radiative correction to the lightest neutral Higgs
mass ($m_h$) induced by the Kaluza-Klein (KK) towers of fermions and sfermions
in a minimal supersymmetric scenario embeded in a 5-dimensional warped
space. The Higgs is again confined to the TeV brane, providing a handle to
address the $\mu$ problem. The KK spectrum of matter supermultiplets is tied
to the explanation of the fermion mass hierarchy . We demonstrate that for a
reasonable choice of extra-dimensional parameters, the KK-induced radiative
correction can enhance the upper limit on $m_h$ by as much as 100 GeV beyond
the 4d limit of 135 GeV. Here the impact is still significant but more modest
as compared to UED scenario, considering the more restrictive constraints on
the RS scenario comming from precision electroweak observables.

In the fourth chapter of this thesis we studied the running of the soft
parameters and the couplings of the minimal supersymmetric standard model
embedded in a flat extra dimension compactified on an $S_1 / \mathbb{Z}_2$
orbifold. In order to keep the theory perturbative at all scales we restricted
the first two generations of fermions to the 3-branes, allowing all other
fields to access the extra dimension. We computed the contributions of the
Kaluza-Klein (KK) towers to the various one-loop $\beta$ functions.  We
demonstrated that radiative electroweak symmetry breaking can be achieved in
this scenario.  We also put constraints on the $m_{0} - M_{{1}/{2}}$ plane of
the theory from various theoretical considerations and experimental
observations. We have incorporated constraints coming from direct LEP search
for supersymmetric particles, Higgs mass limit, anomalous magnetic moment of
the muon $(g-2)_{\mu}$, the $\rho$ parameter, branching ratio of $b
\rightarrow s \gamma $ and WMAP probe for relative dark matter abundance. Our
plots are the first 5d versions of the often displayed 4d $m_0$--$M_{1/2}$
plots. We also study the reasons behind the differences between the 4d and 5d
plots which arises mainly from the effect of the supersymmetric ($N=1$ \&
$N=2$) non-renormalization theorems.